\pdfoutput=1
\documentclass[11pt,a4paper]{article}
\usepackage[margin=2.3cm]{geometry}
\usepackage{amsmath,amssymb,amsthm,mathtools,bm}
\usepackage{booktabs,longtable,enumitem,float,graphicx,array}
\usepackage{etoolbox}
\usepackage[dvipsnames]{xcolor}
\usepackage[most]{tcolorbox}
\usepackage[colorlinks=true,linkcolor=NavyBlue,citecolor=ForestGreen,urlcolor=NavyBlue,bookmarksnumbered=true]{hyperref}
\usepackage[capitalise,nameinlink]{cleveref}
\usepackage[expansion=false]{microtype}

\numberwithin{equation}{section}
\allowdisplaybreaks
\makeatletter
\def\@fnsymbol#1{\ensuremath{\ifcase#1\or *\or \dagger\or \ddagger\or \mathsection\or \mathparagraph\or \|\or **\or \dagger\dagger\or \ddagger\ddagger\else\@ctrerr\fi}}
\makeatother

\pretocmd{\section}{\clearpage}{}{}

\newtheorem{lemma}{Lemma}[section]
\newtheorem{proposition}[lemma]{Proposition}
\newtheorem{definition}[lemma]{Definition}
\newtheorem{remark}[lemma]{Remark}

\tcbset{boxsty/.style={colback=NavyBlue!3,colframe=NavyBlue!60,boxrule=0.6pt,arc=2pt,left=4pt,right=4pt,top=3pt,bottom=3pt,breakable}}
\newtcolorbox{resultbox}[1][]{boxsty,title={#1},fonttitle=\bfseries\small,coltitle=white,colbacktitle=NavyBlue!70}
\newtcolorbox{notebox}[1][]{colback=Goldenrod!6,colframe=Goldenrod!70!black,boxrule=0.5pt,arc=2pt,left=4pt,right=4pt,top=3pt,bottom=3pt,breakable,title={#1},fonttitle=\bfseries\small,coltitle=black,colbacktitle=Goldenrod!25}

\newcommand{\bv}{\bm{v}}
\newcommand{\bu}{\bm{u}}
\newcommand{\bg}{\bm{g}}
\newcommand{\br}{\bm{r}}
\newcommand{\bk}{\bm{k}}
\newcommand{\bx}{\bm{x}}
\newcommand{\by}{\bm{y}}
\newcommand{\bJ}{\bm{J}}
\newcommand{\bQ}{\bm{Q}}
\newcommand{\bJq}{\bm{J}^{q}}
\newcommand{\bJqp}{\bm{J}^{q\prime}}
\newcommand{\btau}{\bm{\tau}}
\newcommand{\bPi}{\bm{\Pi}}
\newcommand{\bW}{\bm{W}}
\newcommand{\bD}{\bm{D}}
\newcommand{\bX}{\bm{X}}
\newcommand{\bY}{\bm{Y}}
\newcommand{\Id}{\mathbb{I}}
\newcommand{\kB}{k_{B}}
\newcommand{\mut}{\tilde{\mu}}
\newcommand{\mub}{\bar{\mu}}
\newcommand{\Lop}{\mathsf{L}}
\newcommand{\Mop}{\mathsf{M}}
\newcommand{\Dop}{\mathsf{D}}
\newcommand{\Cop}{\mathsf{C}}
\newcommand{\Bop}{\mathsf{B}}
\newcommand{\Phiop}{\bm{\Phi}}
\newcommand{\fd}[2]{\frac{\delta #1}{\delta #2}}
\newcommand{\pd}[2]{\frac{\partial #1}{\partial #2}}
\newcommand{\Dt}[1]{\frac{D #1}{D t}}
\newcommand{\divv}{\nabla\!\cdot\!}
\newcommand{\tr}{\operatorname{tr}}
\newcommand{\sym}{\mathrm{s}}
\newcommand{\noise}[1]{\widetilde{#1}}
\newcommand{\eN}{e_{\nabla}}
\newcommand{\sN}{s_{\nabla}}
\newcommand{\fN}{f_{\nabla}}
\newcommand{\EN}{E_{\nabla}}
\newcommand{\SN}{S_{\nabla}}
\newcommand{\FN}{F_{\nabla}}
\newcommand{\dd}{\mathrm{d}}
\newcommand{\avg}[1]{\left\langle #1\right\rangle}

\newcommand{\hs}{\hat{s}}
\newcommand{\hu}{\hat{u}}
\newcommand{\stot}{s_{\rm tot}}
\newcommand{\etot}{\varepsilon_{\rm tot}}
\newcommand{\hQ}{\hat{\bm Q}}
\newcommand{\hQp}{\hat{\bm Q}'}
\newcommand{\hg}{\hat g}
\newcommand{\cdt}{\dot c}
\newcommand{\Pgr}{\Psi_{\nabla\rho}}
\newcommand{\Pgc}{\Psi_{\nabla c}}
\newcommand{\PiB}{\bPi^{\rm B}}
\newcommand{\PiA}{\bPi^{\rm A}}
\newcommand{\PiM}{\bPi^{\rm M}}
\newcommand{\mutB}{\mut_{\rm B}}
\newcommand{\mutA}{\mut_{\rm A}}
\newcommand{\mutM}{\mut_{\rm M}}
\newcommand{\cs}{c_s}
\newcommand{\dvh}{\Delta\hat v}

\title{\textbf{A thermodynamically consistent, fully compressible, fluctuating hydrodynamics model of a Cahn--Hilliard binary fluid mixture}}
\author{Anuj Chaudhri\thanks{E-mail: \href{mailto:chaudhri@alumni.upenn.edu}{\nolinkurl{chaudhri@alumni.upenn.edu}}}\\[4pt]
\small BASIS Independent Fremont, 39706 Mission Blvd, Fremont, CA 94539, USA}
\date{\today}

\begin{document}
\maketitle
\begin{abstract}
\noindent
Diffuse-interface models of the Cahn--Hilliard--Navier--Stokes (CHNS) type are now a standard tool for simulating phase-separating binary fluids. At the mesoscopic scales where interfaces are only a few nanometres to micrometres wide, thermal fluctuations are not a small correction. They nucleate droplets, roughen interfaces, drive capillary waves and accelerate coarsening, and they are represented in Landau--Lifshitz fluctuating hydrodynamics by stochastic fluxes whose strength must be fixed by the fluctuation--dissipation theorem (FDT). Most existing fluctuating CHNS models are incompressible or isothermal, treat the density and the composition variables inconsistently, or add noise to deterministic equations without proving that the result samples the correct equilibrium distribution. The aim of this paper is to derive, from first principles and in a single framework, a complete and thermodynamically consistent set of \emph{fully compressible} fluctuating hydrodynamic equations for a Cahn--Hilliard binary mixture, in both its isothermal and its non-isothermal form, and for both symmetric and non-symmetric mixtures.

The derivation uses the GENERIC formalism of Grmela and \"Ottinger, in which reversible dynamics is generated by the energy through a Poisson operator and irreversible dynamics by the entropy through a friction operator. Following Jeli\'c and \"Ottinger, the friction operator is not postulated but constructed from the fluctuations of the fluxes: one specifies how currents change the slow variables and the second moments of the fast current fluctuations, and evaluates the friction operator through a Green--Kubo formula. In this construction the deterministic dissipative fluxes and the stochastic fluxes are images of the same linear operator and the same correlation matrix, so the FDT holds by construction and every noise amplitude follows directly from a transport coefficient.

We first settle the meaning of the variables. The density in the momentum equation is the total mass density, the Cahn--Hilliard variable is the mass fraction, the conserved quantity is the partial density, and the diffusion flux must be driven by the specific, not the volumetric, exchange chemical potential. We then derive a generalized Korteweg tensor valid for any isotropic square-gradient energy in both density and composition. The isothermal model is generated by the Helmholtz free energy. The non-isothermal model is generated by energy and entropy functionals that include gradient energy and gradient entropy. Its reversible pressure tensor is fixed uniquely by the degeneracy condition, which produces a thermal capillary stress whenever the interfacial free energy has an entropic part. Thermodynamic consistency is proved in full: antisymmetry and the Jacobi identity of the Poisson operator, symmetry and positivity of the friction operator, both degeneracy conditions, local energy conservation including interstitial-working fluxes, a local entropy balance with non-negative production, Soret--Dufour reciprocity, and the Einstein statistics of the linearized stochastic equations.

The final equations are written in conservative flux form, ready for finite-volume discretization. We also derive closed-form static, intermediate and dynamic structure factors for all four models, including Landau--Placzek and Rayleigh--Brillouin spectra, as analytic benchmarks for numerical codes. An appendix shows that the same models follow from classical irreversible thermodynamics, using only the Helmholtz or Gibbs free energy and the entropy production. That route is unique for isothermal flows and admits a gauge freedom in the non-isothermal case, which GENERIC removes. All identities were verified with computer algebra.
\end{abstract}

\clearpage
\tableofcontents

\section{Introduction and background}\label{sec:intro}

\subsection{Motivation}\label{sec:motivation}
When a binary liquid mixture is quenched into its miscibility gap it separates into two coexisting phases. The domains coarsen, and the interfaces between them carry a surface tension that couples back to the flow. Diffuse-interface (phase-field) models describe this process by a smooth order parameter whose free energy contains a square-gradient term, as introduced by van der Waals and by Cahn and Hilliard \cite{CahnHilliard1958,RowlinsonWidom}. Coupled to the Navier--Stokes equations through a reversible capillary (Korteweg) stress, they give the Cahn--Hilliard--Navier--Stokes (CHNS) equations, now a standard tool for multiphase flow with topological changes \cite{AndersonMcFaddenWheeler1998,PadhanPandit2025}.

Two features of real mixtures are often idealized away in CHNS models, and both matter at the mesoscopic scales where diffuse-interface models are most natural.
\begin{itemize}[leftmargin=*]
\item \textbf{Compressibility and energy transport.} The components of a mixture generally have different densities. Demixing then changes the local density and releases or absorbs latent heat. Sound waves, the piston effect and thermocapillary (Marangoni) stresses all involve the coupling between composition, density and temperature. A model that is to capture them must be fully compressible and, in general, non-isothermal.
\item \textbf{Thermal fluctuations.} On scales of nanometres to micrometres, thermal noise is not a small correction. It nucleates droplets, roughens interfaces, excites capillary waves, renormalizes transport coefficients near the critical point and seeds spinodal decomposition. Landau and Lifshitz showed how to include it in hydrodynamics through stochastic fluxes \cite{LandauLifshitzFM,deZarateSengers2006}. The strength of these fluxes is not free: it is fixed by the fluctuation--dissipation theorem (FDT), which guarantees that the equations sample the correct equilibrium (Gibbs--Boltzmann or Einstein) distribution.
\end{itemize}
This paper derives, from first principles and in a single framework, the fully compressible fluctuating hydrodynamic equations of a Cahn--Hilliard binary mixture. We call these equations the Landau--Lifshitz--Cahn--Hilliard--Navier--Stokes (LLCHNS) equations. We derive them for the isothermal and the non-isothermal case, each for a symmetric and a non-symmetric mixture. Thermodynamic consistency, including the FDT, is established completely.

\subsection{Background}\label{sec:background}
\paragraph{Deterministic diffuse-interface hydrodynamics.} The coupling of a conserved order parameter to an incompressible flow is the ``model H'' of dynamic critical phenomena \cite{HohenbergHalperin1977}. For phase-separating fluids it was developed into the incompressible CHNS model, reviewed by Anderson, McFadden and Wheeler \cite{AndersonMcFaddenWheeler1998}. When the two components have different densities the barycentric velocity is no longer solenoidal. Lowengrub and Truskinovsky \cite{LowengrubTruskinovsky1998} derived a quasi-incompressible model in which the pressure enters the chemical potential. Gurtin, Polignone and Vi\~nals \cite{GurtinPolignoneVinals1996} and Abels, Garcke and Gr\"un \cite{AbelsGarckeGrun2012} instead use a volume-averaged velocity that remains solenoidal. Heida, M\'alek and Rajagopal \cite{HeidaMalekRajagopal2012} derived compressible and non-isothermal CHNS--Fourier systems from the maximization of entropy production and identified a new energy flux of interstitial-working type \cite{DunnSerrin1985}. Cherfils \emph{et al.} \cite{Cherfils2019} analysed a compressible isothermal NSCH system mathematically. A recent review by Padhan and Pandit \cite{PadhanPandit2025} surveys laminar, turbulent and active CHNS flows. None of these works includes thermal fluctuations.

\paragraph{Fluctuating hydrodynamics of phase-separating fluids.} Cook \cite{Cook1970} added a conserved noise to the Cahn--Hilliard equation to describe the early stage of spinodal decomposition. For the full hydrodynamic problem, three lines of work are relevant here.
\begin{itemize}[leftmargin=*]
\item \emph{Microscopic and mesoscopic derivations.} Espa\~nol derived the thermohydrodynamics of a van der Waals fluid \cite{Espanol2001}. Espa\~nol and Thieulot derived the hydrodynamic equations of phase-separating mixtures from the microscopic dynamics by projection operators \cite{EspanolThieulot2003}. Onuki formulated a dynamic van der Waals theory with gradient entropy, including two-phase flow in a heat flux \cite{Onuki2005,Onuki2007}.
\item \emph{Numerical fluctuating hydrodynamics.} Donev and co-workers analysed the accuracy of finite-volume schemes for stochastic conservation laws \cite{Donev2010}. They developed low-Mach-number models for diffusively mixing fluids \cite{Donev2014} and linked fluctuating hydrodynamics to dynamic density-functional theory \cite{DonevVandenEijnden2014}. Chaudhri, Bell, Garcia and Donev \cite{ChaudhriPRE2014} formulated the fully compressible fluctuating hydrodynamics of a single-component van der Waals fluid with a Korteweg stress. They solved it with a conservative finite-volume scheme and validated it against static structure factors and capillary-wave spectra.
\item \emph{GENERIC.} The GENERIC formalism of Grmela and \"Ottinger \cite{GrmelaOttinger1997,OttingerGrmela1997,Ottinger2005} writes any thermodynamically consistent evolution as the sum of an energy-generated reversible part and an entropy-generated irreversible part. Its stochastic extension builds the FDT into the structure \cite{Ottinger2005,Ottinger1998}. In her thesis, Jeli\'c \cite{JelicThesis2009} gave a GENERIC formulation of the incompressible Cahn--Hilliard fluid. She also developed, with \"Ottinger, a systematic construction of the friction operator from the fluctuations of fluxes. In unpublished notes \cite{ChaudhriNotes2012}, the present author combined classical irreversible thermodynamics and GENERIC to formulate fluctuating CHNS and van der Waals--Navier--Stokes models. Those notes are the starting point of the present paper, which completes and, in several places, corrects them.
\end{itemize}

\subsection{What is missing}\label{sec:gaps}
Taken together, these works leave several gaps that become essential once a compressible, fluctuating, two-component model is to be simulated.
\begin{enumerate}[leftmargin=*,label=(\roman*)]
\item \textbf{The meaning of the variables.} In a compressible mixture, ``the density'' and ``the concentration'' can refer to several different quantities: total or partial density, mass, mole or volume fraction, barycentric or volume-averaged velocity. Different choices lead to different equations. In particular, driving the diffusion flux by the \emph{volumetric} derivative $\delta F/\delta c$ instead of the \emph{specific} potential $\rho^{-1}\delta F/\delta c$ is harmless for incompressible flow. For compressible flow it destroys the sign of the entropy production and violates the FDT.
\item \textbf{Fully compressible fluctuating CHNS equations with a proof of the FDT.} Existing fluctuating models are either incompressible, single-component \cite{ChaudhriPRE2014}, or postulate the noise without proving that the nonlinear compressible dynamics has the correct equilibrium statistics.
\item \textbf{A non-isothermal fluctuating model.} Such a model needs gradient energy \emph{and} gradient entropy, the correct reversible pressure tensor, interstitial-working energy fluxes, Soret--Dufour cross effects and correlated heat and mass noises. These ingredients have appeared separately \cite{JelicThesis2009,Onuki2007,HeidaMalekRajagopal2012}, but not in one consistent compressible, stochastic model.
\item \textbf{Symmetric versus non-symmetric mixtures.} Barodiffusion, composition--pressure coupling, density--composition gradient coupling and thermal diffusion are absent or odd in a symmetric mixture and generic in a non-symmetric one. A systematic account of how they enter the fluctuating equations is missing.
\item \textbf{Analytic benchmarks.} Finite-volume codes for fluctuating hydrodynamics are validated against exact results for the linearized equations \cite{Donev2010,ChaudhriPRE2014}. For compressible two-component, non-isothermal models, closed-form dynamic structure factors are not available.
\item \textbf{The relation between ``pure thermodynamics'' and GENERIC.} Many diffuse-interface models are derived by classical irreversible thermodynamics (CIT): a free energy, balance laws and a non-negative entropy production. It is not obvious whether this route gives the same equations as GENERIC, in particular when the gradient coefficients depend on temperature.
\end{enumerate}

\subsection{Approach and main results}\label{sec:results}
\paragraph{Method.} The derivation follows three complementary routes.
\begin{itemize}[leftmargin=*]
\item \textbf{GENERIC} \cite{GrmelaOttinger1997,OttingerGrmela1997,Ottinger2005}, in the form developed for the Cahn--Hilliard model by Jeli\'c \cite{JelicThesis2009} (Chapter~5) and used in Section~6 of \cite{ChaudhriNotes2012}. There are four building blocks: the energy $E$, the entropy $S$, a Poisson operator $\Lop$ and a friction operator $\Mop$. We construct them for the compressible mixture and check them against every structural requirement.
\item \textbf{The Jeli\'c--\"Ottinger construction of the friction operator from fluctuations} (\cite{JelicThesis2009}, Chapters~2 and~3; \cite{Ottinger2005,Ottinger1998}). Instead of postulating $\Mop$, we postulate the \emph{structure} of the fast fluctuations of the slow variables:
\begin{itemize}[nosep]
\item divergences of fluctuating currents for conserved densities (Onsager regression);
\item energy fluctuations chosen so that the total energy is not changed (fluctuations live on the energy shell).
\end{itemize}
We then specify the second moments of the underlying Wiener increments and compute $\Mop$ from the Einstein form of the Green--Kubo formula. This makes the relation between noise and dissipation completely transparent. Both are images of the \emph{same} operator $\Phiop^{*}$ and the \emph{same} correlation matrix $\Cop$ (\cref{lem:JO}):
\begin{equation*}
\text{noise: }\ \dd\noise{\bx}=\Phiop^{*}\,\dd\bW,\qquad
\text{dissipation: }\ \Mop\cdot\fd{S}{\bx}=\Phiop^{*}\,\Cop\,\Phiop\Big(\fd{S}{\bx}\Big),\qquad
\avg{\dd\bW\,\dd\bW}=2\kB\,\Cop\,\dd t .
\end{equation*}
\item \textbf{Classical irreversible thermodynamics} (Section~4.1 of \cite{ChaudhriNotes2012}, \cite{deGrootMazur}). In \cref{app:thermo_route} we carry out the ``pure thermodynamics'' route for the most general free energy considered here, using the Helmholtz and Gibbs free energies and the entropy production. We use it as an independent cross-check of the GENERIC results and to clarify what each framework fixes.
\end{itemize}

\paragraph{Main results.}
\begin{enumerate}[leftmargin=*]
\item \textbf{Meaning of the variables} (\cref{sec:kinematics}).
\begin{itemize}[nosep]
\item $\rho=\rho_A+\rho_B$ is the total mass density.
\item $\bv$ is the barycentric (mass-averaged) velocity.
\item $c=\rho_A/\rho$ is the mass fraction.
\item $\rho c=\rho_A$ is the conserved partial density of species $A$.
\end{itemize}
The variable called $c$ in Section~7 of \cite{ChaudhriNotes2012} is $\phi=2c-1=(\rho_A-\rho_B)/\rho$. Rescalings between the two conventions are given in \cref{tab:phi_map}.
\item \textbf{Generalized Korteweg tensor} (\cref{lem:korteweg}). For any isotropic gradient energy $w(\rho,c,\nabla\rho,\nabla c)$, the reversible stress is
\[
\bPi^{w}=[\rho\,\delta W/\delta\rho-w]\Id+\nabla\rho\otimes\partial w/\partial\nabla\rho+\nabla c\otimes\partial w/\partial\nabla c .
\]
This formula contains as special cases the Korteweg stress of \cite{ChaudhriPRE2014} (Eq.~2.14), the reactive stress $\rho\kappa\nabla c\otimes\nabla c$ of \cite{ChaudhriNotes2012} (Eq.~4.16), and the stress of \cite{Cherfils2019}.
\item \textbf{Isothermal LLCHNS equations} (\cref{sec:isothermal}, \cref{res:iso_final}).
\begin{gather*}
\partial_t\rho+\divv(\rho\bv)=0,\\
\partial_t(\rho\bv)+\divv(\rho\bv\bv^{T}+p\Id)=\divv(\bPi^K+\btau+\noise{\btau}),\\
\partial_t(\rho c)+\divv(\rho c\bv)=\divv(M\nabla\mut)-\divv\noise{\bJ},\\
\mut=\pd{f}{c}+\frac1\rho\fd{F_\nabla}{c},\qquad
\avg{\noise{\bJ}\noise{\bJ}}=2\kB T M\,\Id\,\delta\delta .
\end{gather*}
\item \textbf{Non-isothermal LLCHNS equations} (\cref{sec:noniso}, \cref{res:noniso_final}). The energy equation conserves the total energy, gradient energy included, and contains two interstitial-working fluxes:
\begin{itemize}[nosep]
\item a reversible flux $\rho(\divv\bv)\,\partial\eN/\partial\nabla\rho$;
\item an irreversible flux $(\divv\bJ)\,\rho^{-1}\partial\eN/\partial\nabla c$.
\end{itemize}
The heat and diffusion fluxes satisfy Onsager reciprocity, including Soret and Dufour coupling. The noise covariances are
\[
\avg{\noise{\bJ}\noise{\bJ}}=2\kB MT,\qquad
\avg{\noise{\bJ}\noise{\bJ}^{q\prime}}=2\kB M_T T^2,\qquad
\avg{\noise{\bJ}^{q\prime}\noise{\bJ}^{q\prime}}=2\kB\hat\lambda T^2 .
\]
A gradient \emph{entropy} ($\kappa_S\neq0$) produces a thermal capillary stress, the CH analogue of the cross-coupling stress $\bPi^C$ of \cite{ChaudhriPRE2014,Onuki2007}.
\item \textbf{Symmetric versus non-symmetric mixtures} (\cref{sec:symmetry,sec:iso_sym,sec:iso_nonsym,sec:noniso_sym,sec:noniso_nonsym}). Exchange symmetry fixes the parity of every coefficient in the composition variable. In non-symmetric mixtures, barodiffusion, composition--pressure coupling and thermal diffusion appear generically. They are all contained in the single driving force $\nabla(\mut/T)$ together with the Onsager cross-coefficient.
\item \textbf{Dynamic structure factors} (\cref{sec:dsf}). For all four models, the static, intermediate and dynamic structure factors of the linearized equations are given in closed matrix form, $\bm S(k,\omega)=\bm G\bm S(k)+\bm S(k)\bm G^\dagger$, together with:
\begin{itemize}[nosep]
\item explicit Lorentzian, damped-oscillator and Rayleigh--Brillouin forms;
\item hydrodynamic mode speeds and widths;
\item Landau--Placzek intensities.
\end{itemize}
A reference implementation for benchmarking finite-volume codes accompanies the paper.
\item \textbf{The thermodynamic route} (\cref{app:thermo_route}). The same models follow from the Helmholtz free energy and the entropy production alone. The isothermal theory obtained this way is unique and identical to the GENERIC one. In the non-isothermal case, CIT leaves a gauge freedom: terms bilinear in $\nabla T$ and the gradients can be moved between the stress and the fluxes. GENERIC removes this freedom. The Gibbs free energy gives the pressure--temperature, low-Mach and quasi-incompressible forms.
\item \textbf{Consistency checks of earlier work} (\cref{sec:previous}). Three of these change the final equations.
\begin{itemize}[nosep]
\item Driving the compressible CH flux by the volumetric potential $\delta F/\delta c$ (Eqs.~(4.14), (7.1)--(7.7) of \cite{ChaudhriNotes2012}; the $c$-equation of \cite{JelicThesis2009}) violates the FDT by a factor $\rho$ in the concentration structure factor.
\item The trace part of the stochastic stress in Eq.~(7.6) of \cite{ChaudhriNotes2012} has the wrong coefficient. Eq.~(2.18) of \cite{ChaudhriPRE2014} is correct.
\item The energy equation (2.12) of \cite{ChaudhriPRE2014} conserves $\int(\tfrac12\rho v^2+u)$ instead of the total energy (2.2), which includes the gradient energy. The missing term is $-\bPi^K\!:\!\nabla\bv$. It does not affect the linearized equilibrium tests in that paper.
\end{itemize}
\end{enumerate}

\paragraph{What was verified symbolically.} Every identity used in the derivation was verified with computer algebra (SymPy) in two space dimensions, with arbitrary coefficient functions. The scripts are distributed with the paper and listed in \cref{app:verification}. They cover:
\begin{itemize}[nosep]
\item the generalized Korteweg identity and the symmetry of $\bPi^w$;
\item the degeneracy $\Lop\cdot\delta S/\delta\bx=0$ with the entropic pressure tensor;
\item local reversible and irreversible energy conservation, including the interstitial fluxes;
\item the local entropy balance and production;
\item the linear FDT for the most general model treated, and its failure for the volumetric-potential variant;
\item the thermodynamics of the binary van der Waals mixture;
\item all closed-form structure factors;
\item all identities of the thermodynamic route.
\end{itemize}

\paragraph{Plan of the remainder of this section.} The rest of this section collects the background on which both models are built.
\begin{itemize}[nosep]
\item \Cref{sec:kinematics} fixes the variables.
\item \Cref{sec:thermo} summarizes the thermodynamics of a homogeneous binary mixture and defines symmetric and non-symmetric mixtures.
\item \Cref{sec:gradient} adds the interfacial contributions and derives the generalized Korteweg tensor.
\item \Cref{sec:generic} reviews GENERIC and its stochastic version, and formulates the Jeli\'c--\"Ottinger construction as an abstract lemma used throughout.
\item \Cref{sec:organization} describes the organization of the paper and the conventions.
\end{itemize}

\subsection{Kinematics of a binary mixture: what are $\rho$ and $c$?}\label{sec:kinematics}

Before any dynamics can be written down, the variables must be fixed. This subsection answers a question that is easily glossed over: what does the density $\rho$ of the Navier--Stokes equations mean in a compressible CHNS model, and what does the order parameter $c$ of the Cahn--Hilliard equation mean? The answer is not a matter of taste. It is fixed by requiring the equations to be exact balance laws for the conserved quantities: species masses, total momentum and total energy.

\subsubsection{Species and mixture fields}
Consider a mixture of two molecular species $A$ and $B$ with molecular masses $m_A,m_B$. At each point $\br$ and time $t$ we define, for each species $\alpha$:
\begin{itemize}[nosep]
\item the number density $n_\alpha$;
\item the partial mass density $\rho_\alpha=m_\alpha n_\alpha$;
\item the species (mean) velocity $\bv_\alpha$.
\end{itemize}
These are coarse-grained fields: averages over cells that are small compared with the hydrodynamic scales but contain many molecules \cite{EspanolThieulot2003,Espanol2001}. Species masses are conserved separately (no reactions):
\begin{equation}
\partial_t\rho_\alpha+\divv(\rho_\alpha\bv_\alpha)=0,\qquad \alpha=A,B.\label{eq:species_mass}
\end{equation}
The mixture fields are
\begin{equation}
\rho=\rho_A+\rho_B,\qquad \rho\bv=\rho_A\bv_A+\rho_B\bv_B\equiv\bg .\label{eq:rho_v_def}
\end{equation}
So $\rho$ is the \emph{total} mass density, $\bv$ is the \emph{barycentric} (mass-averaged) velocity, and $\bg$ is the total momentum density: the sum of the momenta of both species. The diffusion fluxes relative to the barycentric velocity are
\begin{equation}
\bJ_\alpha=\rho_\alpha(\bv_\alpha-\bv),\qquad \bJ_A+\bJ_B=0,\qquad \bJ\equiv\bJ_A=-\bJ_B .\label{eq:diffusion_flux}
\end{equation}
Adding the two equations \eqref{eq:species_mass} gives the continuity equation with \emph{no} diffusive term:
\begin{equation}
\partial_t\rho+\divv(\rho\bv)=0 .\label{eq:continuity}
\end{equation}
Equation \eqref{eq:continuity} holds only if $\bv$ is the barycentric velocity. With any other mixture velocity (molar-averaged, volume-averaged) a diffusive term $\divv(\cdot)\neq0$ appears.

Total momentum is conserved. If the stress is attributed to the mixture as a whole, as in all CHNS models and in the theory of interacting continua in the reduced form of \cite{HeidaMalekRajagopal2012}, the momentum balance reads
\begin{equation}
\partial_t\bg+\divv(\bg\bv)=-\divv\bPi_{\rm tot},\label{eq:momentum_generic}
\end{equation}
where $\bPi_{\rm tot}$ is the total momentum flux in the barycentric frame. The diffusive contribution $\sum_\alpha\rho_\alpha(\bv_\alpha-\bv)(\bv_\alpha-\bv)$ is quadratic in the slow diffusion velocities. In a hydrodynamic description it is either neglected or absorbed into the stress; it is in any case not an independent variable at this level.

\subsubsection{The composition variable}
The \textbf{mass fraction} of $A$ is
\begin{equation}
c=\frac{\rho_A}{\rho}=\frac{m_An_A}{m_An_A+m_Bn_B}\in[0,1] .\label{eq:c_def}
\end{equation}
This is the definition in Eq.~(6.3) of \cite{ChaudhriNotes2012} and Eq.~(5.1b) of \cite{JelicThesis2009}. Using \eqref{eq:species_mass} for $\alpha=A$ together with \eqref{eq:diffusion_flux}:
\begin{equation}
\partial_t(\rho c)+\divv(\rho c\bv)=-\divv\bJ\qquad\Longleftrightarrow\qquad \rho\Dt{c}=-\divv\bJ .\label{eq:c_balance}
\end{equation}
\begin{resultbox}[Answer to the question ``what are $\rho$ and $c$?'']
\begin{enumerate}[leftmargin=*,nosep]
\item \textbf{$\rho$ is the total mass density $\rho_A+\rho_B$.} It appears in the continuity, momentum and energy equations together with the barycentric velocity $\bv$.
\item \textbf{$c$ is the mass fraction of $A$, a dimensionless concentration, not a partial density.} The conserved quantity is $\rho c=\rho_A$, the partial mass density of $A$. The Cahn--Hilliard equation, written as $\partial_t(\rho c)+\divv(\rho c\bv)=-\divv\bJ$, is exactly the mass balance of species $A$. Its flux $\bJ$ is the diffusion flux of $A$ relative to the barycentric velocity. The concentration $c$ itself is not conserved, $\int c\,\dd^3r\neq$ const, unless $\rho$ is uniform.
\item The variable called $c$ in Section~7 of \cite{ChaudhriNotes2012} (``relative mass density'', $(m_1-m_2)/(m_1+m_2)$) is
\begin{equation}
\phi=2c-1=\frac{\rho_A-\rho_B}{\rho}\in[-1,1],\qquad \partial_t(\rho\phi)+\divv(\rho\phi\bv)=-\divv\bJ_\phi,\qquad \bJ_\phi=2\bJ .\label{eq:phi_def}
\end{equation}
This is the natural variable for symmetric mixtures. It is related to $c$ linearly, so all results transfer through the rescalings of \cref{tab:phi_map}.
\item The thermodynamic driving force for $\bJ$ is the gradient of the potential conjugate to $\rho_A$ at fixed $\rho$. This is the \emph{specific} exchange chemical potential $\mut=\delta F/\delta\rho_A|_{\rho}=\rho^{-1}\delta F/\delta c|_{\rho}$ (\cref{sec:isothermal}). The volumetric derivative $\delta F/\delta c|_\rho=\rho\mut$ is the force conjugate to $c$, not to the conserved $\rho c$.
\end{enumerate}
\end{resultbox}

\subsubsection{Other composition variables and other mixture velocities}
Other composition variables are in use. They agree with $c$ (up to affine maps) only for symmetric mixtures.
\begin{itemize}[leftmargin=*]
\item \textbf{Mole (number) fraction} $x=n_A/n$, with $n=n_A+n_B$:
\begin{equation}
x=\frac{c/m_A}{c/m_A+(1-c)/m_B},\qquad n=\frac{\rho}{\bar m},\qquad \frac1{\bar m}=\frac{c}{m_A}+\frac{1-c}{m_B}.\label{eq:x_def}
\end{equation}
The natural partner velocity is the molar-averaged $\bu_n=(n_A\bv_A+n_B\bv_B)/n$. It is related to the barycentric one by
\begin{equation}
\bu_n=\bv+\frac{\bJ}{n}\Big(\frac1{m_A}-\frac1{m_B}\Big).\label{eq:un_v}
\end{equation}
\item \textbf{Volume fraction} $\varphi_A=\rho_A/\bar\rho_A$, where $\bar\rho_\alpha$ is the intrinsic density of pure $\alpha$; this makes sense for (nearly) incompressible constituents. The partner velocity is the volume-averaged $\bu_{\rm vol}=\varphi_A\bv_A+\varphi_B\bv_B$:
\begin{equation}
\bu_{\rm vol}=\bv+\bJ\Big(\frac{1}{\bar\rho_A}-\frac{1}{\bar\rho_B}\Big) .\label{eq:uvol_v}
\end{equation}
This velocity is solenoidal for incompressible constituents; it is the velocity of the Abels--Garcke--Gr\"un model.
\end{itemize}
For a \emph{symmetric} mixture ($m_A=m_B$, $\bar\rho_A=\bar\rho_B$) we have $x=c=\varphi_A$ and $\bu_n=\bu_{\rm vol}=\bv$, so the distinction disappears. For a \emph{non-symmetric} mixture it matters:
\begin{itemize}[nosep]
\item the CH flux defined relative to $\bu_n$ or $\bu_{\rm vol}$ differs from $\bJ$ by convective terms;
\item the momentum density $\rho\bv$ is not $\rho\bu_n$;
\item mixing two frames (for example, a volume-fraction CH equation advected by the barycentric velocity) breaks mass or momentum conservation at $O(\bJ)$.
\end{itemize}
In a fully compressible model the triple
\[
(\rho,\ \rho\bv,\ \rho c)
\]
is the only one that gives a hyperbolic part in conservation form with the barycentric momentum. This is why it is used throughout this paper. Results for $x$ or $\varphi_A$ follow by the chain rule.

\begin{table}[h]
\centering\small
\caption{Rescalings between the mass-fraction formulation ($c$) and the $\phi=2c-1$ formulation. Because $\partial/\partial c=2\,\partial/\partial\phi$, the conjugate potential halves, the flux doubles, and the gradient coefficient is divided by four. The FDT covariance $2\kB TM$ is form-invariant.}\label{tab:phi_map}
\begin{tabular}{lcc}
\toprule
quantity & $c$-formulation & $\phi$-formulation\\
\midrule
order parameter & $c$ & $\phi=2c-1$\\
conserved density & $\rho c=\rho_A$ & $\rho\phi=\rho_A-\rho_B$\\
specific potential & $\mut=\rho^{-1}\delta F/\delta c$ & $\mut_\phi=\rho^{-1}\delta F/\delta\phi=\tfrac12\mut$\\
flux & $\bJ=-M\nabla\mut$ & $\bJ_\phi=2\bJ=-M_\phi\nabla\mut_\phi$, \ $M_\phi=4M$\\
gradient energy & $\tfrac12\kappa_c|\nabla c|^2$ & $\tfrac12\kappa_\phi|\nabla\phi|^2$, \ $\kappa_\phi=\kappa_c/4$\\
noise & $\avg{\noise{\bJ}\noise{\bJ}}=2\kB TM$ & $\avg{\noise{\bJ}_\phi\noise{\bJ}_\phi}=2\kB TM_\phi$\\
\bottomrule
\end{tabular}
\end{table}

\subsection{Equilibrium thermodynamics of a homogeneous binary mixture}\label{sec:thermo}
With the variables fixed, we next need the local equilibrium thermodynamics of a \emph{homogeneous} mixture. It supplies the pressure, the chemical potentials and the entropy in terms of $(\rho,c)$ and $T$ (or $\varepsilon$), and it is where the distinction between symmetric and non-symmetric mixtures is made precise. The interfacial (gradient) contributions are added in \cref{sec:gradient}.

\subsubsection{Fundamental relation in the variables $(n_A,n_B)$, $(\rho_A,\rho_B)$ and $(\rho,c)$}
Under local equilibrium the internal energy density $\varepsilon$ of a homogeneous mixture at rest satisfies the Gibbs relation
\begin{equation}
\dd\varepsilon=T\,\dd s+\mu_A\,\dd n_A+\mu_B\,\dd n_B=T\,\dd s+\hat\mu_A\,\dd\rho_A+\hat\mu_B\,\dd\rho_B,\qquad \hat\mu_\alpha=\frac{\mu_\alpha}{m_\alpha},\label{eq:gibbs_partial}
\end{equation}
together with the Euler relation and the Gibbs--Duhem relation:
\begin{equation}
\varepsilon=Ts-p+\hat\mu_A\rho_A+\hat\mu_B\rho_B,\qquad \dd p=s\,\dd T+\rho_A\,\dd\hat\mu_A+\rho_B\,\dd\hat\mu_B .\label{eq:euler_GD}
\end{equation}
This is Eq.~(5.4)--(5.6) of \cite{JelicThesis2009} and Eq.~(6.4)--(6.5) of \cite{ChaudhriNotes2012}. Insert $\dd\rho_A=c\,\dd\rho+\rho\,\dd c$ and $\dd\rho_B=(1-c)\,\dd\rho-\rho\,\dd c$:
\begin{equation}
\hat\mu_A\dd\rho_A+\hat\mu_B\dd\rho_B=\mub\,\dd\rho+\rho\mu\,\dd c,\qquad
\mub\equiv c\hat\mu_A+(1-c)\hat\mu_B,\qquad \mu\equiv\hat\mu_A-\hat\mu_B .\label{eq:mubar_mu}
\end{equation}
Here $\mub$ is the specific Gibbs energy and $\mu$ is the \emph{exchange chemical potential} per unit mass. Hence
\begin{gather}
\dd\varepsilon=T\,\dd s+\mub\,\dd\rho+\rho\mu\,\dd c,\qquad p=Ts-\varepsilon+\rho\mub,\label{eq:gibbs_rho_c}\\
\dd p=s\,\dd T+\rho\,\dd\mub-\rho\mu\,\dd c .\label{eq:GD_rho_c}
\end{gather}
In the entropy representation $s=s(\rho,c,\varepsilon)$:
\begin{equation}
\pd{s}{\varepsilon}\Big|_{\rho,c}=\frac1T,\qquad \pd{s}{\rho}\Big|_{c,\varepsilon}=-\frac{\mub}{T},\qquad \pd{s}{c}\Big|_{\rho,\varepsilon}=-\frac{\rho\mu}{T}.\label{eq:s_derivs}
\end{equation}
These coincide with Eqs.~(6.6)--(6.8) of \cite{ChaudhriNotes2012} and Eqs.~(5.8)--(5.9) of \cite{JelicThesis2009}, where Jeli\'c's $\mu'=\rho\mu$. The Gibbs--Duhem relation in entropy form, which is needed for the degeneracy of the Poisson operator, is
\begin{equation}
\dd\Big(\frac pT\Big)=-\varepsilon\,\dd\Big(\frac1T\Big)+\rho\,\dd\Big(\frac{\mub}{T}\Big)-\frac{\rho\mu}{T}\,\dd c .\label{eq:GD_entropy}
\end{equation}
\emph{Proof.} From $s=(\varepsilon+p-\rho\mub)/T$ and \eqref{eq:s_derivs}. \hfill$\square$

It is sometimes convenient to use the two conserved densities $(\rho,\rho_A)$ as independent variables. Since $\rho_B=\rho-\rho_A$,
\begin{equation}
\pd{s}{\rho_A}\Big|_{\rho,\varepsilon}=-\frac{\mu}{T},\qquad \pd{s}{\rho}\Big|_{\rho_A,\varepsilon}=-\frac{\hat\mu_B}{T},\qquad
\dd\Big(\frac pT\Big)=-\varepsilon\,\dd\Big(\frac1T\Big)+\rho\,\dd\Big(\frac{\hat\mu_B}{T}\Big)+\rho_A\,\dd\Big(\frac{\mu}{T}\Big).\label{eq:s_derivs_rhoA}
\end{equation}

\subsubsection{Helmholtz free energy per unit mass}
Let $f(\rho,c,T)=(\varepsilon-Ts)/\rho$ be the specific Helmholtz free energy and $\psi=\rho f$ the free-energy density. Then
\begin{gather}
\dd f=-\hs\,\dd T+\frac{p}{\rho^2}\dd\rho+\mu\,\dd c,\qquad p=\rho^2\pd{f}{\rho}\Big|_{c,T},\qquad \mu=\pd{f}{c}\Big|_{\rho,T},\qquad \hs=-\pd{f}{T}\Big|_{\rho,c},\label{eq:f_derivs}\\
\pd{\psi}{\rho}\Big|_{c,T}=f+\frac p\rho=\mub,\qquad \pd{\psi}{c}\Big|_{\rho,T}=\rho\mu,\qquad
\pd{\psi}{\rho_A}\Big|_{\rho,T}=\mu,\qquad \pd{\psi}{\rho}\Big|_{\rho_A,T}=\hat\mu_B .\label{eq:psi_derivs}
\end{gather}
The relation $p=\rho^2\partial f/\partial\rho$ is the one used in \cite{ChaudhriNotes2012} ($P^{\rm CH}$, Eq.~4.1) and \cite{ChaudhriPRE2014}.

\paragraph{Maxwell relations and partial specific volumes.} With the specific Gibbs function $\hat g(T,p,c)=f+p/\rho$,
\begin{equation}
\dd\hat g=-\hs\,\dd T+\frac1\rho\,\dd p+\mu\,\dd c\ \Longrightarrow\
\pd{\mu}{p}\Big|_{T,c}=\pd{(1/\rho)}{c}\Big|_{T,p}=\hat v_A-\hat v_B,\qquad \pd{\mu}{T}\Big|_{p,c}=-\pd{\hs}{c}\Big|_{T,p},\label{eq:maxwell}
\end{equation}
where $\hat v_\alpha$ are the partial specific volumes. Also, from $\partial^2f/\partial\rho\partial c$,
\begin{equation}
\pd{\mu}{\rho}\Big|_{T,c}=\frac1{\rho^2}\pd{p}{c}\Big|_{T,\rho} .\label{eq:maxwell2}
\end{equation}
The difference of partial specific volumes \eqref{eq:maxwell} measures how strongly composition and pressure are coupled. It controls \emph{barodiffusion}: a pressure gradient drives diffusion through $\nabla\mu\supset(\hat v_A-\hat v_B)\nabla p$ \cite{LandauLifshitzFM,deZarateSengers2006}.

\paragraph{Thermodynamic stability.} Local stability of the homogeneous state at fixed $T$ requires $\psi(\rho,\rho_A)$ to be convex, i.e.\ the Hessian $\partial^2\psi/\partial(\rho,\rho_A)^2$ must be positive definite. Equivalently,
\begin{equation}
\pd{p}{\rho}\Big|_{T,c}>0,\qquad \pd{\mu}{c}\Big|_{T,p}>0,\qquad \Big(\pd{\mu}{c}\Big|_{T,p}=\pd{\mu}{c}\Big|_{T,\rho}-\frac{1}{\rho^2}\frac{(\partial p/\partial c)_{T,\rho}^2}{(\partial p/\partial\rho)_{T,c}}\Big).\label{eq:stability}
\end{equation}
The \emph{material spinodal} of liquid--liquid demixing is $(\partial\mu/\partial c)_{T,p}=0$. Inside it the homogeneous mixture is unstable and the Cahn--Hilliard dynamics produces spinodal decomposition. For the full stability analysis, including liquid--vapour instabilities, one requires positivity of the whole Hessian.

\subsubsection{Symmetric and non-symmetric mixtures}\label{sec:symmetry}
Let $\mathcal P$ denote exchange of the labels $A\leftrightarrow B$. Under $\mathcal P$:
\begin{equation}
c\mapsto 1-c,\quad \phi\mapsto-\phi,\quad \rho,\bv,T,\varepsilon,s,p,\mub\mapsto\text{unchanged},\quad \mu\mapsto-\mu,\quad \bJ\mapsto-\bJ,\quad \nabla c\mapsto-\nabla c .\label{eq:parity}
\end{equation}
\begin{definition}[Symmetric binary mixture]\label{def:symmetric}
A mixture is \emph{symmetric} if:
\begin{enumerate}[nosep]
\item $m_A=m_B$;
\item $f(\rho,c,T)=f(\rho,1-c,T)$;
\item the gradient energy and entropy densities are $\mathcal P$-invariant;
\item all transport coefficients transform as tensors under $\mathcal P$, i.e.\ the Onsager matrix commutes with $\mathcal P$.
\end{enumerate}
Otherwise the mixture is \emph{non-symmetric}.
\end{definition}

\begin{proposition}[Consequences of exchange symmetry]\label{prop:symmetry}
In a symmetric mixture, with $\phi=2c-1$:
\begin{enumerate}[nosep]
\item $f$, $p$, $\hs$, $\hu$, $\mub$ are even in $\phi$, and $\mu$ is odd. Consequently $(\partial p/\partial c)_\rho$, $(\partial\mu/\partial\rho)_c$, $(\partial\mu/\partial T)_{\rho,c}$ and $\hat v_A-\hat v_B$ are all odd in $\phi$. They vanish at $\phi=0$.
\item Coexisting liquid phases have $\phi_\pm=\pm\phi_{\rm eq}$, identical densities, pressures and temperatures, and $\mu_{\rm coex}=0$.
\item The mobility $M$, the viscosities $\eta,\zeta$ and the reduced conductivity $\hat\lambda$ are even in $\phi$. The thermodiffusion coefficient $M_T$, the Onsager cross coefficient between $\bJ$ and the heat flux, is odd.
\item In a quadratic gradient energy $\tfrac12K_{\rho\rho}|\nabla\rho|^2+K_{\rho c}\nabla\rho\cdot\nabla c+\tfrac12K_{cc}|\nabla c|^2$, the coefficients $K_{\rho\rho}$ and $K_{cc}$ are even and $K_{\rho c}$ is odd.
\end{enumerate}
\end{proposition}
\begin{proof}
Each statement follows from invariance of the corresponding building block under \eqref{eq:parity}. For example, $\bJ=L_{11}(\ldots)+L_{1q}\nabla(1/T)$: since $\bJ$ is odd and $\nabla(1/T)$ is even, $L_{1q}$ must be odd. Likewise $K_{\rho c}\nabla\rho\cdot\nabla c$ is invariant only if $K_{\rho c}$ is odd.
\end{proof}

\begin{definition}[Decoupled symmetric model]\label{def:decoupled}
The \emph{decoupled} symmetric model is the special case
\begin{equation}
f(\rho,\phi,T)=f_0(\rho,T)+g(\phi,T),\qquad g(\phi,T)=g(-\phi,T),\label{eq:decoupled}
\end{equation}
with constant $\kappa$ and with $\eta,\zeta,\hat\lambda$ independent of $\phi$. Then $p=\rho^2\partial_\rho f_0$ is independent of composition, $\mu=2g_\phi$ is independent of density, and $\hat v_A=\hat v_B$, so there is no barodiffusion. At linear order the composition mode decouples from the sound mode. In the non-isothermal case it remains coupled to temperature through $g_T$, i.e.\ through the heat of demixing.
\end{definition}
The model of Section~7 of \cite{ChaudhriNotes2012} (Eq.~7.11) is exactly of this type, with $f_0(\rho)=a^2\ln\rho$ (isothermal ideal gas, $p=\rho a^2$) and $g(\phi)=\frac{\alpha}{4}(\phi^2-\beta/\alpha)^2$.

\subsubsection{Examples of free energies}\label{sec:examples_f}
\paragraph{(a) Landau double well (symmetric, decoupled).}
\begin{equation}
g(\phi,T)=\frac{A(T)}{2}\phi^2+\frac{B}{4}\phi^4,\qquad A(T)=a_0(T-T_c),\qquad \phi_{\rm eq}^2=-\frac{A}{B}\ (T<T_c).\label{eq:landau}
\end{equation}
The specific entropy and energy of mixing are $\hs_{\rm mix}=-g_T=-\tfrac12a_0\phi^2$ and $\hu_{\rm mix}=g-Tg_T=-\tfrac12a_0T_c\phi^2+\tfrac14B\phi^4$. Demixing therefore releases (latent) heat; this is important in the non-isothermal model. The form $\frac\alpha4(\phi^2-\beta/\alpha)^2$ used in Section~7 of \cite{ChaudhriNotes2012} corresponds to $A=-\beta$, $B=\alpha$ (up to a constant).

\paragraph{(b) Regular solution and Flory--Huggins.}
For a lattice mixture of molecules occupying $N_A$ and $N_B$ sites, in terms of the volume fraction $\varphi$ (per site of volume $v_0$):
\[
\frac{\psi_{\rm mix}v_0}{\kB T}=\frac{\varphi}{N_A}\ln\varphi+\frac{1-\varphi}{N_B}\ln(1-\varphi)+\chi\varphi(1-\varphi).
\]
With $N_A=N_B$ this is symmetric (regular solution). With $N_A\neq N_B$ (polymer blends) it is non-symmetric. To use it in the present framework one expresses $\varphi$ in terms of $(\rho,c)$ through \cref{sec:kinematics}.

\paragraph{(c) Binary van der Waals mixture (one-fluid mixing rules).}
This model connects directly to the single-component van der Waals fluid of \cite{ChaudhriPRE2014} and contains liquid--vapour and liquid--liquid separation. With thermal wavelengths $\Lambda_\alpha\propto T^{-1/2}$, the free-energy density is
\begin{equation}
\psi(n_A,n_B,T)=\kB T\sum_{\alpha}n_\alpha\Big[\ln\frac{n_\alpha\Lambda_\alpha^3}{1-b_An_A-b_Bn_B}-1\Big]-\sum_{\alpha,\beta}a_{\alpha\beta}n_\alpha n_\beta .\label{eq:vdw_mix}
\end{equation}
It gives (verified symbolically, \cref{app:verification})
\begin{gather}
p=\frac{n\kB T}{1-b_An_A-b_Bn_B}-\sum_{\alpha\beta}a_{\alpha\beta}n_\alpha n_\beta,\label{eq:vdw_p}\\
\mu_\alpha=\kB T\ln\frac{n_\alpha\Lambda_\alpha^3}{1-b_An_A-b_Bn_B}+\frac{\kB T\,b_\alpha n}{1-b_An_A-b_Bn_B}-2\sum_\beta a_{\alpha\beta}n_\beta,\label{eq:vdw_mu}\\
\varepsilon=\tfrac32n\kB T-\sum_{\alpha\beta}a_{\alpha\beta}n_\alpha n_\beta,\qquad c_v=\tfrac32 n\kB .\label{eq:vdw_u}
\end{gather}
In the variables $(\rho,c)$: $n_A=\rho c/m_A$, $n_B=\rho(1-c)/m_B$, $\hat\mu_\alpha=\mu_\alpha/m_\alpha$, $\mu=\hat\mu_A-\hat\mu_B$. With $n_B=0$ this reduces to Eqs.~(2.4)--(2.7) of \cite{ChaudhriPRE2014}.

\emph{Symmetric limit.} Take $m_A=m_B=m$, $b_A=b_B=b$, $a_{AA}=a_{BB}=a$, $a_{AB}=a'$. Then $x=c$, and
\begin{equation}
f=f_{\rm vdW}(\rho,T)+\frac{\kB T}{m}\big[c\ln c+(1-c)\ln(1-c)\big]+\frac{2(a-a')\rho}{m^2}c(1-c).\label{eq:vdw_sym}
\end{equation}
Expanding in $\phi$ gives a Landau form:
\begin{equation}
f=f_{\rm vdW}+\text{const}+\frac{\kB}{2m}\big(T-T_c(\rho)\big)\phi^2+\frac{\kB T}{12m}\phi^4+O(\phi^6),\qquad \kB T_c(\rho)=\frac{(a-a')\rho}{m}.\label{eq:vdw_landau}
\end{equation}
Because the consolute temperature depends on density, even the symmetric van der Waals mixture is \emph{not} decoupled in the sense of \cref{def:decoupled}. One finds $(\partial p/\partial c)_\rho=2(a-a')(\rho/m)^2(1-2c)=-2(a-a')(\rho/m)^2\phi$, which is odd in $\phi$ as \cref{prop:symmetry} requires. The decoupled model is the idealization $\partial T_c/\partial\rho=0$.

\emph{Gradient coefficients.} A van der Waals square-gradient energy $\tfrac12\sum_{\alpha\beta}\kappa_{\alpha\beta}\nabla n_\alpha\cdot\nabla n_\beta$, with $\kappa_{\alpha\beta}$ given by Rayleigh-type moments of the attractive potentials (Eq.~4.1 of \cite{ChaudhriPRE2014}), becomes in the variables $(\rho,c)$
\begin{equation}
\begin{aligned}
K_{\rho\rho}&=\frac{c^2\kappa_{AA}}{m_A^2}+\frac{2c(1-c)\kappa_{AB}}{m_Am_B}+\frac{(1-c)^2\kappa_{BB}}{m_B^2},\\
K_{\rho c}&=\rho\Big[\frac{c\,\kappa_{AA}}{m_A^2}+\frac{(1-2c)\kappa_{AB}}{m_Am_B}-\frac{(1-c)\kappa_{BB}}{m_B^2}\Big],\\
K_{cc}&=\rho^2\Big[\frac{\kappa_{AA}}{m_A^2}-\frac{2\kappa_{AB}}{m_Am_B}+\frac{\kappa_{BB}}{m_B^2}\Big].
\end{aligned}\label{eq:vdw_K}
\end{equation}
In the symmetric limit ($\kappa_{AA}=\kappa_{BB}=\kappa$, $\kappa_{AB}=\kappa'$), $K_{cc}=2\rho^2(\kappa-\kappa')/m^2$ is even, $K_{\rho c}=\rho(2c-1)(\kappa-\kappa')/m^2=\rho\phi(\kappa-\kappa')/m^2$ is odd, and $K_{\rho\rho}$ is even, consistent with \cref{prop:symmetry}. In general, therefore, a mixture whose density varies across the interface has gradient energy in \emph{both} $\rho$ and $c$. This is the situation that Cahn and Hilliard explicitly excluded (``\ldots the density remaining constant across the interface, or \ldots the gradient energy coefficient for the density variation is negligible'', quoted in \cite{HeidaMalekRajagopal2012}). The formalism below therefore allows a general gradient energy $\eN(\rho,c,\nabla\rho,\nabla c)$.

\paragraph{(d) Ideal mixing and the mobility--diffusivity relation.}
For an ideal mixture, $\mu_\alpha=\mu_\alpha^0(T,p)+\kB T\ln x_\alpha$. At fixed $(T,p)$,
\begin{equation}
\pd{\mu}{c}\Big|_{T,p}=\frac{\kB T\,\bar m}{m_Am_B\,c(1-c)},\label{eq:ideal_dmudc}
\end{equation}
which reduces to $\kB T/(m\,c(1-c))$ when $m_A=m_B$. Fick's law $\bJ=-\rho D\nabla c$ (at uniform $T,p$) and $\bJ=-M\nabla\mu$ then give
\begin{equation}
M=\rho D\Big(\pd{\mu}{c}\Big)^{-1}_{T,p}=\frac{\rho D\,m_Am_B\,c(1-c)}{\kB T\,\bar m}.\label{eq:M_D}
\end{equation}
This is the physical origin of the degenerate mobility $M\propto c(1-c)\propto(1-\phi^2)$ used in \cite{ChaudhriNotes2012} (Section~7.3).

\subsection{Interfacial (square-gradient) contributions and the generalized Korteweg tensor}\label{sec:gradient}
A homogeneous free energy cannot describe an interface. The square-gradient theory of van der Waals and of Cahn and Hilliard \cite{CahnHilliard1958,RowlinsonWidom} adds terms quadratic in the gradients of the fields, which penalize sharp variations and give the interface a finite width and a surface tension. In a non-isothermal theory one must distinguish the gradient \emph{energy} from the gradient \emph{entropy}. This subsection introduces both, and derives the one identity, \cref{lem:korteweg}, from which every reversible capillary stress in the paper follows.

\subsubsection{Gradient energy, gradient entropy and gradient free energy}
Following van der Waals, Cahn and Hilliard, and \cite{RowlinsonWidom,JelicThesis2009,Onuki2007}, the energy and entropy of a non-uniform mixture contain square-gradient corrections:
\begin{equation}
\EN=\int\eN(\rho,c,\nabla\rho,\nabla c)\,\dd^3r,\qquad \SN=\int\sN(\rho,c,\nabla c)\,\dd^3r .\label{eq:EN_SN}
\end{equation}
The densities are isotropic quadratic forms:
\begin{equation}
\eN=\tfrac12K^E_{\rho\rho}|\nabla\rho|^2+K^E_{\rho c}\nabla\rho\cdot\nabla c+\tfrac12\kappa_E|\nabla c|^2,\qquad \sN=\tfrac12\kappa_S|\nabla c|^2,\label{eq:eN_sN}
\end{equation}
with coefficients that may depend on $(\rho,c)$ but not on $\varepsilon$. Their difference at temperature $T$ is the gradient free energy:
\begin{equation}
\fN=\eN-T\sN,\qquad \kappa\equiv\kappa_E-T\kappa_S .\label{eq:fN}
\end{equation}
\begin{itemize}[leftmargin=*]
\item \emph{Why both energy and entropy?} If the gradient coefficient $\kappa(T)$ of the free energy depends on temperature, thermodynamics requires $\kappa_S=-\partial\kappa/\partial T$ and $\kappa_E=\kappa-T\partial\kappa/\partial T$ \cite{RowlinsonWidom}. Jeli\'c \cite{JelicThesis2009} (Eq.~5.3), following Onuki \cite{Onuki2007} and Gonnella \emph{et al.}, keeps both parts, with $\kappa_E>0$ and $\kappa_S<0$. Espa\~nol and Thieulot \cite{EspanolThieulot2003} and Ref.~\cite{ChaudhriPRE2014} put the whole gradient term in the energy ($\kappa_S=0$). The isothermal theory needs only $\kappa$. The non-isothermal theory needs both, and they produce different physics (\cref{sec:noniso}).
\item \emph{Why does $\sN$ not depend on $\nabla\rho$?} With $\sN=\sN(\rho,c,\nabla c)$, the Poisson operator constructed below satisfies the Jacobi identity: it becomes the canonical Lie--Poisson operator in the variables $(\rho,\bg,\sigma,c)$, where $\sigma=s+\sN$ (\cref{app:jacobi}). If $\sN$ depended on $\nabla\rho$, this property would be lost. Density gradients are therefore admitted only in the energy, as in \cite{ChaudhriPRE2014,Espanol2001}. In the isothermal theory this restriction is immaterial.
\item \emph{CH-only case.} The pure Cahn--Hilliard case is $K^E_{\rho\rho}=K^E_{\rho c}=0$. Two common choices of $\kappa$ are a constant (``volumetric'') coefficient, as in \cite{Cherfils2019} and Jeli\'c's $\kappa_E$, and a coefficient proportional to $\rho$, $\kappa=\rho\hat\kappa$, as in the specific free energy $f(\rho,c,\nabla c)\ni\tfrac12\hat\kappa|\nabla c|^2$ of \cite{ChaudhriNotes2012} (Eq.~4.13) and \cite{LowengrubTruskinovsky1998}. The two choices give different reversible stresses (\cref{sec:korteweg_cases}). Both are thermodynamically admissible; they are different constitutive models.
\end{itemize}

\subsubsection{Functional derivatives}
For a density $w(\rho,c,\nabla\rho,\nabla c)$ with $W=\int w$, the functional derivatives at fixed other fields are
\begin{equation}
\fd{W}{\rho}\Big|_c=\pd{w}{\rho}-\divv\pd{w}{\nabla\rho},\qquad \fd{W}{c}\Big|_\rho=\pd{w}{c}-\divv\pd{w}{\nabla c}.\label{eq:fd_def}
\end{equation}
We use the notation
\begin{equation}
\mu_E\equiv\fd{\EN}{c}\Big|_\rho,\qquad \mu_S\equiv\fd{\SN}{c}\Big|_\rho,\qquad e^\rho\equiv\fd{\EN}{\rho}\Big|_c,\qquad s^\rho\equiv\fd{\SN}{\rho}\Big|_c=\pd{\sN}{\rho}.\label{eq:muE_muS}
\end{equation}
These are volumetric potentials (energy or entropy per unit volume per unit change of $c$). For constant coefficients and the CH-only case,
\[
\mu_E=-\kappa_E\nabla^2c,\qquad \mu_S=-\kappa_S\nabla^2c,\qquad \mu_E-T\mu_S=-\kappa\nabla^2c .
\]
For $\kappa=\rho\hat\kappa$ and isothermal conditions, $\delta\FN/\delta c=-\divv(\rho\hat\kappa\nabla c)$, which is the term in Eq.~(4.14) of \cite{ChaudhriNotes2012}.

\subsubsection{The generalized Korteweg identity}
\begin{lemma}[Generalized Korteweg tensor]\label{lem:korteweg}
Let $w(\rho,c,\nabla\rho,\nabla c)$ be a scalar invariant under rotations, and $W=\int w\,\dd^3r$. Then
\begin{equation}
\rho\nabla\fd{W}{\rho}\Big|_c-\fd{W}{c}\Big|_\rho\nabla c=\divv\bPi^{w},\label{eq:korteweg_identity}
\end{equation}
\begin{equation}
\bPi^{w}=\Big[\rho\fd{W}{\rho}\Big|_c-w\Big]\Id+\nabla\rho\otimes\pd{w}{\nabla\rho}+\nabla c\otimes\pd{w}{\nabla c}.\label{eq:Piw}
\end{equation}
Rotational invariance makes $\bPi^w$ symmetric. For a quadratic form, $\nabla\rho\otimes w_{\nabla\rho}+\nabla c\otimes w_{\nabla c}=K_{\rho\rho}\nabla\rho\nabla\rho+K_{\rho c}(\nabla\rho\nabla c+\nabla c\nabla\rho)+K_{cc}\nabla c\nabla c$.
\end{lemma}
\begin{proof}
Write $w_\rho=\partial w/\partial\rho$, $w_{\rho,j}=\partial w/\partial(\partial_j\rho)$, and similarly for $c$. Then
\[
\partial_iw=w_\rho\partial_i\rho+w_c\partial_ic+w_{\rho,j}\partial_i\partial_j\rho+w_{c,j}\partial_i\partial_jc .
\]
Compute
\begin{align*}
\partial_j\Pi^w_{ji}&=\partial_i\Big(\rho\fd W\rho\Big)-\partial_iw+\partial_j(\partial_i\rho\,w_{\rho,j})+\partial_j(\partial_ic\,w_{c,j})\\
&=\partial_i\rho\,\fd W\rho+\rho\,\partial_i\fd W\rho-w_\rho\partial_i\rho-w_c\partial_ic-w_{\rho,j}\partial_i\partial_j\rho-w_{c,j}\partial_i\partial_jc\\
&\quad+\partial_i\partial_j\rho\,w_{\rho,j}+\partial_i\rho\,\partial_jw_{\rho,j}+\partial_i\partial_jc\,w_{c,j}+\partial_ic\,\partial_jw_{c,j}\\
&=\partial_i\rho\Big(\fd W\rho-w_\rho+\partial_jw_{\rho,j}\Big)+\rho\,\partial_i\fd W\rho-\partial_ic\big(w_c-\partial_jw_{c,j}\big)
=\rho\,\partial_i\fd W\rho-\partial_ic\,\fd Wc ,
\end{align*}
using \eqref{eq:fd_def}. The tensor $\bPi^w$ is symmetric because for an isotropic $w$ the antisymmetric part of $\nabla\rho\otimes w_{\nabla\rho}+\nabla c\otimes w_{\nabla c}$ is the generator of infinitesimal rotations acting on $w$, which vanishes.
\end{proof}

\Cref{lem:korteweg} is the only identity needed to obtain all reversible capillary stresses. It is used three times below:
\begin{itemize}[nosep]
\item with $w=\fN$ (isothermal stress);
\item with $w=\eN$ (energetic stress);
\item with $w=\sN$ (entropic pressure tensor and degeneracy).
\end{itemize}

\subsubsection{Special cases}\label{sec:korteweg_cases}
\begin{enumerate}[label=(\roman*),leftmargin=*]
\item \textbf{$w=\tfrac12\kappa|\nabla c|^2$, $\kappa$ constant} \cite{Cherfils2019,JelicThesis2009}:
\begin{equation}
\bPi^w=-\tfrac12\kappa|\nabla c|^2\Id+\kappa\nabla c\otimes\nabla c,\qquad \divv\bPi^w=\kappa\nabla^2c\,\nabla c .\label{eq:K_const}
\end{equation}
\item \textbf{$w=\tfrac12\rho\hat\kappa|\nabla c|^2$, $\hat\kappa$ constant} (specific gradient energy, \cite{ChaudhriNotes2012} Eq.~4.13):
\begin{equation}
\bPi^w=\rho\hat\kappa\,\nabla c\otimes\nabla c .\label{eq:K_rho}
\end{equation}
This is exactly the reactive stress of Eq.~(4.16) in \cite{ChaudhriNotes2012} and of Lowengrub--Truskinovsky \cite{LowengrubTruskinovsky1998}.
\item \textbf{$w=\tfrac12\kappa(\rho,c)|\nabla c|^2$} (general CH-only):
\begin{equation}
\bPi^w=-\tfrac12\big(\kappa-\rho\,\partial_\rho\kappa\big)|\nabla c|^2\Id+\kappa\nabla c\otimes\nabla c .\label{eq:K_general}
\end{equation}
Cases (i) and (ii) correspond to $\rho\partial_\rho\kappa=0$ and $\rho\partial_\rho\kappa=\kappa$.
\item \textbf{$w=\tfrac12\kappa|\nabla\rho|^2$, $\kappa$ constant} (single-component van der Waals fluid):
\begin{equation}
\bPi^w=-\Big(\kappa\rho\nabla^2\rho+\tfrac12\kappa|\nabla\rho|^2\Big)\Id+\kappa\nabla\rho\otimes\nabla\rho .\label{eq:K_vdw}
\end{equation}
With $\bPi=P\Id-\bPi^K$, this reproduces Eq.~(2.14) of \cite{ChaudhriPRE2014}, $\bPi^K=-\bPi^w$, and also $\divv\bPi^w=\kappa\rho\nabla\nabla^2\rho$ (their Eq.~2.16).
\end{enumerate}

\subsubsection{Isothermal equilibrium and surface tension}\label{sec:equilibrium_interface}
Equilibrium at fixed $T$ minimizes $F=\int(\psi+\fN)$ at fixed total masses $\int\rho$ and $\int\rho_A$. Since $\delta\rho_A=\rho\,\delta c$ at fixed $\rho$, the Euler--Lagrange equations are
\begin{equation}
\fd{F}{\rho_A}\Big|_\rho=\frac{1}{\rho}\fd{F}{c}\Big|_\rho=\mut=\lambda_A,\qquad \fd F\rho\Big|_c=\lambda_B+c\,\lambda_A,\label{eq:EL}
\end{equation}
where $\lambda_A,\lambda_B$ are constant Lagrange multipliers. Applying \cref{lem:korteweg} to $\psi+\fN$, with the bulk part handled by the isothermal Gibbs--Duhem relation \eqref{eq:GD_rho_c}, gives
\begin{equation}
\rho\nabla\fd F\rho\Big|_c-\fd Fc\Big|_\rho\nabla c=\divv\big(p\Id+\bPi^{\fN}\big)\equiv\divv\bPi .\label{eq:iso_identity}
\end{equation}
At equilibrium the left side is $\rho\nabla(\lambda_B+c\lambda_A)-\rho\lambda_A\nabla c=0$. Chemical equilibrium (uniform $\mut$) therefore implies mechanical equilibrium, $\divv\bPi=0$, as it must.

\paragraph{Planar interface, CH-only, decoupled symmetric model.} Take the limit of a stiff bulk equation of state, so that $\rho\simeq\rho_0$, with $g=\frac A2\phi^2+\frac B4\phi^4$, $A<0$, and constant $\kappa_\phi$. Then $\kappa_\phi\phi''=\rho_0g'(\phi)$, with first integral $\tfrac12\kappa_\phi\phi'^2=\rho_0[g(\phi)-g(\phi_{\rm eq})]$. The solution is
\begin{equation}
\phi(z)=\phi_{\rm eq}\tanh\frac{z}{\ell},\qquad \ell=\frac1{\phi_{\rm eq}}\sqrt{\frac{2\kappa_\phi}{\rho_0B}}=\sqrt{\frac{2\kappa_\phi}{\rho_0|A|}} .\label{eq:tanh}
\end{equation}
The mechanical (Bakker) definition of the surface tension, $\sigma=\int(\Pi_{zz}-\Pi_{xx})\,\dd z$, gives with \eqref{eq:K_const}
\begin{equation}
\sigma=\int_{-\infty}^{\infty}\kappa_\phi\,\phi'^2\,\dd z=\frac{2\sqrt2}{3}\sqrt{\kappa_\phi\rho_0B}\;\phi_{\rm eq}^3=\frac{2\sqrt2}{3}\frac{\sqrt{\kappa_\phi\rho_0}\,|A|^{3/2}}{B}.\label{eq:sigma}
\end{equation}
For $\rho_0=1$ this agrees with Eq.~(4.4) of \cite{JelicThesis2009}. The equality $\Pi_{zz}-\Pi_{xx}=\kappa\phi'^2$ holds exactly for any equation of state, by \eqref{eq:Piw}.

\subsection{GENERIC, stochastic GENERIC, and the Jeli\'c--\"Ottinger construction of the friction operator}\label{sec:generic}
The last ingredient is the dynamical framework. GENERIC (General Equation for the Non-Equilibrium Reversible--Irreversible Coupling) organizes any thermodynamically consistent evolution into a reversible part generated by the energy and an irreversible part generated by the entropy. Its stochastic version adds noise whose covariance is fixed by the friction operator. We review both and then formulate the Jeli\'c--\"Ottinger construction as an abstract lemma, \cref{lem:JO}, which will be applied to the isothermal model in \cref{sec:isothermal} and to the non-isothermal model in \cref{sec:noniso}. The isothermal reduction of GENERIC is given at the beginning of \cref{sec:isothermal}.

\subsubsection{Deterministic GENERIC}
Let $\bx$ be a complete set of slow variables. For hydrodynamic fields, $\bx(\br)$ is a vector of fields. GENERIC \cite{GrmelaOttinger1997,OttingerGrmela1997,Ottinger2005} postulates
\begin{equation}
\frac{\dd\bx}{\dd t}=\Lop(\bx)\cdot\fd{E}{\bx}+\Mop(\bx)\cdot\fd{S}{\bx},\label{eq:generic}
\end{equation}
with four building blocks: the energy $E$, the entropy $S$, the Poisson operator $\Lop$ and the friction operator $\Mop$. Here ``$\cdot$'' denotes contraction over components \emph{and} integration over the second spatial argument, $(\Lop\cdot\bm a)(\br)=\int\Lop(\br,\br')\bm a(\br')\,\dd^3r'$. For local operators we use the single-variable ``$\br$-notation'' of \cite{JelicThesis2009}, with differential operators acting on everything to their right. The building blocks must satisfy:
\begin{enumerate}[label=(G\arabic*),leftmargin=*]
\item $\Lop$ is antisymmetric, $\int\bm a\cdot\Lop\cdot\bm b=-\int\bm b\cdot\Lop\cdot\bm a$, and the bracket $\{A,B\}=\int\fd A\bx\cdot\Lop\cdot\fd B\bx$ satisfies the Jacobi identity.
\item $\Mop$ is symmetric (Onsager--Casimir) and positive semi-definite, $\int\bm a\cdot\Mop\cdot\bm a\ge0$.
\item Degeneracy: $\Lop\cdot\fd S\bx=0$ (entropy is a Casimir of the reversible dynamics) and $\Mop\cdot\fd E\bx=0$ (the irreversible dynamics conserves energy).
\end{enumerate}
From (G1)--(G3), for an isolated system,
\begin{equation}
\frac{\dd E}{\dd t}=0,\qquad \frac{\dd S}{\dd t}=\int\fd S\bx\cdot\Mop\cdot\fd S\bx\ \ge\ 0 .\label{eq:generic_consequences}
\end{equation}
Other conserved quantities $I$ (masses, momentum) are Casimirs of $\Lop$ in the sense $\Lop\cdot\delta I/\delta\bx=0$, or are conserved because both $\Lop\cdot\delta E/\delta\bx$ and $\Mop\cdot\delta S/\delta\bx$ have divergence form in the corresponding components. This form comes from the conservation-law structure of the fluctuations (\cref{sec:JO}).

\subsubsection{Stochastic GENERIC and the fluctuation--dissipation theorem}\label{sec:sgeneric}
Thermal fluctuations are added through the It\^o stochastic differential equation \cite{Ottinger2005} (Section~1.2.6 there), \cite{GrmelaOttinger1997}:
\begin{equation}
\dd\bx=\Big[\Lop\cdot\fd E\bx+\Mop\cdot\fd S\bx+\kB\,\pd{}{\bx}\cdot\Mop\Big]\dd t+\Bop\cdot\dd\bW_t,\qquad \Bop\cdot\Bop^{T}=2\kB\,\Mop ,\label{eq:sgeneric}
\end{equation}
together with
\begin{equation}
\fd E\bx\cdot\Bop=0,\qquad \pd{}{\bx}\cdot\Big(\Lop\cdot\fd E\bx\Big)=0 .\label{eq:sgeneric_conditions}
\end{equation}
Here $\bW_t$ is a vector of independent Wiener processes. The second relation in \eqref{eq:sgeneric} is the fluctuation--dissipation theorem in GENERIC form: the covariance of the noise equals $2\kB$ times the friction operator. The first condition in \eqref{eq:sgeneric_conditions} states that the noise does not change the energy (it acts on the energy shell). The second is a Liouville-type condition: the reversible flow is incompressible in state space. Since $\partial_{x_i}(\Lop_{ij}E_{x_j})=(\partial_{x_i}\Lop_{ij})E_{x_j}$ by antisymmetry of $\Lop$ and symmetry of the Hessian of $E$, it is a condition on $\Lop$ alone. It holds formally for the Lie--Poisson operators of hydrodynamics \cite{Ottinger2005,Espanol2001}. The corresponding Fokker--Planck equation is
\begin{equation}
\partial_tP=-\pd{}{\bx}\cdot\Big[\Big(\Lop\cdot\fd E\bx+\Mop\cdot\fd S\bx\Big)P\Big]+\kB\,\pd{}{\bx}\cdot\Mop\cdot\pd{}{\bx}P .\label{eq:FP}
\end{equation}
This is Eq.~(5.22) of \cite{ChaudhriNotes2012}. The It\^o drift $\kB\partial_{\bx}\cdot\Mop$ in \eqref{eq:sgeneric} is precisely what makes \eqref{eq:FP} have the diffusion term in this ``$\Mop\,\partial P$'' form.

\begin{proposition}[Equilibrium distribution]\label{prop:Peq}
Under (G1)--(G3) and \eqref{eq:sgeneric_conditions}, the distribution
\begin{equation}
P_{\rm eq}(\bx)\propto\exp\Big(\frac{S(\bx)}{\kB}\Big)\,\delta\big(E(\bx)-E_0\big)\prod_I\delta\big(I(\bx)-I_0\big)\label{eq:Peq}
\end{equation}
is stationary. The product runs over all other conserved quantities $I$.
\end{proposition}
\begin{proof}
Write $E_{\bx}=\delta E/\delta\bx$, etc. The probability current is $\bm{\mathcal J}=(\Lop\cdot E_{\bx}+\Mop\cdot S_{\bx})P-\kB\Mop\cdot\partial_{\bx}P$. Differentiating \eqref{eq:Peq},
\[
\partial_{\bx}P_{\rm eq}=\frac{S_{\bx}}{\kB}P_{\rm eq}+E_{\bx}\,(\cdots)+\sum_I I_{\bx}\,(\cdots),
\]
where the dots are derivatives of the $\delta$-functions. Since $\Mop\cdot E_{\bx}=0$, and $\Mop\cdot I_{\bx}=0$ for the conserved densities by construction, we get $\kB\Mop\cdot\partial_{\bx}P_{\rm eq}=\Mop\cdot S_{\bx}P_{\rm eq}$, hence $\bm{\mathcal J}=(\Lop\cdot E_{\bx})P_{\rm eq}$. Then
\[
\partial_{\bx}\cdot\bm{\mathcal J}=\big[\partial_{\bx}\cdot(\Lop\cdot E_{\bx})\big]P_{\rm eq}+(\Lop\cdot E_{\bx})\cdot\partial_{\bx}P_{\rm eq}.
\]
The first term vanishes by \eqref{eq:sgeneric_conditions}. In the second, $(\Lop\cdot E_{\bx})\cdot S_{\bx}=-E_{\bx}\cdot\Lop\cdot S_{\bx}=0$ by degeneracy, $(\Lop\cdot E_{\bx})\cdot E_{\bx}=0$ by antisymmetry, and $(\Lop\cdot E_{\bx})\cdot I_{\bx}=-\dot I|_{\rm rev}=0$ because $I$ is conserved by the reversible dynamics.
\end{proof}
The Gibbs factor $\exp(S/\kB)$ on the energy shell is the Einstein formula. Linearized around a homogeneous state, it predicts the static structure factors. In \cref{sec:iso_linear,sec:noniso_linear} we show that the linearized versions of our final equations reproduce them exactly.

\paragraph{About the thermal drift $\kB\partial_{\bx}\cdot\Mop$.} For local field theories this term is formally proportional to derivatives of $\delta(\mathbf 0)$. It is ill-defined in the continuum but finite and computable after spatial discretization, where it is $O(\kB/\Delta V)$. Landau--Lifshitz fluctuating hydrodynamics, including \cite{ChaudhriPRE2014}, conventionally omits it. Equivalently, the noise is interpreted so that the discretized dynamics has the correct Gibbs--Boltzmann invariant measure; see \cite{DonevVandenEijnden2014,Espanol2001,Ottinger2005}. We write it explicitly in the general equations and drop it in the final ``Landau--Lifshitz'' form. It vanishes identically when $\Mop$ does not depend on the state, and it never affects the linearized theory.

\subsubsection{The Jeli\'c--\"Ottinger construction: friction from fluctuations}\label{sec:JO}
In GENERIC the friction matrix has a Green--Kubo representation (\cite{JelicThesis2009} Eqs.~2.22--2.28 and 3.12--3.13; \cite{Ottinger1998}):
\begin{equation}
\Mop_{jk}(\bx)=\frac1{\kB}\int_0^{\tau}\avg{\dot\Pi^{\rm f}_j(t)\,\dot\Pi^{\rm f}_k(0)}_{\bx}\dd t=\frac{1}{2\kB\tau}\avg{\Delta_\tau\Pi^{\rm f}_j\,\Delta_\tau\Pi^{\rm f}_k}_{\bx}.\label{eq:GK}
\end{equation}
Here $\Delta_\tau\Pi^{\rm f}$ is the fast (fluctuating) part of the change of the slow variables over an intermediate time $\tau$, shorter than the slow scales and longer than the fast ones. For variables of mixed time-reversal parity the left side carries the factor $\tfrac12[1+\epsilon_j\epsilon_k]$; all fluxes considered here involve variables of the same parity in each correlated pair.

Following \cite{JelicThesis2009} (Section~3.2.3), the construction proceeds in five steps:
\begin{enumerate}[label=\textbf{Step \arabic*.},leftmargin=*]
\item Express $\Mop$ by \eqref{eq:GK}.
\item Choose densities of extensive (preferably conserved) quantities as variables, so that fluctuations are \emph{current} fluctuations.
\item Make a structural ansatz for $\Delta_\tau\bx^{\rm f}$ in the spirit of Onsager's regression hypothesis. Fluctuations of conserved densities are divergences of fluctuating currents, $\Delta_\tau\bx^{\rm f}=-\divv\bW$, where the $\bW$ are Wiener increments over $\tau$. The energy fluctuation is chosen so that $\int\fd E\bx\cdot\Delta_\tau\bx^{\rm f}=0$, i.e.\ the fluctuations stay on the energy shell.
\item Specify second moments, local in space:
\begin{equation}
\avg{W^{(\alpha)}(\br)W^{(\beta)}(\br')}=2\kB\tau\,\Cop^{(\alpha\beta)}(\br)\,\delta(\br-\br'),\label{eq:W_moments}
\end{equation}
where $\Cop$ is symmetric, positive semi-definite, and respects the Curie principle (tensors of different rank do not couple in an isotropic fluid).
\item Compute $\Mop$ from \eqref{eq:GK}.
\end{enumerate}
The following lemma, which makes the Jeli\'c--\"Ottinger construction abstract, is used for both the isothermal and the non-isothermal models.

\begin{lemma}[Friction and noise from a common flux map]\label{lem:JO}
Suppose the fluctuation ansatz has the form
\begin{equation}
\int\bm a\cdot\Delta_\tau\bx^{\rm f}\,\dd^3r=\int\bW\cdot\Phiop(\bm a)\,\dd^3r\qquad\text{for every test field }\bm a,\label{eq:Phi_def}
\end{equation}
where $\Phiop$ is a linear (differential) operator mapping test fields to ``thermodynamic forces''. That is, $\Delta_\tau\bx^{\rm f}=\Phiop^{*}\bW$, with $\Phiop^*$ the formal adjoint. Suppose also that $\bW$ has the moments \eqref{eq:W_moments}. Then:
\begin{enumerate}[label=(\alph*),nosep]
\item $\Mop=\Phiop^{*}\Cop\,\Phiop$, i.e.\ $\int\bm a\cdot\Mop\cdot\bm b=\int\Phiop(\bm a)\cdot\Cop\cdot\Phiop(\bm b)$.
\item $\Mop$ is symmetric, and positive semi-definite if $\Cop$ is.
\item $\Mop\cdot\fd E\bx=0$ if $\Phiop\big(\fd E\bx\big)=0$.
\item The irreversible dynamics is $\Mop\cdot\fd S\bx=\Phiop^{*}\bY$ with fluxes $\bY=\Cop\cdot\bX$ and forces $\bX=\Phiop\big(\fd S\bx\big)$. The entropy production is $\dot S=\int\bX\cdot\Cop\cdot\bX\ge0$.
\item The noise $\Bop\cdot\dd\bW_t\equiv\Phiop^{*}\dd\bW$ with $\avg{\dd\bW\,\dd\bW}=2\kB\Cop\,\dd t$ satisfies the FDT $\Bop\Bop^T=2\kB\Mop$ \emph{identically}. It also satisfies $\fd E\bx\cdot\Bop=0$ whenever (c) holds.
\end{enumerate}
\end{lemma}
\begin{proof}
(a)
\begin{align*}
\int\bm a\cdot\Mop\cdot\bm b&=\frac1{2\kB\tau}\avg{\Big(\int\bm a\cdot\Delta\bx^{\rm f}\Big)\Big(\int\bm b\cdot\Delta\bx^{\rm f}\Big)}\\
&=\frac1{2\kB\tau}\iint\Phiop(\bm a)(\br)\cdot\avg{\bW(\br)\bW(\br')}\cdot\Phiop(\bm b)(\br')\,\dd^3r\,\dd^3r'
=\int\Phiop(\bm a)\cdot\Cop\cdot\Phiop(\bm b)\,\dd^3r .
\end{align*}
(b) and (c) follow immediately from (a). (d) Set $\bm a$ arbitrary and $\bm b=\fd S\bx$ in (a). (e) The covariance of $\Phiop^*\dd\bW$ is $\Phiop^*(2\kB\Cop\,\dd t)\Phiop=2\kB\Mop\,\dd t$, and $\int\fd E\bx\cdot\Phiop^*\dd\bW=\int\dd\bW\cdot\Phiop(\fd E\bx)=0$.
\end{proof}

\begin{resultbox}[How dissipation and noise are related (general statement)]
In the Jeli\'c--\"Ottinger construction a \emph{single} operator $\Phiop$ (the kinematics of how currents change the state) and a \emph{single} matrix $\Cop$ (the strengths and correlations of the elementary current fluctuations) determine everything:
\begin{equation*}
\underbrace{\Phiop^*\,\Cop\,\Phiop(\delta S/\delta\bx)}_{\text{deterministic dissipation}},\qquad
\underbrace{\Phiop^*\,\dd\bW,\ \ \avg{\dd\bW\dd\bW}=2\kB\Cop\,\dd t}_{\text{stochastic forcing}} .
\end{equation*}
In the flux--force language: the deterministic fluxes are $\bY=\Cop\cdot\bX$, with $\Cop$ the Onsager matrix. The stochastic fluxes have covariance $2\kB\Cop$, i.e.\ $2\kB\times$\,(Onsager matrix)\,$\times\delta(\br-\br')\delta(t-t')$. This is the Landau--Lifshitz prescription, now \emph{derived}. The temperature factors in the familiar amplitudes ($2\kB T\eta$, $2\kB\lambda T^2$, $2\kB TM$) come from the entropy-based forces $\bX$, which contain $1/T$, $\nabla(1/T)$ and $\nabla(\mut/T)$.
\end{resultbox}

\subsection{Organization of the paper and conventions}\label{sec:organization}
\paragraph{Organization.} The four models treated are:
\begin{center}
\begin{tabular}{lll}
\toprule
 & symmetric mixture & non-symmetric mixture\\
\midrule
isothermal & model I-S, \cref{sec:iso_sym} & model I-N, \cref{sec:iso_nonsym}\\
non-isothermal & model N-S, \cref{sec:noniso_sym} & model N-N, \cref{sec:noniso_nonsym}\\
\bottomrule
\end{tabular}
\end{center}
\begin{itemize}[leftmargin=*]
\item \Cref{sec:isothermal} derives the isothermal model. It starts from the Helmholtz free-energy functional, defines all derived quantities, and constructs the reversible and irreversible dynamics with GENERIC. It assembles the deterministic and stochastic parts into the final equations and specializes them to symmetric and non-symmetric mixtures.
\item \Cref{sec:noniso} does the same for the non-isothermal model. It begins with a summary of what changes relative to the isothermal case and shows that its isothermal limit reproduces \cref{sec:isothermal}.
\item \Cref{sec:summary} summarizes the four models side by side.
\item \Cref{sec:previous} compares the results with earlier work, point by point.
\item \Cref{sec:dsf} derives the dynamic structure factors that serve as analytic benchmarks.
\item \Cref{sec:conclusions} concludes.
\end{itemize}
The appendices contain the proofs and the list of symbolic checks (\cref{app:proofs}), remarks on discretization (\cref{app:numerics}), and the derivation of all models from the free energy and the entropy production alone (\cref{app:thermo_route}). Throughout, the final equations are written in the conservative flux form of \cite{ChaudhriPRE2014}:
\[
\partial_t\bm U=-\divv\bm F_H-\divv\bm F_K-\divv\bm F_D-\divv\bm F_S ,
\]
with hyperbolic, capillary (Korteweg), dissipative and stochastic fluxes.

\paragraph{Conventions.}
\begin{itemize}[leftmargin=*,nosep]
\item Gradients: $(\nabla\bv)_{ij}=\partial_iv_j$; $(\divv\bm A)_i=\partial_jA_{ji}$; $\bm A\!:\!\bm B=A_{ij}B_{ij}$; summation over repeated indices.
\item $\bm A^{\sym}=\tfrac12(\bm A+\bm A^T)$. $\bD=(\nabla\bv)^{\sym}$ is the rate-of-strain tensor, and $\mathring{\bD}=\bD-\tfrac13(\tr\bD)\Id$.
\item Material derivative: $D/Dt=\partial_t+\bv\cdot\nabla$.
\item Densities per unit volume carry no hat. Quantities per unit mass (specific quantities) carry a hat ($\hs,\hu$), except the specific Helmholtz free energy $f$ and the chemical potentials $\mu$, $\mub$, $\mut$, which are always per unit mass unless stated otherwise.
\item White noise: $\avg{W_\alpha(\br,t)W_\beta(\br',t')}=\delta_{\alpha\beta}\delta(\br-\br')\delta(t-t')$.
\item Following \cite{ChaudhriPRE2014}, the Korteweg tensor $\bPi^K$ enters the momentum flux with a minus sign: $\bPi=p\Id-\bPi^K$.
\end{itemize}

\begin{table}[H]
\centering\small
\caption{Principal symbols.}\label{tab:symbols}
\begin{tabular}{>{\raggedright\arraybackslash}p{5.4cm}>{\raggedright\arraybackslash}p{10.0cm}}
\toprule
$n_\alpha$, $m_\alpha$, $\rho_\alpha=m_\alpha n_\alpha$ & number density, molecular mass, partial mass density of species $\alpha\in\{A,B\}$\\
$\rho=\rho_A+\rho_B$, $\bv$, $\bg=\rho\bv$ & total mass density, barycentric velocity, momentum density\\
$c=\rho_A/\rho$, $\phi=2c-1$ & mass fraction of $A$, relative mass-density difference\\
$\bJ=\rho_A(\bv_A-\bv)$ & diffusion (mass) flux of $A$ relative to $\bv$\\
$\varepsilon$, $s$ & local internal energy and entropy densities (no gradient parts)\\
$\eN$, $\sN$, $\fN=\eN-T\sN$ & gradient (interfacial) energy, entropy, free-energy densities\\
$e_{\rm tot}=\tfrac12\rho v^2+\varepsilon+\eN$ & total energy density ($=\rho E$ in the flux form)\\
$f(\rho,c,T)$, $\psi=\rho f$ & specific Helmholtz free energy, free-energy density\\
$\hat\mu_\alpha=\mu_\alpha/m_\alpha$ & chemical potential of $\alpha$ per unit mass\\
$\mub=c\hat\mu_A+(1-c)\hat\mu_B$, $\mu=\hat\mu_A-\hat\mu_B$ & mean (specific Gibbs) and exchange chemical potentials\\
$\mu_E=\delta\EN/\delta c$, $\mu_S=\delta\SN/\delta c$ & gradient energy/entropy chemical potentials (per volume)\\
$\mut=\mu+(\mu_E-T\mu_S)/\rho$ & generalized specific exchange potential\\
$p$ & local thermodynamic pressure\\
$\bPi$, $\bPi^K$ & reversible momentum flux ($=p\Id-\bPi^K$), Korteweg tensor\\
$\btau$, $\bJq$, $\bJqp=\bJq-\mut\bJ$, $\bQ=-\bJq$ & viscous stress, heat flux, reduced heat flux, heat flux in the sign convention of \cite{ChaudhriPRE2014}\\
$\eta,\zeta,\hat\lambda,\lambda,M,M_T$ & viscosities, conductivities, mobility, thermodiffusion coefficient\\
$\kappa_E,\kappa_S,\kappa=\kappa_E-T\kappa_S$ & energetic, entropic and free-energy gradient coefficients\\
$\Lop,\Mop,\Dop$ & Poisson, friction (non-isothermal), friction (isothermal) operators\\
\bottomrule
\end{tabular}
\end{table}

\section{The isothermal, fully compressible LLCHNS model}\label{sec:isothermal}
We first consider a mixture in contact with a heat bath at temperature $T_0$, with heat conduction fast enough that the temperature is uniform and constant, $T\equiv T_0$. This is the setting of most diffuse-interface simulations and of the fluctuating van der Waals model of \cite{ChaudhriPRE2014}. The temperature is then not a dynamical variable, and the whole dynamics is generated by a single functional, the Helmholtz free energy. The derivation proceeds in the following steps.
\begin{enumerate}[nosep]
\item Reduce GENERIC to its isothermal form (\cref{sec:iso_generic}).
\item Write down the full Helmholtz free-energy functional of the compressible mixture and define all quantities derived from it (\cref{sec:iso_F}).
\item Construct the Poisson operator and obtain the reversible dynamics, including the Korteweg stress (\cref{sec:iso_poisson}).
\item Construct the friction operator from the fluctuations of the currents and obtain the irreversible dynamics (\cref{sec:iso_friction}).
\item Prove thermodynamic consistency (\cref{sec:iso_consistency}).
\item Derive the noise terms from the FDT (\cref{sec:iso_fdt}).
\item Assemble the final equations (\cref{sec:iso_final}).
\item Specialize to symmetric and non-symmetric mixtures and compare them (\cref{sec:iso_sym,sec:iso_nonsym,sec:iso_compare}).
\end{enumerate}
The section ends with the quasi-incompressible limit (\cref{sec:QI}).

\subsection{Isothermal GENERIC}\label{sec:iso_generic}
For a system in contact with a heat bath at temperature $T_0$, with heat conduction fast enough that $T\equiv T_0$, the thermal variable can be eliminated. The variables are then $\by$ (here $\by=(\rho,\bg,c)$), and the single generator is the Helmholtz free energy $F=E-T_0S$, expressed in terms of $\by$ at $T=T_0$. Two observations make this precise.
\begin{itemize}[nosep]
\item By the thermodynamic identity $(\partial s/\partial y)_{\varepsilon}=-T^{-1}(\partial\psi/\partial y)_T$ for the non-thermal variables, the $\by$-components of the irreversible part $\Mop\cdot\fd S\bx$, evaluated at uniform $T=T_0$, take the form $-\Dop\cdot\fd F\by$. Here $\Dop=\Phiop_y^*(\Cop/T_0)\Phiop_y$ is built from the $\by$-part of the flux map (shown explicitly in \cref{sec:iso_limit}).
\item The reversible part $\Lop\cdot\fd E\bx$, restricted to $\by$, becomes $\Lop_{yy}\cdot\fd F\by$ because $\Lop\cdot\fd S\bx=0$.
\end{itemize}
Hence
\begin{equation}
\dd\by=\Big[\Lop\cdot\fd F\by-\Dop\cdot\fd F\by+\kB T_0\pd{}{\by}\cdot\Dop\Big]\dd t+\noise{\Bop}\cdot\dd\bW_t,\qquad
\noise{\Bop}\noise{\Bop}^T=2\kB T_0\,\Dop,\qquad \Dop\equiv\frac{\Mop_{yy}}{T_0},\label{eq:iso_sgeneric}
\end{equation}
with stationary distribution $P_{\rm eq}\propto\exp[-F(\by)/(\kB T_0)]\prod_I\delta(I-I_0)$. Global consistency requires $\Lop$ antisymmetric and $\Dop$ symmetric and positive semi-definite. Then
\begin{equation}
\frac{\dd F}{\dd t}=-\int\fd F\by\cdot\Dop\cdot\fd F\by\le0 .\label{eq:F_decrease}
\end{equation}
No degeneracy conditions are needed, because energy is exchanged with the bath. In the rest of this section we construct $\Lop$ and $\Dop$ directly for the isothermal model. In \cref{sec:iso_limit} we verify that they are the $T\equiv T_0$ reduction of the non-isothermal building blocks of \cref{sec:noniso}.

\subsection{State variables and the Helmholtz free-energy functional}\label{sec:iso_F}
The state variables are the conserved total mass density, the momentum density and the mass fraction:
\begin{equation}
\by=\big(\rho(\br),\ \bg(\br),\ c(\br)\big).\label{eq:y_iso}
\end{equation}
This mirrors the choice $\bx=(\rho,\bg,\varepsilon,c)$ of \cite{JelicThesis2009,ChaudhriNotes2012} with the thermal variable removed. The generator is the Helmholtz free energy at the bath temperature $T_0$:
\begin{equation}
F[\by]=\int\Big[\frac{|\bg|^2}{2\rho}+\psi(\rho,c;T_0)+\fN(\rho,c,\nabla\rho,\nabla c;T_0)\Big]\dd^3r,\qquad \psi=\rho f .\label{eq:F_iso}
\end{equation}
Its functional derivatives follow from \eqref{eq:psi_derivs} and \eqref{eq:fd_def}:
\begin{equation}
\fd F\rho\Big|_{\bg,c}=-\tfrac12|\bv|^2+\mut_\rho,\qquad \fd F\bg=\bv,\qquad \fd Fc\Big|_{\rho,\bg}=\rho\mut ,\label{eq:F_derivs}
\end{equation}
where we define
\begin{equation}
\boxed{\ \mut\equiv\mu+\frac1\rho\fd{\FN}{c}\Big|_\rho=\pd fc+\frac1\rho\Big(\pd{\fN}{c}-\divv\pd{\fN}{\nabla c}\Big),\qquad \mut_\rho\equiv\mub+\fd{\FN}{\rho}\Big|_c .\ }\label{eq:mut_iso}
\end{equation}
For the CH-only case with $\fN=\tfrac12\kappa(\rho,c)|\nabla c|^2$:
\begin{equation}
\mut=\pd fc+\frac1\rho\Big[\tfrac12\partial_c\kappa\,|\nabla c|^2-\divv(\kappa\nabla c)\Big]
\ \xrightarrow{\ \kappa\ \text{const}\ }\ \pd fc-\frac{\kappa}{\rho}\nabla^2c
\ \xrightarrow{\ \kappa=\rho\hat\kappa\ }\ \pd fc-\frac{1}{\rho}\divv(\rho\hat\kappa\nabla c).\label{eq:mut_cases}
\end{equation}

\paragraph{Summary of the quantities derived from $F$.} For later reference, every quantity that enters the isothermal equations is a derivative of the two densities $\psi(\rho,c;T_0)$ and $\fN(\rho,c,\nabla\rho,\nabla c;T_0)$:
\begin{itemize}[leftmargin=*,nosep]
\item the thermodynamic pressure $p=\rho^2\partial f/\partial\rho|_{c}=\rho\,\partial\psi/\partial\rho-\psi$;
\item the specific Gibbs energy $\mub=\partial\psi/\partial\rho|_c$ and the exchange potential $\mu=\partial f/\partial c|_\rho$ of the homogeneous mixture, \eqref{eq:psi_derivs};
\item the generalized potentials $\mut_\rho=\mub+\delta\FN/\delta\rho|_c$ and $\mut=\mu+\rho^{-1}\delta\FN/\delta c|_\rho$, \eqref{eq:mut_iso};
\item the Korteweg tensor $\bPi^K=-\bPi^{\fN}$, obtained from \cref{lem:korteweg} with $w=\fN$ in \cref{sec:iso_poisson};
\item the interstitial-working fluxes $\rho(\divv\bv)\partial\fN/\partial\nabla\rho$ and $-\dot c\,\partial\fN/\partial\nabla c$, which appear in the local free-energy balance (\cref{sec:iso_consistency}).
\end{itemize}
The transport coefficients $\eta$, $\zeta$ and $M$ are additional constitutive input. They enter only through the friction operator.

\begin{notebox}[Why the specific potential $\mut$ and not the volumetric $\delta F/\delta c$]
In the variables $(\rho,\rho_A=\rho c)$, a variation $\delta\rho_A$ at fixed $\rho$ is $\delta c=\delta\rho_A/\rho$. Hence
\begin{equation}
\fd{F}{\rho_A}\Big|_\rho=\frac1\rho\fd Fc\Big|_\rho=\mut .\label{eq:mut_rhoA}
\end{equation}
$\mut$ is therefore the potential conjugate to the \emph{conserved} density $\rho_A$. In a homogeneous system it reduces to the exchange chemical potential per unit mass, $\hat\mu_A-\hat\mu_B$. The volumetric derivative $\delta F/\delta c=\rho\mut$ is what \cite{ChaudhriNotes2012} (Eq.~4.14) calls $\mu^{C}$ and Jeli\'c \cite{JelicThesis2009} calls $\mu'-\kappa\nabla^2c$. It is the conjugate of $c$, which is not conserved. For an incompressible ($\rho=$ const) mixture the two differ by a constant factor that can be absorbed into the mobility. For a compressible mixture they do not, and only $\nabla\mut$ gives a dissipative diffusion flux with a sign-definite entropy production and the correct FDT (\cref{sec:iso_consistency,sec:iso_linear}). The compressible model of Cherfils \emph{et al.} \cite{Cherfils2019} (their Eqs.~1.1c--d, $\rho\mu=\rho\partial_cf-\Delta c$, $\partial_t(\rho c)+\ldots=\Delta\mu$) uses the specific potential, in agreement with the present result. The same conclusion follows from the free-energy imbalance alone, without GENERIC (\cref{app:C-isothermal}).
\end{notebox}

\subsection{Reversible dynamics: the Poisson operator}\label{sec:iso_poisson}
The Poisson operator is the hydrodynamic Lie--Poisson operator for a fluid that carries an advected scalar $c$ \cite{Ottinger2005,BerisEdwards1994,MorrisonGreene1980}. Its action on a test vector $\bm a=(a_\rho,\bm a_g,a_c)$ is
\begin{equation}
\begin{aligned}
(\Lop\cdot\bm a)_\rho&=-\divv(\rho\,\bm a_g),\\
(\Lop\cdot\bm a)_{\bg}&=-\rho\nabla a_\rho-\divv(\bm a_g\otimes\bg)-(\nabla\bm a_g)\cdot\bg+a_c\nabla c,\\
(\Lop\cdot\bm a)_c&=-\bm a_g\cdot\nabla c .
\end{aligned}\label{eq:L_iso_action}
\end{equation}
In index notation, $[\divv(\bm a\otimes\bg)]_i=\partial_j(a_jg_i)$ and $[(\nabla\bm a)\cdot\bg]_i=g_j\partial_ia_j$. In the compact $\br$-notation of \cite{JelicThesis2009} (Eq.~5.12) and \cite{ChaudhriNotes2012} (Eq.~6.18), this is the $3\times3$ operator obtained by deleting the energy row and column:
\begin{equation}
\Lop(\br)=-\begin{pmatrix}0&\nabla\rho&0\\ \rho\nabla&[\nabla\bg+\bg\nabla]^T&-(\nabla c)\\ 0&(\nabla c)&0\end{pmatrix}.\label{eq:L_iso_matrix}
\end{equation}
The associated bracket is
\begin{equation}
\{A,B\}=\int\Big[\rho\big(B_{\bg}\cdot\nabla A_\rho-A_{\bg}\cdot\nabla B_\rho\big)+\bg\cdot\big((B_{\bg}\cdot\nabla)A_{\bg}-(A_{\bg}\cdot\nabla)B_{\bg}\big)+\nabla c\cdot\big(B_cA_{\bg}-A_cB_{\bg}\big)\Big]\dd^3r .\label{eq:bracket_iso}
\end{equation}
\begin{itemize}[leftmargin=*,nosep]
\item \emph{Antisymmetry} is manifest in \eqref{eq:bracket_iso}.
\item \emph{Jacobi identity.} \eqref{eq:bracket_iso} is the Lie--Poisson bracket on the dual of the semidirect product of the Lie algebra of vector fields with the scalar functions $\rho$ (a density) and $c$ (a scalar) \cite{MorrisonGreene1980,HolmMarsdenRatiuWeinstein1985}. It satisfies Jacobi.
\item \emph{Liouville condition} \eqref{eq:sgeneric_conditions} holds formally \cite{Espanol2001,Ottinger2005}.
\end{itemize}

\paragraph{Reversible equations.} With $\bm a=\delta F/\delta\by$ from \eqref{eq:F_derivs}, the kinetic part $-\rho\nabla(-\tfrac12v^2)=(\nabla\bv)\cdot\bg$ cancels the term $-(\nabla\bv)\cdot\bg$, and
\begin{equation}
\partial_t\rho\big|_{\rm rev}=-\divv(\rho\bv),\qquad
\partial_t\bg\big|_{\rm rev}=-\divv(\rho\bv\bv)-\rho\nabla\mut_\rho+\rho\mut\nabla c,\qquad
\partial_tc\big|_{\rm rev}=-\bv\cdot\nabla c .\label{eq:rev_iso_raw}
\end{equation}
Apply the isothermal Gibbs--Duhem relation \eqref{eq:GD_rho_c}, $\nabla p=\rho\nabla\mub-\rho\mu\nabla c$, to the bulk part, and \cref{lem:korteweg} with $w=\fN$ to the gradient part:
\begin{equation}
\rho\nabla\mut_\rho-\rho\mut\nabla c=\divv\big(p\Id+\bPi^{\fN}\big).\label{eq:iso_force_identity}
\end{equation}
Therefore
\begin{equation}
\partial_t\bg\big|_{\rm rev}=-\divv\big(\rho\bv\bv+p\Id-\bPi^K\big),\qquad
\bPi^K\equiv-\bPi^{\fN}=-\Big[\rho\fd{\FN}\rho\Big|_c-\fN\Big]\Id-\nabla\rho\otimes\pd{\fN}{\nabla\rho}-\nabla c\otimes\pd{\fN}{\nabla c}.\label{eq:Korteweg_iso}
\end{equation}
\begin{resultbox}[Isothermal capillary (Korteweg) stress; sign convention of Chaudhri et al.\ (2014): $\bPi=p\Id-\bPi^K$]
\begin{align}
\text{CH, general }\kappa(\rho,c):&\quad \bPi^K=\tfrac12(\kappa-\rho\partial_\rho\kappa)|\nabla c|^2\Id-\kappa\nabla c\otimes\nabla c,\label{eq:PiK_general}\\
\text{CH, }\kappa\text{ const}:&\quad \bPi^K=\tfrac12\kappa|\nabla c|^2\Id-\kappa\nabla c\otimes\nabla c,\qquad \divv\bPi^K=-\kappa\nabla^2c\,\nabla c,\label{eq:PiK_const}\\
\text{CH, }\kappa=\rho\hat\kappa:&\quad \bPi^K=-\rho\hat\kappa\nabla c\otimes\nabla c,\label{eq:PiK_rho}\\
\text{vdW, }\tfrac12\kappa|\nabla\rho|^2:&\quad \bPi^K=\big(\kappa\rho\nabla^2\rho+\tfrac12\kappa|\nabla\rho|^2\big)\Id-\kappa\nabla\rho\otimes\nabla\rho\quad\text{(Eq.~2.14 of \cite{ChaudhriPRE2014})}.\label{eq:PiK_vdw}
\end{align}
The reversible force can always be written in the ``potential'' (non-conservative) form, convenient for numerics:
\begin{equation}
-\divv(p\Id-\bPi^K)=-\rho\nabla\mut_\rho+\rho\mut\nabla c .\label{eq:potential_form}
\end{equation}
For constant $\rho$ (incompressible limit) the alternatives $\rho\mut\nabla c$ and $-\rho c\nabla\mut$ differ by $\nabla(\rho c\mut)$, which is absorbed into the pressure. This reproduces the capillary forces $\mu_c\nabla c$ of \cite{JelicThesis2009} (Eq.~4.7) and $-\phi\nabla\mu$ of \cite{PadhanPandit2025} (Eq.~3.4).
\end{resultbox}

\paragraph{Reversible free-energy balance.} Using $\partial_t\psi|_{\rm rev}=\mub\,\partial_t\rho+\rho\mu\,\partial_tc=-\divv(\psi\bv)-p\divv\bv$ (from $\psi=\rho\mub-p$) and \cref{lem:korteweg}, the reversible dynamics conserves the free-energy density $\psi_{\rm tot}=\tfrac12\rho v^2+\psi+\fN$ locally:
\begin{equation}
\partial_t\psi_{\rm tot}\big|_{\rm rev}+\divv\Big[\psi_{\rm tot}\bv+\bPi\cdot\bv+\rho(\divv\bv)\pd{\fN}{\nabla\rho}\Big]=0 .\label{eq:F_local_rev}
\end{equation}
The last flux is the Dunn--Serrin interstitial working \cite{DunnSerrin1985}. It is present only if the gradient energy depends on $\nabla\rho$. (This is the isothermal version of the identity verified in \cref{app:verification}, check~1.)

\subsection{Irreversible dynamics from fluctuations (Jeli\'c--\"Ottinger)}\label{sec:iso_friction}
\paragraph{Steps 2--3: fluctuation ansatz.} The total mass is transported only convectively, since $\bv$ is barycentric. Its fast fluctuation therefore vanishes. Momentum and species-$A$ mass fluctuate by divergences of fluctuating currents:
\begin{equation}
\Delta\rho^{\rm f}=0,\qquad \Delta\bg^{\rm f}=-\divv\bW^{\tau},\qquad \Delta c^{\rm f}=-\frac1\rho\divv\bW^{J}\quad(\text{i.e. }\Delta(\rho c)^{\rm f}=-\divv\bW^J).\label{eq:fluct_iso}
\end{equation}
$\bW^\tau$ is a symmetric tensor Wiener increment (fluctuating momentum flux; symmetry ensures angular-momentum conservation) and $\bW^J$ is a vector Wiener increment (fluctuating diffusion flux). The factor $1/\rho$ in $\Delta c^{\rm f}$ is \emph{forced} by conservation of $\int\rho c$. It is absent in the incompressible treatment of \cite{JelicThesis2009}. Pairing with a test field $\bm a$ and integrating by parts:
\begin{equation}
\int\bm a\cdot\Delta\by^{\rm f}=\int\Big[\bW^\tau:(\nabla\bm a_g)^{\sym}+\bW^J\cdot\nabla\Big(\frac{a_c}{\rho}\Big)\Big]
\quad\Longrightarrow\quad
\Phiop_{\rm iso}(\bm a)=\Big((\nabla\bm a_g)^{\sym},\ \nabla\frac{a_c}{\rho}\Big).\label{eq:Phi_iso}
\end{equation}

\paragraph{Step 4: second moments.} Locality, isotropy and the Curie principle (no coupling between the rank-2 $\bW^\tau$ and the rank-1 $\bW^J$) give
\begin{equation}
\begin{aligned}
\avg{W^\tau_{ij}(\br)W^\tau_{kl}(\br')}&=2\kB T_0\tau\Big[\eta\big(\delta_{ik}\delta_{jl}+\delta_{il}\delta_{jk}-\tfrac23\delta_{ij}\delta_{kl}\big)+\zeta\delta_{ij}\delta_{kl}\Big]\delta(\br-\br'),\\
\avg{W^J_i(\br)W^J_j(\br')}&=2\kB T_0\tau\,M\,\delta_{ij}\,\delta(\br-\br'),\qquad \avg{W^\tau_{ij}W^J_k}=0 .
\end{aligned}\label{eq:moments_iso}
\end{equation}
Here $\eta,\zeta,M\ge0$ may depend on $(\rho,c)$. With $\kB\to\kB T_0$ in \cref{lem:JO} (isothermal normalization, \eqref{eq:iso_sgeneric}), the correlation matrix is $\bar\Cop={\rm diag}(\bar\Cop^{\tau\tau},M\Id)$ with $\bar\Cop^{\tau\tau}_{ijkl}=\eta(\delta_{ik}\delta_{jl}+\delta_{il}\delta_{jk}-\tfrac23\delta_{ij}\delta_{kl})+\zeta\delta_{ij}\delta_{kl}$.

\paragraph{Step 5: the isothermal friction operator.} By \cref{lem:JO}, $\Dop=\Phiop_{\rm iso}^*\bar\Cop\,\Phiop_{\rm iso}$. Explicitly, the only non-zero blocks are
\begin{equation}
\Dop_{\bg\bg}\cdot\bm b=-\divv\Big[\eta\big(\nabla\bm b+(\nabla\bm b)^T\big)+\big(\zeta-\tfrac23\eta\big)(\divv\bm b)\Id\Big],\qquad
\Dop_{cc}\,b=-\frac1\rho\divv\Big[M\nabla\Big(\frac b\rho\Big)\Big].\label{eq:D_iso}
\end{equation}
In $\br$-notation, $\Dop_{cc}=-\rho^{-1}\nabla\cdot M\nabla\rho^{-1}$, the compressible generalization of $\Mop^{\rm dif}_{cc}=-\nabla\cdot MT\nabla$ in Eq.~(5.20b) of \cite{JelicThesis2009} and Eq.~(6.23) of \cite{ChaudhriNotes2012}. Properties:
\begin{itemize}[leftmargin=*,nosep]
\item $\Dop$ is symmetric and positive semi-definite (\cref{lem:JO}).
\item $\Dop_{\bg\bg}\cdot\bm b=0$ for rigid motions $\bm b=\bm b_0+\bm\omega\times\br$, so total momentum and angular momentum are not dissipated.
\item $\rho\,(\Dop_{cc}b)$ is a divergence, so $\int\rho c$ is conserved.
\item $\Dop_{cc}(\rho\,\lambda)=0$ for constant $\lambda$, so uniform $\mut$ is stationary.
\end{itemize}

\paragraph{Irreversible equations.} The forces are $\Phiop_{\rm iso}(\delta F/\delta\by)=(\bD,\nabla\mut)$ with $\bD=(\nabla\bv)^{\sym}$, and $-\Dop\cdot\delta F/\delta\by$ gives
\begin{gather}
\partial_t\rho\big|_{\rm irr}=0,\qquad \partial_t\bg\big|_{\rm irr}=\divv\btau,\qquad \rho\,\partial_tc\big|_{\rm irr}=-\divv\bJ,\label{eq:irr_iso}\\
\btau=2\eta\mathring{\bD}+\zeta(\divv\bv)\Id=\eta\big[\nabla\bv+(\nabla\bv)^T\big]+\big(\zeta-\tfrac23\eta\big)(\divv\bv)\Id,\qquad
\bJ=-M\nabla\mut .\label{eq:tau_J_iso}
\end{gather}

\subsection{Thermodynamic consistency of the isothermal model}\label{sec:iso_consistency}
\paragraph{Global.} By antisymmetry of $\Lop$ and \eqref{eq:F_decrease},
\begin{equation}
\frac{\dd F}{\dd t}=-\int\fd F\by\cdot\Dop\cdot\fd F\by=-\int\Big[\bD:\bar\Cop^{\tau\tau}:\bD+M|\nabla\mut|^2\Big]
=-\int\Big[2\eta\,\mathring{\bD}:\mathring{\bD}+\zeta(\divv\bv)^2+M|\nabla\mut|^2\Big]\le0 .\label{eq:dFdt_iso}
\end{equation}
This is the isothermal statement of the second law, $T_0\dot S_{\rm univ}=-\dot F\ge0$. Equality holds if and only if the motion is rigid ($\mathring{\bD}=0$, $\divv\bv=0$) and $\mut$ is uniform, i.e.\ at the equilibria of \cref{sec:equilibrium_interface}.

\paragraph{Local.} Adding the irreversible contributions to \eqref{eq:F_local_rev}:
\begin{align*}
\partial_t(\tfrac12\rho v^2)\big|_{\rm irr}&=\bv\cdot\divv\btau=\divv(\btau\cdot\bv)-\btau:\nabla\bv,\\
\partial_t(\psi+\fN)\big|_{\rm irr}&=\fd{(\Psi+\FN)}{c}\partial_tc\big|_{\rm irr}+\divv\Big(\pd{\fN}{\nabla c}\partial_tc\big|_{\rm irr}\Big)
=-\mut\,\divv\bJ-\divv\Big(\pd{\fN}{\nabla c}\frac{\divv\bJ}{\rho}\Big).
\end{align*}
Hence the complete local balance is
\begin{equation}
\partial_t\psi_{\rm tot}+\divv\Big[\psi_{\rm tot}\bv+\bPi\cdot\bv+\rho(\divv\bv)\pd{\fN}{\nabla\rho}-\btau\cdot\bv+\mut\bJ+\frac{\divv\bJ}{\rho}\pd{\fN}{\nabla c}\Big]
=-\Big[\btau:\nabla\bv+M|\nabla\mut|^2\Big]\le0 .\label{eq:F_local_full}
\end{equation}
The term $\rho^{-1}(\divv\bJ)\partial\fN/\partial\nabla c=-\dot c\,\partial\fN/\partial\nabla c$ is the irreversible interstitial working flux. It is the analogue of the term ``$-\rho\Phi\,\dd c/\dd t$'' of Eq.~(4.17) in \cite{ChaudhriNotes2012} and of $\partial_z\hat\varepsilon\,\divv\bm j_1$ in Eq.~(43) of \cite{HeidaMalekRajagopal2012}.

\paragraph{The volumetric alternative is not consistent.} Replace $\bJ=-M\nabla\mut$ by $\bJ'=-M'\nabla(\rho\mut)$, the form of Eqs.~(4.14) and (7.1)--(7.7) of \cite{ChaudhriNotes2012}. The diffusive part of $\dot F$ becomes
\[
\int\bJ'\cdot\nabla\mut=-\int M'\nabla(\rho\mut)\cdot\nabla\mut=-\int M'\big(\rho|\nabla\mut|^2+\mut\nabla\rho\cdot\nabla\mut\big),
\]
whose sign is not fixed. Wherever $\nabla\rho\neq0$, for example across a liquid--liquid interface between phases of different density, free energy can be produced.

\subsection{Noise terms from the fluctuation--dissipation theorem}\label{sec:iso_fdt}
By \cref{lem:JO}(e), the noise of \eqref{eq:iso_sgeneric} is $\noise\Bop\cdot\dd\bW_t=\Phiop^*_{\rm iso}\dd\bW$. In continuum form it is the divergence of stochastic fluxes $\noise\btau$ (momentum) and $\noise\bJ$ (species-$A$ mass), white in space and time. The covariances are fixed by the \emph{same} coefficients as the deterministic fluxes \eqref{eq:tau_J_iso}:
\begin{equation}
\boxed{
\begin{aligned}
\avg{\noise\tau_{ij}(\br,t)\noise\tau_{kl}(\br',t')}&=2\kB T_0\Big[\eta\big(\delta_{ik}\delta_{jl}+\delta_{il}\delta_{jk}\big)+\big(\zeta-\tfrac23\eta\big)\delta_{ij}\delta_{kl}\Big]\delta(\br-\br')\delta(t-t'),\\
\avg{\noise J_i(\br,t)\noise J_j(\br',t')}&=2\kB T_0\,M\,\delta_{ij}\,\delta(\br-\br')\delta(t-t'),\qquad
\avg{\noise\tau_{ij}\noise J_k}=0 .
\end{aligned}}\label{eq:noise_iso}
\end{equation}
\textbf{Explicit construction} \cite{ChaudhriPRE2014}, Eq.~(2.18). Let $\mathcal W^v_{ij}$ and $\mathcal W^c_i$ be independent white-noise fields, $\avg{\mathcal W^v_{ij}\mathcal W^v_{kl}}=\delta_{ik}\delta_{jl}\delta\delta$, $\avg{\mathcal W^c_i\mathcal W^c_j}=\delta_{ij}\delta\delta$, and $\noise{\mathcal W}^v=(\mathcal W^v+\mathcal W^{vT})/\sqrt2$. Then
\begin{equation}
\noise\btau=\sqrt{2\eta\kB T_0}\,\noise{\mathcal W}^v+\Big(\sqrt{\frac{\zeta\kB T_0}{3}}-\frac{\sqrt{2\eta\kB T_0}}{3}\Big)\tr\big(\noise{\mathcal W}^v\big)\Id,\qquad
\noise\bJ=\sqrt{2M\kB T_0}\,\bm{\mathcal W}^c .\label{eq:noise_construction_iso}
\end{equation}
\emph{Proof} that \eqref{eq:noise_construction_iso} has covariance \eqref{eq:noise_iso}. Write $\noise\btau=a\noise{\mathcal W}^v+b\tr(\noise{\mathcal W}^v)\Id$. Using $\avg{\noise{\mathcal W}^v_{ij}\noise{\mathcal W}^v_{kl}}=\delta_{ik}\delta_{jl}+\delta_{il}\delta_{jk}$, $\avg{\noise{\mathcal W}^v_{ij}\tr\noise{\mathcal W}^v}=2\delta_{ij}$ and $\avg{(\tr\noise{\mathcal W}^v)^2}=6$:
\[
\avg{\noise\tau_{ij}\noise\tau_{kl}}=a^2(\delta_{ik}\delta_{jl}+\delta_{il}\delta_{jk})+(4ab+6b^2)\delta_{ij}\delta_{kl}.
\]
With $a^2=2\eta\kB T_0$ and $b=\sqrt{\zeta\kB T_0/3}-a/3$, one gets $4ab+6b^2=2\zeta\kB T_0-\tfrac23a^2=2\kB T_0(\zeta-\tfrac23\eta)$. \hfill$\square$

\noindent For $\zeta=0$ the trace coefficient is $-\sqrt{2\eta\kB T_0}/3$, not the value $+\sqrt{2\eta\kB T/3}$ printed in Eq.~(7.6) of \cite{ChaudhriNotes2012}. The latter gives $4ab+6b^2=2\eta\kB T(2+4/\sqrt3)$ instead of $-\tfrac43\eta\kB T$ (\cref{app:verification}, check~5). The covariance stated in Eq.~(7.10) of \cite{ChaudhriNotes2012} is correct.

\paragraph{Thermal drift and invariant measure.} When $M$, $\eta$ or $\zeta$ depend on the state, \eqref{eq:iso_sgeneric} contains the drift $\kB T_0\,\partial_{\by}\cdot\Dop$. For example, for $M=M(c)$ its $c$-component is formally $\kB T_0\rho^{-1}\divv[M'(c)\nabla(\rho^{-1}\delta(\mathbf 0))]$. Following standard practice \cite{ChaudhriPRE2014,DonevVandenEijnden2014} it is omitted in the continuum equations. It must be reinstated (or the noise interpreted kinetically) in spatial discretizations with strongly state-dependent mobility, if the discrete dynamics is to sample $P_{\rm eq}\propto\exp(-F/\kB T_0)$ exactly.

\subsection{Assembling the isothermal LLCHNS equations}\label{sec:iso_final}
We now put the pieces together. The isothermal stochastic GENERIC equation \eqref{eq:iso_sgeneric} is
\[
\dd\by=\underbrace{\Lop\cdot\fd F\by\,\dd t}_{\text{reversible, \cref{sec:iso_poisson}}}\ \underbrace{-\ \Dop\cdot\fd F\by\,\dd t}_{\text{dissipative, \cref{sec:iso_friction}}}\ +\ \underbrace{\Phiop_{\rm iso}^*\,\dd\bW}_{\text{noise, \cref{sec:iso_fdt}}} ,
\]
with the thermal drift omitted as discussed above. Component by component:
\begin{itemize}[leftmargin=*,nosep]
\item \emph{mass:} $\Lop$ gives $-\divv(\rho\bv)$; $\Dop$ and the noise give nothing, because $\Delta\rho^{\rm f}=0$;
\item \emph{momentum:} $\Lop$ gives the Euler flux $\rho\bv\bv+p\Id-\bPi^K$ by \eqref{eq:Korteweg_iso}; $\Dop$ gives $\divv\btau$; the noise gives $\divv\noise\btau$;
\item \emph{composition:} $\Lop$ gives advection $-\bv\cdot\nabla c$; $\Dop$ gives $\rho^{-1}\divv(M\nabla\mut)$; the noise gives $-\rho^{-1}\divv\noise\bJ$.
\end{itemize}
Multiplying the composition equation by $\rho$ and combining it with the continuity equation gives the conservation law for $\rho c$. The result is the following model.

\begin{resultbox}[Isothermal, fully compressible fluctuating CHNS (LLCHNS) equations]\label{res:iso_final}
In the conservative form of \cite{ChaudhriPRE2014} (their Eq.~3.1):
\begin{equation}
\frac{\partial\bm U}{\partial t}=-\divv\bm F_H-\divv\bm F_K-\divv\bm F_D-\divv\bm F_S,\qquad
\bm U=\begin{pmatrix}\rho\\ \rho\bv\\ \rho c\end{pmatrix},\label{eq:U_iso}
\end{equation}
\begin{equation}
\bm F_H=\begin{pmatrix}\rho\bv\\ \rho\bv\bv^T+p\Id\\ \rho c\,\bv\end{pmatrix},\quad
\bm F_K=-\begin{pmatrix}0\\ \bPi^K\\ 0\end{pmatrix},\quad
\bm F_D=-\begin{pmatrix}0\\ \btau\\ M\nabla\mut\end{pmatrix},\quad
\bm F_S=-\begin{pmatrix}0\\ \noise\btau\\ -\noise\bJ\end{pmatrix}.\label{eq:fluxes_iso}
\end{equation}
Written out:
\begin{align}
&\partial_t\rho+\divv(\rho\bv)=0,\label{eq:iso_mass}\\
&\partial_t(\rho\bv)+\divv(\rho\bv\bv^T)+\nabla p=\divv\big(\bPi^K+\btau+\noise\btau\big),\label{eq:iso_mom}\\
&\partial_t(\rho c)+\divv(\rho c\bv)=\divv\big(M\nabla\mut\big)-\divv\noise\bJ .\label{eq:iso_c}
\end{align}
Constitutive relations:
\begin{itemize}[leftmargin=*,nosep]
\item $p=\rho^2\partial f/\partial\rho|_{c,T_0}$;
\item $\mut=\partial f/\partial c+\rho^{-1}\delta\FN/\delta c|_\rho$, from \eqref{eq:mut_iso}--\eqref{eq:mut_cases};
\item $\bPi^K$ from \eqref{eq:Korteweg_iso}--\eqref{eq:PiK_vdw};
\item $\btau=\eta[\nabla\bv+\nabla\bv^T]+(\zeta-\frac23\eta)(\divv\bv)\Id$;
\item noises from \eqref{eq:noise_iso}--\eqref{eq:noise_construction_iso}.
\end{itemize}
The fields $f(\rho,c,T_0)$, $\kappa(\rho,c)$, $\eta(\rho,c)$, $\zeta(\rho,c)$, $M(\rho,c)\ge0$ are arbitrary constitutive input. Thermodynamic consistency is guaranteed by \eqref{eq:dFdt_iso}--\eqref{eq:F_local_full}, and the FDT by \cref{lem:JO}.
\end{resultbox}

\paragraph{Special cases.}
\begin{itemize}[leftmargin=*,nosep]
\item $\kappa$ constant (volumetric):
\[
\divv\bPi^K=-\kappa\nabla^2c\nabla c,\qquad \mut=\partial_cf-\kappa\nabla^2c/\rho .
\]
This is the deterministic model of \cite{Cherfils2019}.
\item $\kappa=\rho\hat\kappa$ (specific):
\[
\divv\bPi^K=-\divv(\rho\hat\kappa\nabla c\otimes\nabla c),\qquad \mut=\partial_cf-\rho^{-1}\divv(\rho\hat\kappa\nabla c).
\]
This is the deterministic model of \cite{ChaudhriNotes2012} (Eqs.~4.13--4.16) with the flux corrected to $\bJ=-M\nabla\mut$, i.e.\ $M\nabla(\mu^C/\rho)$ in the notation there.
\item In the variable $\phi=2c-1$, use \cref{tab:phi_map}: $\partial_t(\rho\phi)+\divv(\rho\phi\bv)=\divv(M_\phi\nabla\mut_\phi)-\divv\noise\bJ_\phi$, with $\avg{\noise\bJ_\phi\noise\bJ_\phi}=2\kB T_0M_\phi$.
\end{itemize}

\subsection{Symmetric mixtures: model I-S}\label{sec:iso_sym}
\paragraph{General symmetric mixture.} By \cref{prop:symmetry}:
\begin{itemize}[nosep]
\item $p(\rho,\phi)$ and $\eta,\zeta,M$ are even in $\phi$;
\item $\mut$ is odd;
\item composition--pressure couplings ($\partial p/\partial\phi$, $\partial\mut/\partial\rho$, $K_{\rho c}$) are odd and vanish at $\phi=0$.
\end{itemize}
The equations of \cref{res:iso_final} keep their general form. Around the critical composition $\phi_0=0$ the composition fluctuations decouple from the sound modes at linear order. Off-critical (for example inside the coexisting phases, $\phi=\pm\phi_{\rm eq}$) they do not, unless the model is decoupled.

\paragraph{Decoupled symmetric model in the variable $\phi$.} Take \cref{def:decoupled} with $f=f_0(\rho)+g(\phi)$ at $T_0$, constant $\kappa_\phi$, and even $M_\phi(\phi)$, $\eta$, $\zeta$. Then $\mut_\phi=g'(\phi)-(\kappa_\phi/\rho)\nabla^2\phi$, and \cref{res:iso_final} becomes
\begin{resultbox}[Isothermal, symmetric, decoupled compressible LLCHNS]
\begin{align}
&\partial_t\rho+\divv(\rho\bv)=0,\label{eq:sym_iso_mass}\\
&\partial_t(\rho\bv)+\divv(\rho\bv\bv^T)+\nabla p_0(\rho)=\divv(\btau+\noise\btau)-\divv\Big[\kappa_\phi\Big(\nabla\phi\otimes\nabla\phi-\tfrac12|\nabla\phi|^2\Id\Big)\Big],\label{eq:sym_iso_mom}\\
&\partial_t(\rho\phi)+\divv(\rho\phi\bv)=\divv\Big[M_\phi\nabla\Big(g'(\phi)-\frac{\kappa_\phi}{\rho}\nabla^2\phi\Big)\Big]-\divv\noise\bJ_\phi,\label{eq:sym_iso_phi}\\
&\avg{\noise J_{\phi,i}(\br,t)\noise J_{\phi,j}(\br',t')}=2\kB T_0M_\phi(\phi)\,\delta_{ij}\delta(\br-\br')\delta(t-t'),\quad \noise\btau\ \text{as in \eqref{eq:noise_iso}}.\nonumber
\end{align}
\end{resultbox}
\begin{itemize}[leftmargin=*,nosep]
\item With $p_0=\rho a^2$ (i.e.\ $f_0=a^2\ln\rho$) and $g=\frac\alpha4(\phi^2-\beta/\alpha)^2$, this is the FCHNS model of \cite{ChaudhriNotes2012} (Section~7.4, ``expanded form''), with three changes.
\begin{enumerate}[nosep]
\item The diffusion flux is driven by $\nabla\mut_\phi$ (specific), not by $\nabla\mu^C=\nabla(\rho\mut_\phi)$.
\item The volumetric gradient coefficient gives the stress $-\divv[\kappa_\phi(\nabla\phi\nabla\phi-\tfrac12|\nabla\phi|^2\Id)]$. If one prefers the specific choice of \cite{ChaudhriNotes2012} $\kappa_\phi=\rho\hat\kappa$, the stress is $-\divv(\rho\hat\kappa\nabla\phi\nabla\phi)$ and $\mut_\phi=g'-\rho^{-1}\divv(\rho\hat\kappa\nabla\phi)$.
\item The trace part of $\noise\btau$ is as in \eqref{eq:noise_construction_iso}.
\end{enumerate}
\item The mobility may be constant or degenerate. The choice $M_\phi=M_0(1-\phi^2)$ of \cite{ChaudhriNotes2012} (Section~7.3) is the symmetric case of \eqref{eq:M_D}. With a state-dependent $M_\phi$ the noise is multiplicative. See the remark on the thermal drift in \cref{sec:iso_fdt}.
\end{itemize}

\subsubsection{Linear fluctuations and structure factors (symmetric decoupled case)}\label{sec:iso_linear}
Linearize \eqref{eq:sym_iso_mass}--\eqref{eq:sym_iso_phi} about the homogeneous state $(\rho_0,\phi_0,\bv=0)$, with $c_T^2=p_0'(\rho_0)$ (isothermal sound speed), $g''_0=g''(\phi_0)$, $M_0=M_\phi(\phi_0)$ and $\nu_l=(\tfrac43\eta+\zeta)/\rho_0$. In Fourier space ($\nabla\to i\bk$), the composition mode obeys the Ornstein--Uhlenbeck equation
\begin{equation}
\partial_t\delta\phi_{\bk}=-\Gamma_k\,\delta\phi_{\bk}-\frac{i\bk\cdot\noise\bJ_{\phi,\bk}}{\rho_0},\qquad
\Gamma_k=\frac{M_0k^2}{\rho_0^2}\big(\rho_0g''_0+\kappa_\phi k^2\big).\label{eq:OU_phi}
\end{equation}
The longitudinal momentum and density obey a damped wave equation driven by $k_ik_j\noise\tau_{ij}$. The two sets are independent. The static and dynamic structure factors are
\begin{align}
S_{\phi\phi}(k)&=\avg{|\delta\phi_{\bk}|^2}=\frac{2\kB T_0M_0k^2/\rho_0^2}{2\Gamma_k}=\frac{\kB T_0}{\rho_0g''_0+\kappa_\phi k^2},\qquad
S_{\phi\phi}(k,\omega)=\frac{2\kB T_0M_0k^2/\rho_0^2}{\omega^2+\Gamma_k^2},\label{eq:Sphiphi}\\
S_{\rho\rho}(k)&=\frac{\rho_0\kB T_0}{c_T^2},\qquad
S_{\rho\rho}(k,\omega)=\frac{2\kB T_0\rho_0\nu_lk^4}{(c_T^2k^2-\omega^2)^2+(\nu_lk^2\omega)^2},\qquad S_{\rho\phi}=0 .\label{eq:Srhorho}
\end{align}
\textbf{FDT check.} The thermodynamic (Gibbs) prediction follows from $P_{\rm eq}\propto\exp(-F/\kB T_0)$ at fixed total masses. The Hessian of $\Psi-\lambda_B\int\rho-\lambda_A\int\rho_A$ in $(\delta\rho,\delta\phi)$ is diagonal with entries $c_T^2/\rho_0$ and $\rho_0g''_0+\kappa_\phi k^2$. The Lagrange-multiplier term cancels the $\psi_{\rho\phi}\delta\rho\delta\phi$ cross term; see \eqref{eq:Hk} below. This gives exactly \eqref{eq:Sphiphi}--\eqref{eq:Srhorho}.

\emph{The volumetric variant fails the check.} If instead the flux is $-M'\nabla(\rho\mut_\phi)$ with noise $2\kB T_0M'$, then $\Gamma_k$ acquires an extra factor $\rho_0$ while the noise does not. The result is $S_{\phi\phi}=\kB T_0/[\rho_0(\rho_0g''_0+\kappa_\phi k^2)]$, off by a factor $\rho_0$. This is the structure factor $\propto1/\rho_o^2$ of Eq.~(7.15) of \cite{ChaudhriNotes2012} (there with $\kappa_\phi=\rho\hat\kappa$). It agrees with the Gibbs value only in units where $\rho_0=1$ (\cref{app:verification}, check~4).

\textbf{Spinodal decomposition (Cahn--Cook).} For $g_0''<0$ the deterministic part of \eqref{eq:OU_phi} is unstable for $k<k_c$, with $k_c^2=-\rho_0g''_0/\kappa_\phi$. The fastest-growing mode is $k_m=k_c/\sqrt2$. The noise term is the Cook term \cite{Cook1970}: it seeds the instability and, through the nonlinear terms, renormalizes the coarsening.

\subsection{Non-symmetric mixtures: model I-N}\label{sec:iso_nonsym}
For a non-symmetric mixture all equations of \cref{res:iso_final} hold as written, with general $f(\rho,c,T_0)$, a general gradient energy (in particular with $K_{\rho c}\neq0$, \eqref{eq:vdw_K}), and composition-dependent $\eta,\zeta,M$. Three physical couplings, all absent or odd in the symmetric case, now appear generically. Each is contained in the general equations without additional assumptions.

\paragraph{(i) Barodiffusion.} By \eqref{eq:maxwell}, at fixed $T$,
\begin{equation}
\nabla\mu=\pd\mu c\Big|_{T,p}\nabla c+(\hat v_A-\hat v_B)\nabla p .\label{eq:baro}
\end{equation}
Hence the diffusion flux of \eqref{eq:iso_c} is
\begin{equation}
\bJ=-M\nabla\mut=-\rho D\Big[\nabla c+\frac{k_p}{p}\nabla p\Big]-M\nabla\Big(\frac1\rho\fd{\FN}c\Big),\qquad
\rho D=M\pd\mu c\Big|_{T,p},\qquad \frac{k_p}{p}=\frac{\hat v_A-\hat v_B}{(\partial\mu/\partial c)_{T,p}} .\label{eq:J_baro}
\end{equation}
This is the Landau--Lifshitz form \cite{LandauLifshitzFM} (Secs.~58--59) with the barodiffusion ratio $k_p$, plus the Cahn--Hilliard term. The barodiffusion coefficient is \emph{not} an independent transport coefficient: it is fixed by thermodynamics. It vanishes for density-matched components.

\paragraph{(ii) Composition--pressure coupling.} The pressure $p(\rho,c)$ depends on composition, with $(\partial p/\partial c)_\rho=\rho^2(\partial\mu/\partial\rho)_c$ by \eqref{eq:maxwell2}. A composition gradient at fixed density therefore produces a bulk force $-\nabla p$, in addition to the capillary force.

\paragraph{(iii) Coexisting phases of different density.} The coexisting phases of different density are separated by an interface across which both $\rho$ and $c$ vary. The density profile is not slaved to $c$; it is determined by \eqref{eq:EL}, with gradient energy in both variables if $K_{\rho\rho},K_{\rho c}\neq0$.

\paragraph{Explicit equations for the binary van der Waals mixture.} With \eqref{eq:vdw_mix}--\eqref{eq:vdw_K} and $n_A=\rho c/m_A$, $n_B=\rho(1-c)/m_B$:
\begin{gather}
p=\frac{\rho\kB T_0/\bar m}{1-b_An_A-b_Bn_B}-\sum_{\alpha\beta}a_{\alpha\beta}n_\alpha n_\beta,\qquad
\mu=\frac{\mu_A}{m_A}-\frac{\mu_B}{m_B},\nonumber\\
\fN=\tfrac12K_{\rho\rho}|\nabla\rho|^2+K_{\rho c}\nabla\rho\cdot\nabla c+\tfrac12K_{cc}|\nabla c|^2 .\label{eq:vdw_iso_closure}
\end{gather}
The Korteweg tensor follows from \eqref{eq:Korteweg_iso}:
\begin{equation}
\begin{aligned}
\bPi^K=&-\Big[\rho\Big(\tfrac12\partial_\rho K_{\rho\rho}|\nabla\rho|^2+\partial_\rho K_{\rho c}\nabla\rho\cdot\nabla c+\tfrac12\partial_\rho K_{cc}|\nabla c|^2-\divv(K_{\rho\rho}\nabla\rho+K_{\rho c}\nabla c)\Big)-\fN\Big]\Id\\
&-K_{\rho\rho}\nabla\rho\nabla\rho-K_{\rho c}(\nabla\rho\nabla c+\nabla c\nabla\rho)-K_{cc}\nabla c\nabla c ,
\end{aligned}\label{eq:PiK_vdwmix}
\end{equation}
and the generalized exchange potential is
\begin{equation}
\mut=\mu+\frac1\rho\Big[\tfrac12\partial_cK_{\rho\rho}|\nabla\rho|^2+\partial_cK_{\rho c}\nabla\rho\cdot\nabla c+\tfrac12\partial_cK_{cc}|\nabla c|^2-\divv(K_{\rho c}\nabla\rho+K_{cc}\nabla c)\Big].\label{eq:mut_vdwmix}
\end{equation}
For $n_B\to0$ (pure $A$) the momentum equation reduces to that of \cite{ChaudhriPRE2014} at constant temperature.

\paragraph{Linear fluctuations: the full $2\times2$ static structure factor.} About a homogeneous state $(\rho_0,c_0)$ the relevant equilibrium Hessian is that of $\Psi-\lambda_B\int\rho-\lambda_A\int\rho_A$, where the multipliers enforce conservation of the masses. In the variables $(\delta\rho,\delta c)$, using $\delta\rho_A=c_0\delta\rho+\rho_0\delta c$, \eqref{eq:psi_derivs} and \eqref{eq:maxwell2}, it is
\begin{equation}
H(k)=\begin{pmatrix}
\dfrac1{\rho_0}\Big(\pd p\rho\Big)_{c}+K_{\rho\rho}k^2 & \dfrac1{\rho_0}\Big(\pd pc\Big)_{\rho}+K_{\rho c}k^2\\[8pt]
\dfrac1{\rho_0}\Big(\pd pc\Big)_{\rho}+K_{\rho c}k^2 & \rho_0\Big(\pd\mu c\Big)_{\rho}+K_{cc}k^2
\end{pmatrix}.\label{eq:Hk}
\end{equation}
(Without the multiplier term, the off-diagonal element would contain the spurious extra piece $\mu_0=\partial^2\psi/\partial\rho\partial c-\rho_0(\partial\mu/\partial\rho)_c$.) The static structure factors are $S(k)=\kB T_0H(k)^{-1}$. In particular, at $k\to0$,
\begin{equation}
S_{cc}(0)=\frac{\kB T_0}{\rho_0(\partial\mu/\partial c)_{T,p}},\qquad
S_{\rho\rho}(0)=\frac{\rho_0\kB T_0}{(\partial p/\partial\rho)_{T,\mu}},\qquad
S_{\rho c}(0)=-\frac{\kB T_0\,(\partial p/\partial c)_\rho/\rho_0}{(\partial\mu/\partial c)_\rho(\partial p/\partial\rho)_{T,\mu}} .\label{eq:S0_nonsym}
\end{equation}
These are the classical fluctuation formulas for binary mixtures \cite{LandauLifshitzSP,deZarateSengers2006}. They follow from $\det H(0)=(\partial p/\partial\rho)_c(\partial\mu/\partial c)_p=(\partial\mu/\partial c)_\rho(\partial p/\partial\rho)_\mu$.

\emph{Dynamics.} The linearized equations couple density, longitudinal momentum and composition:
\begin{equation}
\begin{aligned}
\partial_t\delta\rho&=-i\bk\cdot\delta\bg,\\
\partial_t\delta\rho_A&=-ic_0\bk\cdot\delta\bg-Mk^2\,\delta\mut-i\bk\cdot\noise\bJ\qquad\big(\text{i.e. }-\divv\bJ\text{ with }\bJ=-M\nabla\delta\mut\big),\\
\partial_t\delta\bg&=-i\bk\Big[\delta p+\rho_0k^2(K_{\rho\rho}\delta\rho+K_{\rho c}\delta c)\Big]-\nu_l k^2\delta\bg_\parallel-\nu_tk^2\delta\bg_\perp+i\bk\cdot\noise\btau,
\end{aligned}\label{eq:lin_nonsym_iso}
\end{equation}
with $\delta\mut=(\partial\mu/\partial\rho)_c\delta\rho+(\partial\mu/\partial c)_\rho\delta c+\rho_0^{-1}k^2(K_{\rho c}\delta\rho+K_{cc}\delta c)$, $\nu_t=\eta/\rho_0$, and $\nu_l=(\frac43\eta+\zeta)/\rho_0$. The barodiffusion enters through the term $(\partial\mu/\partial\rho)_c\,\delta\rho=\rho_0^{-2}(\partial p/\partial c)_\rho\,\delta\rho$ in $\delta\mut$. We verified symbolically (\cref{app:verification}, check~4) that for \emph{arbitrary} $(\partial^2\psi)$, $K_{ab}$, $\eta,\zeta,M$, the stationary covariance of \eqref{eq:lin_nonsym_iso} with the noises \eqref{eq:noise_iso} is exactly $\kB T_0H^{-1}$. The FDT is therefore satisfied for non-symmetric mixtures as well. The general reason is given in \cref{app:linear_fdt}.

\subsection{Symmetric versus non-symmetric mixtures: what changes}\label{sec:iso_compare}
The general equations of \cref{res:iso_final} are the same for both classes of mixture. The difference lies entirely in the constitutive input, and through it in which couplings survive. \Cref{tab:iso_compare} collects the differences.
\begin{table}[H]
\centering\small
\caption{Isothermal model: symmetric (I-S) versus non-symmetric (I-N) mixtures. ``Odd'' means odd in $\phi=2c-1$, hence zero at the critical composition $\phi=0$.}\label{tab:iso_compare}
\renewcommand{\arraystretch}{1.2}
\begin{tabular}{>{\raggedright\arraybackslash}p{4.2cm}>{\raggedright\arraybackslash}p{5.1cm}>{\raggedright\arraybackslash}p{5.7cm}}
\toprule
ingredient & symmetric (I-S) & non-symmetric (I-N)\\
\midrule
free energy & $f(\rho,\phi)$ even in $\phi$; decoupled case $f_0(\rho)+g(\phi)$ & general $f(\rho,c)$\\
pressure & $p(\rho,\phi)$ even; $p_0(\rho)$ if decoupled & $p(\rho,c)$; $(\partial p/\partial c)_\rho=\rho^2(\partial\mu/\partial\rho)_c\neq0$\\
exchange potential & $\mut_\phi$ odd & $\mut$ with no parity\\
gradient energy & $K_{\rho\rho},K_{cc}$ even, $K_{\rho c}$ odd; usually only $\tfrac12\kappa_\phi|\nabla\phi|^2$ & all three $K_{ab}$ generic\\
barodiffusion & $k_p\propto\hat v_A-\hat v_B$ odd; zero if decoupled & $k_p=p(\hat v_A-\hat v_B)/(\partial\mu/\partial c)_{T,p}$, \eqref{eq:J_baro}\\
coexisting phases & equal density, $\phi_\pm=\pm\phi_{\rm eq}$, $\mu_{\rm coex}=0$ & different densities; $\rho$ and $c$ both vary across the interface\\
linear modes at $\phi_0=0$ & composition decouples from sound & composition couples to sound through $p_c$ and $K_{\rho c}$\\
density spectrum & Brillouin doublet only & additional central (barodiffusive) peak, \eqref{eq:dsf_IN_I}\\
static structure factor & $S_{\rho\phi}=0$ & full $2\times2$ matrix $\kB T_0H(k)^{-1}$, \eqref{eq:Hk}\\
quasi-incompressible limit & trivial ($\hat v_A=\hat v_B$, $\divv\bv=0$) & $\divv\bv\neq0$; Lowengrub--Truskinovsky (\cref{sec:QI})\\
noise & $2\kB T_0M_\phi$, $M_\phi$ even & $2\kB T_0M$, $M$ with no parity\\
\bottomrule
\end{tabular}
\end{table}
In both cases the noise is fixed by the same rule: covariance $2\kB T_0$ times the coefficient of the corresponding deterministic flux. A non-symmetric mixture therefore needs no new kind of noise. The new physics comes entirely from the thermodynamic couplings, which the noise inherits through the dynamics.

\subsection{Quasi-incompressible limit}\label{sec:QI}
Suppose the pure constituents are incompressible with intrinsic densities $\bar\rho_A\neq\bar\rho_B$ and ideal volume of mixing, so that $1/\rho=c/\bar\rho_A+(1-c)/\bar\rho_B$. Then the continuity equation and $\rho\,Dc/Dt=-\divv(\bJ+\noise\bJ)$ give
\begin{equation}
\divv\bv=\Big(\frac1{\bar\rho_B}-\frac1{\bar\rho_A}\Big)\divv(\bJ+\noise\bJ).\label{eq:QI_divv}
\end{equation}
The barycentric velocity is \emph{not} solenoidal: diffusion, including its fluctuations, creates volume. The pressure becomes a Lagrange multiplier. The specific Gibbs free energy acquires the term $p/\rho$, so the exchange potential acquires $(\hat v_A-\hat v_B)p=(1/\bar\rho_A-1/\bar\rho_B)p$. This is the quasi-incompressible model of Lowengrub and Truskinovsky \cite{LowengrubTruskinovsky1998}, now with consistent noise. Two simplifications are often seen:
\begin{itemize}[nosep]
\item Using a solenoidal velocity together with a variable density $\rho(\phi)$, as in Eq.~(3.4) of \cite{PadhanPandit2025}, is consistent only if that velocity is the volume-averaged velocity \eqref{eq:uvol_v} (Abels--Garcke--Gr\"un). The momentum equation then acquires the additional flux $\bv\otimes\bJ$-type terms of that model.
\item Treating it as the barycentric velocity violates \eqref{eq:QI_divv}.
\end{itemize}
A derivation of this limit from the Gibbs free energy, with the pressure entering the exchange potential as $\mut_{\rm QI}=\rho^{-1}\delta F/\delta c+p\,\dvh$, is given in \cref{app:C-QI}. The low-Mach model of \cite{Donev2014}, in which the thermodynamic pressure is a parameter rather than a field, is derived in \cref{app:C-lowmach}.

\section{The non-isothermal, fully compressible LLCHNS model}\label{sec:noniso}
We now remove the heat bath. The temperature becomes a field, the internal energy becomes a dynamical variable, and the mixture can exchange heat internally: demixing releases latent heat, compression heats the fluid, and temperature gradients drive both stresses and diffusion. The derivation follows the same steps as in \cref{sec:isothermal}. We first list what changes (\cref{sec:noniso_changes}). We then write down the energy, entropy and free-energy functionals and define all functional derivatives (\cref{sec:noniso_functionals}), and derive the reversible and irreversible dynamics with GENERIC (\cref{sec:noniso_poisson,sec:noniso_friction}). Thermodynamic consistency (\cref{sec:noniso_consistency,sec:compare_CIT}) and the noise terms (\cref{sec:noniso_fdt}) come next, followed by the final equations (\cref{sec:noniso_final,sec:noniso_primitive}). We check that the isothermal limit reproduces \cref{sec:isothermal} (\cref{sec:iso_limit}). Finally we specialize to symmetric and non-symmetric mixtures (\cref{sec:noniso_sym,sec:noniso_nonsym,sec:noniso_compare}) and study the linear fluctuations (\cref{sec:noniso_linear}).

\subsection{What changes relative to the isothermal model}\label{sec:noniso_changes}
\begin{table}[H]
\centering\small
\caption{The isothermal and non-isothermal constructions compared. Every item in the right column is derived in this section.}\label{tab:iso_vs_noniso}
\renewcommand{\arraystretch}{1.2}
\begin{tabular}{>{\raggedright\arraybackslash}p{3.3cm}>{\raggedright\arraybackslash}p{5.3cm}>{\raggedright\arraybackslash}p{6.4cm}}
\toprule
 & isothermal (\cref{sec:isothermal}) & non-isothermal (this section)\\
\midrule
variables & $(\rho,\bg,c)$ & $(\rho,\bg,\varepsilon,c)$; $\varepsilon$ excludes the gradient energy\\
generators & one: $F=E-T_0S$ & two: $E$ and $S$\\
gradient terms & $\fN=\eN-T_0\sN$ (only $\kappa$ matters) & $\eN$ in $E$ and $\sN$ in $S$ separately ($\kappa_E$ and $\kappa_S$ matter)\\
Poisson operator & $3\times3$ Lie--Poisson & $4\times4$, with the entropic pressure tensor $\bPi^{(S)}$ fixed by degeneracy\\
reversible stress & $\bPi^K=-\bPi^{\fN}$ & $\bPi^K=-\bPi^{\eN}+T\bPi^{\sN}$; thermal capillary stress $\propto\kappa_S\nabla T$\\
structural conditions & antisymmetry, Jacobi, $\Dop\ge0$ & in addition both degeneracies, $\Lop\cdot\delta S/\delta\bx=0$ and $\Mop\cdot\delta E/\delta\bx=0$\\
fluctuating currents & momentum, mass & momentum, mass, heat; energy fluctuations on the energy shell\\
thermodynamic forces & $\bD$, $\nabla\mut$ & $\bD/T$, $\nabla(1/T)$, $\nabla(\mut/T)$\\
dissipative fluxes & $\btau$, $\bJ=-M\nabla\mut$ & $\btau$, $\bJ=-M\nabla\mut-M_T\nabla T$, $\bJqp=-\hat\lambda\nabla T-TM_T\nabla\mut$\\
energy flux & --- & convective, pressure work, heat, Dufour, interstitial working\\
noise & $2\kB T_0\times$ (coefficients) & $2\kB\times$ Onsager matrix: $2\kB TM$, $2\kB T^2M_T$, $2\kB T^2\hat\lambda$\\
second law & $\dot F\le0$ & $\dot S\ge0$, with a local entropy balance\\
equilibrium measure & $\exp(-F/\kB T_0)$ & $\exp(S/\kB)$ on the energy shell\\
\bottomrule
\end{tabular}
\end{table}
Three of these changes deserve emphasis, because they have no isothermal counterpart.
\begin{itemize}[leftmargin=*]
\item \textbf{The pressure tensor in the Poisson operator is not free.} It is fixed by the requirement that the reversible dynamics conserve the entropy (degeneracy). It contains the gradient \emph{entropy} but not the gradient \emph{energy}. The full Korteweg stress then emerges automatically.
\item \textbf{Energy fluctuations must live on the energy shell.} The fluctuating stress and the fluctuating diffusion flux change the kinetic and gradient energies. The internal energy must compensate exactly, which fixes the structure of the friction operator and of the noise in the energy equation.
\item \textbf{Heat and mass currents are correlated.} The Onsager cross coefficient between the heat and diffusion fluxes is allowed by symmetry. It produces the Soret and Dufour effects and a correlated pair of noises.
\end{itemize}

\subsection{State variables and the energy, entropy and free-energy functionals}\label{sec:noniso_functionals}
Following \cite{JelicThesis2009} (Eq.~5.2) and \cite{ChaudhriNotes2012} (Section~6), the state variables are
\begin{equation}
\bx=\big(\rho(\br),\ \bg(\br),\ \varepsilon(\br),\ c(\br)\big).\label{eq:x_noniso}
\end{equation}
Here $\varepsilon$ is the \emph{local} internal energy density: the thermal plus short-range interaction energy of a homogeneous fluid element, \emph{excluding} the gradient energy. The two generators are
\begin{equation}
E[\bx]=\int\Big[\frac{|\bg|^2}{2\rho}+\varepsilon+\eN(\rho,c,\nabla\rho,\nabla c)\Big]\dd^3r,\qquad
S[\bx]=\int\Big[s(\rho,c,\varepsilon)+\sN(\rho,c,\nabla c)\Big]\dd^3r .\label{eq:ES_noniso}
\end{equation}
With \eqref{eq:s_derivs} and \eqref{eq:muE_muS}, their functional derivatives are
\begin{equation}
\fd E\bx=\begin{pmatrix}-\tfrac12|\bv|^2+e^\rho\\ \bv\\ 1\\ \mu_E\end{pmatrix},\qquad
\fd S\bx=\begin{pmatrix}-\mub/T+s^\rho\\ \bm0\\ 1/T\\ -\rho\mu/T+\mu_S\end{pmatrix}.\label{eq:dEdS_noniso}
\end{equation}
For constant $\kappa_E,\kappa_S$ and the CH-only case these are Eqs.~(5.10a,b) of \cite{JelicThesis2009}: $\mu_E=-\kappa_E\nabla^2c$, $\mu_S=-\kappa_S\nabla^2c$, and $e^\rho=s^\rho=0$. Compared with Eqs.~(6.9)--(6.10) of \cite{ChaudhriNotes2012}, the $\rho$-component of $\delta E/\delta\bx$ acquires $e^\rho=\tfrac12\partial_\rho K|\nabla c|^2$ whenever the gradient coefficient depends on $\rho$, e.g.\ for $K=\rho\hat\kappa$.

\paragraph{The free-energy functional at a given temperature field.} Although the non-isothermal dynamics needs $E$ and $S$ separately, it is useful to keep the Helmholtz functional of \cref{sec:isothermal} in view. At a given (not necessarily uniform) temperature field $T(\br)$, define
\begin{equation}
F_T[\rho,c]=\int\Big[\psi(\rho,c,T)+\eN(\rho,c,\nabla\rho,\nabla c)-T\,\sN(\rho,c,\nabla c)\Big]\dd^3r,\qquad \psi=\rho f(\rho,c,T).\label{eq:FT_noniso}
\end{equation}
Its bulk part generates the local thermodynamics through \eqref{eq:f_derivs}. Its gradient part has free-energy coefficient $\kappa=\kappa_E-T\kappa_S$. All quantities that enter the non-isothermal equations can be expressed through the derivatives of $E$, $S$ and $F_T$:
\begin{itemize}[leftmargin=*,nosep]
\item $T$, $p$, $\mub$, $\mu$ from the local entropy $s(\rho,c,\varepsilon)$, \eqref{eq:s_derivs} and \eqref{eq:gibbs_rho_c};
\item the gradient potentials $\mu_E=\delta\EN/\delta c$, $\mu_S=\delta\SN/\delta c$, $e^\rho=\delta\EN/\delta\rho$, $s^\rho=\delta\SN/\delta\rho$, \eqref{eq:muE_muS};
\item the generalized exchange potential $\mut=\mu+(\mu_E-T\mu_S)/\rho$. It is $\rho^{-1}\delta F_T/\delta c$ evaluated with $T$ held \emph{outside} the functional derivative of the gradient entropy, \eqref{eq:forces_noniso};
\item the energetic and entropic Korteweg tensors $\bPi^{\eN}$ and $\bPi^{\sN}$ from \cref{lem:korteweg}.
\end{itemize}
Whether $T$ is held inside or outside the functional derivative is immaterial when $T$ is uniform or when $\kappa_S=0$. In general it is not, and \cref{app:thermo_route} shows that classical irreversible thermodynamics alone cannot decide between the two. GENERIC selects the form given here.

\subsection{Reversible dynamics: Poisson operator and entropic pressure tensor}\label{sec:noniso_poisson}
The Poisson operator is that of \cite{JelicThesis2009} (Eqs.~5.11--5.12), with an entropic pressure tensor $\bPi^{(S)}$ yet to be determined. Its action on $\bm a=(a_\rho,\bm a_g,a_\varepsilon,a_c)$ is
\begin{equation}
\begin{aligned}
(\Lop\cdot\bm a)_\rho&=-\divv(\rho\bm a_g),\\
(\Lop\cdot\bm a)_{\bg}&=-\rho\nabla a_\rho-\divv(\bm a_g\otimes\bg)-(\nabla\bm a_g)\cdot\bg-\varepsilon\nabla a_\varepsilon-\divv\big(\bPi^{(S)}a_\varepsilon\big)+a_c\nabla c,\\
(\Lop\cdot\bm a)_\varepsilon&=-\divv(\varepsilon\bm a_g)-\bPi^{(S)}:\nabla\bm a_g,\\
(\Lop\cdot\bm a)_c&=-\bm a_g\cdot\nabla c .
\end{aligned}\label{eq:L_noniso_action}
\end{equation}
In $\br$-notation,
\begin{equation}
\Lop(\br)=-\begin{pmatrix}0&\nabla\rho&0&0\\ \rho\nabla&[\nabla\bg+\bg\nabla]^T&\varepsilon\nabla+\nabla\cdot\bPi^{(S)}&-(\nabla c)\\ 0&\nabla\varepsilon+\bPi^{(S)}\cdot\nabla&0&0\\ 0&(\nabla c)&0&0\end{pmatrix}.\label{eq:L_noniso_matrix}
\end{equation}
Antisymmetry holds by construction. For the $(\varepsilon,\bg)$ pair,
\[
\int b_\varepsilon(\Lop\cdot\bm a)_\varepsilon=\int\big[\varepsilon\,\bm a_g\cdot\nabla b_\varepsilon+\bm a_g\cdot\divv(\bPi^{(S)}b_\varepsilon)\big]=-\int\bm a_g\cdot(\Lop\cdot\bm b)_{\bg}\big|_{b_\varepsilon\text{-part}} .
\]
The other pairs are as in \cref{sec:iso_poisson}.

\begin{proposition}[Degeneracy fixes the pressure tensor]\label{prop:PiS}
$\Lop\cdot\fd S\bx=0$ holds if and only if, up to an additive tensor $\bm X$ with $\divv(\bm X/T)=0$,
\begin{equation}
\boxed{\ \bPi^{(S)}=p\,\Id-T\,\bPi^{\sN},\qquad \bPi^{\sN}=\Big[\rho\pd{\sN}{\rho}-\sN\Big]\Id+\nabla c\otimes\pd{\sN}{\nabla c}.\ }\label{eq:PiS}
\end{equation}
For $\sN=\tfrac12\kappa_S(\rho,c)|\nabla c|^2$:
\begin{equation}
\bPi^{(S)}=\Big[p+\tfrac12T\big(\kappa_S-\rho\partial_\rho\kappa_S\big)|\nabla c|^2\Big]\Id-T\kappa_S\nabla c\otimes\nabla c .\label{eq:PiS_CH}
\end{equation}
For constant $\kappa_S$ this is Eq.~(5.14) of \cite{JelicThesis2009}.
\end{proposition}
\begin{proof}
With $\delta S/\delta\bx$ from \eqref{eq:dEdS_noniso}, the $\rho$, $\varepsilon$ and $c$ components of $\Lop\cdot\delta S/\delta\bx$ vanish because $\delta S/\delta\bg=0$. The $\bg$-component is
\[
-\rho\nabla\Big(-\frac{\mub}T+s^\rho\Big)-\varepsilon\nabla\frac1T-\divv\frac{\bPi^{(S)}}{T}+\Big(-\frac{\rho\mu}T+\mu_S\Big)\nabla c .
\]
By the Gibbs--Duhem relation \eqref{eq:GD_entropy}, the bulk terms equal $\nabla(p/T)$. By \cref{lem:korteweg} with $w=\sN$ (where $\delta\SN/\delta\rho=s^\rho$ and $\delta\SN/\delta c=\mu_S$), the gradient terms equal $-\divv\bPi^{\sN}$. Hence the $\bg$-component is $\divv[(p/T)\Id-\bPi^{\sN}-\bPi^{(S)}/T]$. It vanishes if and only if $\bPi^{(S)}/T=(p/T)\Id-\bPi^{\sN}$, modulo divergence-free tensors.
\end{proof}

\begin{remark}[The pressure in $\Lop$ is the \emph{entropic} one]\label{rem:PiS}
The gradient \emph{energy} does not enter the Poisson operator at all, which is independent of $E$. If $\kappa_S=0$, the pressure in $\Lop$ is just the local thermodynamic $p$. In Eqs.~(6.16)--(6.18) of \cite{ChaudhriNotes2012} the pressure in $L_{gu},L_{ug}$ is replaced by $P^{\rm CH}-\tfrac12K|\nabla c|^2$, so that the term $\Lop_{gc}\cdot\delta E/\delta c$ produces the desired Korteweg stress. That choice violates the degeneracy $\Lop\cdot\delta S/\delta\bx=0$ when $\kappa_S=0$. It is also unnecessary: with the correct $\bPi^{(S)}$, the full Korteweg stress emerges from $\Lop\cdot\delta E/\delta\bx$ automatically, through \cref{lem:korteweg} applied to $\eN$ (see \eqref{eq:Pi_total}).
\end{remark}

\paragraph{Jacobi identity.} In the variables $(\rho,\bg,\sigma,c)$ with $\sigma=s+\sN$ (total entropy density), $\Lop$ becomes the canonical Lie--Poisson operator with $\sigma$ an advected density. It therefore satisfies the Jacobi identity, which is invariant under changes of variables. The proof is in \cref{app:jacobi}. It requires $\sN$ to be independent of $\nabla\rho$, which explains the restriction made in \cref{sec:gradient}.

\paragraph{Reversible equations.} Take $\bm a=\delta E/\delta\bx$. In the $\bg$-row:
\begin{itemize}[nosep]
\item the kinetic terms cancel as before;
\item $-\varepsilon\nabla(1)=0$;
\item $-\rho\nabla e^\rho+\mu_E\nabla c=-\divv\bPi^{\eN}$ by \cref{lem:korteweg}.
\end{itemize}
Hence
\begin{equation}
\begin{aligned}
\partial_t\rho\big|_{\rm rev}&=-\divv(\rho\bv),\qquad
\partial_t\bg\big|_{\rm rev}=-\divv(\rho\bv\bv)-\divv\bPi,\\
\partial_t\varepsilon\big|_{\rm rev}&=-\divv(\varepsilon\bv)-\bPi^{(S)}:\nabla\bv,\qquad
\partial_tc\big|_{\rm rev}=-\bv\cdot\nabla c,
\end{aligned}\label{eq:rev_noniso}
\end{equation}
with the total reversible momentum flux
\begin{equation}
\boxed{\ \bPi=\bPi^{(S)}+\bPi^{\eN}=p\,\Id+\bPi^{\eN}-T\,\bPi^{\sN}\equiv p\,\Id-\bPi^K,\qquad \bPi^K=-\bPi^{\eN}+T\,\bPi^{\sN}.\ }\label{eq:Pi_total}
\end{equation}
This is Eq.~(5.16) of \cite{JelicThesis2009}, generalized to coefficients that depend on $(\rho,c)$ and to density gradients. For the CH-only case,
\begin{equation}
\bPi^K=\tfrac12\Big[(\kappa_E-\rho\partial_\rho\kappa_E)-T(\kappa_S-\rho\partial_\rho\kappa_S)\Big]|\nabla c|^2\Id-(\kappa_E-T\kappa_S)\nabla c\otimes\nabla c .\label{eq:PiK_noniso_CH}
\end{equation}
For constant coefficients, its divergence is (cf.\ Eq.~5.17 of \cite{JelicThesis2009})
\begin{equation}
\divv\bPi=\nabla p+\kappa\,\nabla^2c\,\nabla c+\Big(\tfrac12\kappa_S|\nabla c|^2\Id-\kappa_S\nabla c\otimes\nabla c\Big)\cdot\nabla T,\qquad \kappa=\kappa_E-T\kappa_S .\label{eq:divPi_noniso}
\end{equation}
\begin{notebox}[Thermal capillary stress: CH analogue of $\bPi^C$ of \cite{ChaudhriPRE2014,Onuki2007}]
The last term of \eqref{eq:divPi_noniso} is a stress proportional to $\nabla T$. It exists only if the interfacial free energy has an entropic part ($\kappa_S\neq0$, equivalently $\partial\kappa/\partial T\neq0$). It is the Cahn--Hilliard counterpart of the ``cross-coupling'' stress $\bPi^C$ introduced in \cite{ChaudhriPRE2014} following Onuki. Here its form is \emph{derived}: it follows uniquely from the degeneracy condition, with no additional assumption. If the gradient term is purely energetic ($\kappa_S=0$, as in \cite{ChaudhriPRE2014,Espanol2001}), there is no such stress. This is consistent with the observation in \cite{ChaudhriPRE2014} that the Dunn--Serrin cross term vanishes for constant $\kappa$. Physically, $\tfrac12\kappa_S|\nabla c|^2\Id-\kappa_S\nabla c\nabla c$ is (minus) the entropic part of the interfacial stress. Its contraction with $\nabla T$ is the part of the thermocapillary (Marangoni) force that comes from the temperature dependence of the gradient coefficient. The remaining part comes from the temperature dependence of the bulk free energy, through $p$ and $\mu$.
\end{notebox}

\paragraph{Reversible local energy and entropy balances.} The following identities were verified symbolically (\cref{app:verification}, check~1):
\begin{gather}
\partial_te_{\rm tot}\big|_{\rm rev}+\divv\Big[e_{\rm tot}\bv+\bPi\cdot\bv+\rho(\divv\bv)\pd{\eN}{\nabla\rho}\Big]=0,\qquad e_{\rm tot}=\tfrac12\rho v^2+\varepsilon+\eN,\label{eq:E_local_rev}\\
\partial_t\sigma\big|_{\rm rev}+\divv(\sigma\bv)=0,\qquad \sigma=s+\sN .\label{eq:S_local_rev}
\end{gather}
The energy identity \eqref{eq:E_local_rev} holds for \emph{any} fields $p$ and $T$ in $\bPi^{(S)}$; it is a consequence of antisymmetry alone. The entropy identity \eqref{eq:S_local_rev} requires the thermodynamic $p,T$, i.e.\ the degeneracy. The Dunn--Serrin interstitial flux $\rho(\divv\bv)\partial\eN/\partial\nabla\rho$ \cite{DunnSerrin1985} is present only for density-gradient energies. For a Cahn--Hilliard gradient energy in $c$ alone, the reversible energy flux is simply $e_{\rm tot}\bv+\bPi\cdot\bv$.

\subsection{Irreversible dynamics from fluctuations}\label{sec:noniso_friction}
\paragraph{Fluctuation ansatz and the energy shell.} As in \eqref{eq:fluct_iso},
\begin{equation}
\Delta\rho^{\rm f}=0,\qquad \Delta\bg^{\rm f}=-\divv\bW^\tau,\qquad \Delta c^{\rm f}=-\frac1\rho\divv\bW^J,\qquad \Delta\varepsilon^{\rm f}=-\divv\bW^q+\Delta\varepsilon^{\rm f}_{\rm comp}.\label{eq:fluct_noniso}
\end{equation}
Here $\bW^q$ is the fluctuating heat current (a Wiener increment). $\Delta\varepsilon^{\rm f}_{\rm comp}$ must be chosen so that the fluctuations do not change the total energy (Step~3 of \cref{sec:JO}). Using \eqref{eq:dEdS_noniso},
\[
\int\fd E\bx\cdot\Delta\bx^{\rm f}=\int\Big[\bv\cdot(-\divv\bW^\tau)+\Delta\varepsilon^{\rm f}+\mu_E\Big(-\frac1\rho\divv\bW^J\Big)\Big]
=\int\Big[\bW^\tau:\nabla\bv+\Delta\varepsilon^{\rm f}_{\rm comp}+\bW^J\cdot\nabla\frac{\mu_E}{\rho}\Big].
\]
Following \cite{JelicThesis2009}, we compensate locally: the internal energy absorbs the viscous work of the fluctuating stress and the change of gradient energy caused by the fluctuating diffusion,
\begin{equation}
\boxed{\ \Delta\varepsilon^{\rm f}=-\divv\bW^q-\bW^\tau:\nabla\bv+\frac{\mu_E}{\rho}\divv\bW^J=-\divv\bW^q-\bW^\tau:\nabla\bv-\mu_E\,\Delta c^{\rm f}.\ }\label{eq:deps_f}
\end{equation}
Any other admissible choice differs from \eqref{eq:deps_f} by a divergence, which can be absorbed into a redefinition of $\bW^q$. Pairing with a test field $\bm a$:
\begin{equation}
\int\bm a\cdot\Delta\bx^{\rm f}=\int\Big[\bW^\tau:\big((\nabla\bm a_g)^{\sym}-a_\varepsilon\bD\big)+\bW^q\cdot\nabla a_\varepsilon+\bW^J\cdot\nabla\frac{a_c-a_\varepsilon\mu_E}{\rho}\Big],\label{eq:pairing_noniso}
\end{equation}
\begin{equation}
\Phiop(\bm a)=\Big((\nabla\bm a_g)^{\sym}-a_\varepsilon\bD,\ \ \nabla a_\varepsilon,\ \ \nabla\frac{a_c-a_\varepsilon\mu_E}{\rho}\Big),\qquad \bD=(\nabla\bv)^{\sym}.\label{eq:Phi_noniso}
\end{equation}
Then $\Phiop(\delta E/\delta\bx)=(\bD-\bD,\ \nabla1,\ \nabla(\mu_E-\mu_E)/\rho)=\bm0$. By \cref{lem:JO}(c),(e), the degeneracy $\Mop\cdot\delta E/\delta\bx=0$ and energy conservation by the noise, $\delta E/\delta\bx\cdot\Bop=0$, are therefore guaranteed.

\paragraph{Second moments.} Locality, isotropy and the Curie principle give
\begin{equation}
\begin{aligned}
\avg{W^\tau_{ij}W^\tau_{kl}}&=2\kB\tau\,T\Big[\eta\big(\delta_{ik}\delta_{jl}+\delta_{il}\delta_{jk}-\tfrac23\delta_{ij}\delta_{kl}\big)+\zeta\delta_{ij}\delta_{kl}\Big]\delta(\br-\br'),\\
\avg{W^q_iW^q_j}&=2\kB\tau\,L_{qq}\delta_{ij}\delta(\br-\br'),\qquad
\avg{W^q_iW^J_j}=2\kB\tau\,L_{q1}\delta_{ij}\delta(\br-\br'),\\
\avg{W^J_iW^J_j}&=2\kB\tau\,L_{11}\delta_{ij}\delta(\br-\br'),
\end{aligned}\label{eq:moments_noniso}
\end{equation}
with $\avg{\bW^\tau\bW^q}=\avg{\bW^\tau\bW^J}=0$. The matrix $\begin{pmatrix}L_{qq}&L_{q1}\\L_{q1}&L_{11}\end{pmatrix}$ must be positive semi-definite. The cross-correlation $L_{q1}$ between fluctuating heat and mass currents is allowed by symmetry (both are polar vectors). It produces thermal diffusion (Soret) and its Onsager-reciprocal Dufour effect. In Jeli\'c's electromagnetic example it is the analogue of the thermoelectric correlation $\Cop^{(qj)}$ (Eq.~3.15 of \cite{JelicThesis2009}).

\paragraph{The friction operator.} By \cref{lem:JO}(a), $\Mop=\Phiop^*\Cop\Phiop$. Writing $\bm T^\eta[\bm b]\equiv T[\eta(\nabla\bm b+\nabla\bm b^T)+(\zeta-\frac23\eta)(\divv\bm b)\Id]$ and $\btau_T\equiv T[2\eta\mathring{\bD}+\zeta(\divv\bv)\Id]$, the non-zero blocks in $\br$-notation are
\begin{equation}
\begin{aligned}
\Mop_{\bg\bg}\cdot\bm b&=-\divv\bm T^\eta[\bm b],\qquad
\Mop_{\bg\varepsilon}\,b=\divv(\btau_Tb),\qquad
\Mop_{\varepsilon\bg}\cdot\bm b=-\btau_T:\nabla\bm b,\\
\Mop_{\varepsilon\varepsilon}\,b&=T\big[2\eta\mathring\bD:\mathring\bD+\zeta(\divv\bv)^2\big]b-\divv(L_{qq}\nabla b)\\
&\quad+\divv\Big(L_{q1}\nabla\frac{\mu_Eb}{\rho}\Big)+\frac{\mu_E}{\rho}\divv(L_{q1}\nabla b)-\frac{\mu_E}{\rho}\divv\Big(L_{11}\nabla\frac{\mu_Eb}{\rho}\Big),\\
\Mop_{\varepsilon c}\,b&=-\divv\Big(L_{q1}\nabla\frac b\rho\Big)+\frac{\mu_E}\rho\divv\Big(L_{11}\nabla\frac b\rho\Big),\\
\Mop_{c\varepsilon}\,b&=-\frac1\rho\divv(L_{q1}\nabla b)+\frac1\rho\divv\Big(L_{11}\nabla\frac{\mu_Eb}\rho\Big),\\
\Mop_{cc}\,b&=-\frac1\rho\divv\Big(L_{11}\nabla\frac b\rho\Big).
\end{aligned}\label{eq:M_noniso}
\end{equation}
The $\rho$-row and $\rho$-column vanish. With $\rho\to1$, $L_{q1}=0$, $L_{11}=MT$, $L_{qq}=\lambda^qT^2$ and $\mu_E=-\kappa_E\nabla^2c$, these reduce \emph{exactly} to Eqs.~(5.19)--(5.20) of \cite{JelicThesis2009} and Eqs.~(6.21)--(6.23) of \cite{ChaudhriNotes2012}. The factors $1/\rho$ are the compressible generalization required by the conservation of $\int\rho c$. The cross terms with $L_{q1}$ are new.

\paragraph{Forces, fluxes and irreversible equations.} Evaluating \eqref{eq:Phi_noniso} on $\delta S/\delta\bx$ \eqref{eq:dEdS_noniso}, and using $(S_c-\mu_E/T)/\rho=-\mut/T$, gives the thermodynamic forces
\begin{equation}
\bX=\Phiop\Big(\fd S\bx\Big)=\Big(-\frac{\bD}{T},\ \nabla\frac1T,\ -\nabla\frac{\mut}{T}\Big),\qquad
\boxed{\ \mut\equiv\mu+\frac{\mu_E-T\mu_S}{\rho}=\mu+\frac1\rho\fd{(\EN-T\SN)}{c}\Big|_{\rho,T}.\ }\label{eq:forces_noniso}
\end{equation}
At uniform temperature, $\mut$ coincides with the isothermal potential \eqref{eq:mut_iso} with $\fN=\eN-T\sN$. The fluxes are $\bY=\Cop\cdot\bX$:
\begin{equation}
\btau=2\eta\mathring\bD+\zeta(\divv\bv)\Id,\qquad
\bJq=L_{qq}\nabla\frac1T-L_{q1}\nabla\frac{\mut}{T},\qquad
\bJ=L_{1q}\nabla\frac1T-L_{11}\nabla\frac{\mut}{T},\qquad L_{1q}=L_{q1}.\label{eq:fluxes_noniso}
\end{equation}
Applying $\Phiop^*$, i.e.\ reading off the coefficients of $\bm a$ in $\int\bm a\cdot\Mop\cdot\delta S/\delta\bx=\int\Phiop(\bm a)\cdot\bY$, gives
\begin{equation}
\partial_t\rho\big|_{\rm irr}=0,\qquad
\partial_t\bg\big|_{\rm irr}=\divv\btau,\qquad
\partial_t\varepsilon\big|_{\rm irr}=\btau:\nabla\bv-\divv\bJq+\frac{\mu_E}{\rho}\divv\bJ,\qquad
\rho\,\partial_tc\big|_{\rm irr}=-\divv\bJ .\label{eq:irr_noniso}
\end{equation}
This is Eq.~(5.21) of \cite{JelicThesis2009}, generalized to compressible flow (factors of $\rho$), gradient entropy in $\mut$, and Soret/Dufour coupling.

\paragraph{Onsager form and phenomenological coefficients.} The flux--force pairs are $(\btau,\nabla\bv/T)$, $(\bJq,\nabla(1/T))$ and $(\bJ,-\nabla(\mut/T))$. The Onsager matrix of the vector processes is exactly the correlation matrix $\Cop$; reciprocity $L_{q1}=L_{1q}$ is the symmetry of $\Cop$. It is convenient to introduce the \emph{reduced heat flux} $\bJqp=\bJq-\mut\bJ$ and the phenomenological coefficients $M$ (mobility), $M_T$ (thermodiffusion) and $\hat\lambda$:
\begin{equation}
\bJ=-M\nabla\mut-M_T\nabla T,\qquad \bJqp=-\hat\lambda\nabla T-TM_T\nabla\mut .\label{eq:pheno}
\end{equation}
Comparison with \eqref{eq:fluxes_noniso} gives the dictionary
\begin{equation}
L_{11}=MT,\qquad L_{1q}=T^2M_T+\mut MT,\qquad L_{qq}=\hat\lambda T^2+2\mut M_TT^2+\mut^2MT .\label{eq:dictionary}
\end{equation}
Positive semi-definiteness of $\Cop$ is equivalent to
\begin{equation}
\eta\ge0,\qquad \zeta\ge0,\qquad M\ge0,\qquad \lambda\equiv\hat\lambda-\frac{TM_T^2}{M}\ge0 ,\label{eq:positivity}
\end{equation}
where $\lambda$ is the thermal conductivity measured at zero diffusion flux. For $M_T=0$, $\hat\lambda=\lambda$, and the familiar $L_{qq}=\lambda T^2$ is recovered for a single-component fluid ($\bJ=0$).

\emph{Landau--Lifshitz form.} Expanding $\nabla\mu=(\partial\mu/\partial c)_{T,p}\nabla c+(\hat v_A-\hat v_B)\nabla p+(\partial\mu/\partial T)_{c,p}\nabla T$ and writing $\mut_\nabla\equiv(\mu_E-T\mu_S)/\rho$:
\begin{equation}
\bJ=-\rho D\Big[\nabla c+\frac{k_T}{T}\nabla T+\frac{k_p}{p}\nabla p\Big]-M\nabla\mut_\nabla,\qquad
\rho D=M\Big(\pd\mu c\Big)_{T,p},\qquad k_T=T\frac{(\partial\mu/\partial T)_{c,p}+M_T/M}{(\partial\mu/\partial c)_{T,p}},\label{eq:LL_form}
\end{equation}
with $k_p$ as in \eqref{eq:J_baro}. This is the form of \cite{LandauLifshitzFM} (Sec.~58) and \cite{deZarateSengers2006}, augmented by the Cahn--Hilliard term. In \eqref{eq:LL_form}, only $D$ and the Soret part $M_T$ of $k_T$ are independent transport coefficients; $k_p$ and the thermodynamic part of $k_T$ are fixed by the equation of state.

\subsection{Thermodynamic consistency of the non-isothermal model}\label{sec:noniso_consistency}
\paragraph{Global.} By construction (\cref{lem:JO}, \cref{prop:PiS}, \cref{app:jacobi}): (G1)--(G3) hold, $\dd E/\dd t=0$, and
\begin{equation}
\frac{\dd S}{\dd t}=\int\bX\cdot\Cop\cdot\bX\,\dd^3r=\int\Big[\frac{2\eta\mathring\bD:\mathring\bD+\zeta(\divv\bv)^2}{T}+L_{qq}\Big|\nabla\frac1T\Big|^2-2L_{q1}\nabla\frac1T\cdot\nabla\frac\mut T+L_{11}\Big|\nabla\frac\mut T\Big|^2\Big]\dd^3r\ \ge0 .\label{eq:dSdt_noniso}
\end{equation}

\paragraph{Local energy conservation.} Combining \eqref{eq:E_local_rev} with \eqref{eq:irr_noniso}, the change of $\eN$ caused by the irreversible change of $c$ is
\[
\partial_t\eN|_{\rm irr}=\mu_E\,\partial_tc|_{\rm irr}+\divv\big(\partial_{\nabla c}\eN\,\partial_tc|_{\rm irr}\big).
\]
The term $\mu_E\partial_tc|_{\rm irr}=-(\mu_E/\rho)\divv\bJ$ cancels the corresponding term in $\partial_t\varepsilon|_{\rm irr}$. The result (verified symbolically, \cref{app:verification}, check~2) is
\begin{equation}
\boxed{\ \partial_te_{\rm tot}+\divv\Big[e_{\rm tot}\bv+\bPi\cdot\bv+\rho(\divv\bv)\pd\eN{\nabla\rho}-\btau\cdot\bv+\bJq+\frac{\divv\bJ}{\rho}\pd\eN{\nabla c}\Big]=0 .\ }\label{eq:E_local_full}
\end{equation}
The last term is the irreversible interstitial working flux $-\dot c\,\partial\eN/\partial\nabla c$. It coincides with the ``new'' energy-flux term $\partial_z\hat\varepsilon\,\divv\bm j_1$ found by Heida, M\'alek and Rajagopal (\cite{HeidaMalekRajagopal2012}, Eqs.~43 and~51) and with the term $-\rho\Phi\,\dd c/\dd t$ in Eq.~(4.17) of \cite{ChaudhriNotes2012}.

\paragraph{Local entropy balance.} With $\sigma=s+\sN$ and \eqref{eq:irr_noniso} (verified symbolically, \cref{app:verification}, check~2):
\begin{gather}
\partial_t\sigma+\divv\big(\sigma\bv+\bJ_s\big)=\Sigma\ \ge0,\qquad
\bJ_s=\frac{\bJq-\mut\bJ}{T}+\frac{\divv\bJ}{\rho}\pd\sN{\nabla c}=\frac{\bJqp}{T}+\frac{\divv\bJ}{\rho}\pd\sN{\nabla c},\label{eq:entropy_flux}\\
\Sigma=\frac{\btau:\nabla\bv}{T}+\bJq\cdot\nabla\frac1T-\bJ\cdot\nabla\frac\mut T
=\frac{2\eta\mathring\bD:\mathring\bD+\zeta(\divv\bv)^2}{T}+\frac{\hat\lambda|\nabla T|^2}{T^2}+\frac{2M_T}{T}\nabla T\cdot\nabla\mut+\frac MT|\nabla\mut|^2 .\label{eq:Sigma}
\end{gather}
The quadratic form in $(\nabla T,\nabla\mut)$ is positive semi-definite if and only if $M\hat\lambda\ge TM_T^2$, i.e.\ $\lambda\ge0$, consistent with \eqref{eq:positivity}. The entropy flux contains an entropic interstitial term, present only if $\kappa_S\neq0$. The integral of $\Sigma$ equals \eqref{eq:dSdt_noniso}.

\paragraph{Equilibrium.} $\Sigma=0$ requires $\mathring\bD=0$, $\divv\bv=0$ (rigid motion), $\nabla T=0$ and $\nabla\mut=0$. These are exactly the stationarity conditions of $S$ at fixed $E$ and masses, i.e.\ of \cref{sec:equilibrium_interface} with $\fN=\eN-T\sN$.

\subsection{Cross-check with classical irreversible thermodynamics}\label{sec:compare_CIT}
The derivation of Section~4.1 of \cite{ChaudhriNotes2012} uses $\rho\,\dd u/\dd t=-\divv\bm q+\bm{\mathcal T}:\nabla\bv$, where $u$ is the specific internal energy \emph{including} the gradient free energy and $\kappa_S=0$. It is equivalent to the present result under the identifications
\[
\bm q_{\rm notes}=\bJq+\frac{\divv\bJ}{\rho}\pd\eN{\nabla c},\qquad \mu_{\rm notes}=\mut,\qquad \rho\Phi=\pd\eN{\nabla c}.
\]
Indeed, Eq.~(4.17) of \cite{ChaudhriNotes2012}, $\bm q-\bJ\mu=-\lambda\nabla T-\rho\Phi\,\dd c/\dd t$, becomes $\bJqp=-\lambda\nabla T$ using $\dd c/\dd t=-\divv\bJ/\rho$. Eq.~(4.18), $\bm q_s=\bm q/T+\rho\Phi\dot c/T-\mu\bJ/T$, becomes $\bJ_s=\bJqp/T$. Both agree with \eqref{eq:pheno} and \eqref{eq:entropy_flux} for $M_T=0$, $\kappa_S=0$. The ``$\mu$'' that emerges from Eq.~(4.11) of the notes is $f_c-\rho^{-1}\divv(\rho\Phi)$, which is the \emph{specific} $\mut$. It is not the volumetric $\mu^{C}$ subsequently used in Eq.~(4.14). \Cref{app:thermo_route} carries the classical route through for the general free energy of this paper, with temperature-dependent gradient coefficients and density gradients. It shows that the GENERIC model coincides with the entropy-representation version of that route (route~A of \cref{app:C-routeA}).

\subsection{Noise terms from the fluctuation--dissipation theorem}\label{sec:noniso_fdt}
By \cref{lem:JO}(e), the stochastic terms are $\Phiop^*\dd\bW$ with $\avg{\dd\bW\dd\bW}=2\kB\Cop\,\dd t$. In continuum form, add stochastic fluxes $\noise\btau$, $\noise\bJ^q$, $\noise\bJ$ \emph{wherever the corresponding deterministic fluxes appear} in \eqref{eq:irr_noniso}:
\begin{equation}
\partial_t\bg\ni\divv\noise\btau,\qquad
\partial_t\varepsilon\ni\noise\btau:\nabla\bv-\divv\noise\bJ^q+\frac{\mu_E}\rho\divv\noise\bJ,\qquad
\rho\,\partial_tc\ni-\divv\noise\bJ .\label{eq:noise_placement}
\end{equation}
The covariances are $2\kB\times$ the Onsager matrix:
\begin{equation}
\boxed{
\begin{aligned}
\avg{\noise\tau_{ij}(\br,t)\noise\tau_{kl}(\br',t')}&=2\kB T\big[\eta(\delta_{ik}\delta_{jl}+\delta_{il}\delta_{jk})+(\zeta-\tfrac23\eta)\delta_{ij}\delta_{kl}\big]\delta(\br-\br')\delta(t-t'),\\
\avg{\noise J_i\,\noise J_j}&=2\kB MT\,\delta_{ij}\delta\delta,\qquad
\avg{\noise J_i\,\noise J^{q\prime}_j}=2\kB M_TT^2\,\delta_{ij}\delta\delta,\\
\avg{\noise J^{q\prime}_i\,\noise J^{q\prime}_j}&=2\kB\hat\lambda T^2\,\delta_{ij}\delta\delta,\\
\noise\bJ^q&=\noise\bJ^{q\prime}+\mut\,\noise\bJ,\qquad \avg{\noise\btau\,\noise\bJ}=\avg{\noise\btau\,\noise\bJ^{q\prime}}=0 .
\end{aligned}}\label{eq:noise_noniso}
\end{equation}
The covariance of $(\noise\bJ^q,\noise\bJ)$ implied by \eqref{eq:noise_noniso} is $2\kB(L_{qq},L_{q1},L_{11})$ with \eqref{eq:dictionary}, as required. \textbf{Explicit construction.} With independent white noises $\noise{\mathcal W}^v$, $\bm{\mathcal W}^c$, $\bm{\mathcal W}^E$:
\begin{equation}
\noise\btau\ \text{as in \eqref{eq:noise_construction_iso} with }T_0\to T,\qquad
\noise\bJ=\sqrt{2\kB MT}\,\bm{\mathcal W}^c,\qquad
\noise\bJ^{q\prime}=\frac{TM_T}{M}\noise\bJ+\sqrt{2\kB\lambda T^2}\,\bm{\mathcal W}^E .\label{eq:noise_construction_noniso}
\end{equation}
\emph{Check.} $\avg{\noise\bJ^{q\prime}\noise\bJ^{q\prime}}=2\kB[(TM_T/M)^2MT+\lambda T^2]=2\kB T^2[\lambda+TM_T^2/M]=2\kB\hat\lambda T^2$, and $\avg{\noise\bJ\noise\bJ^{q\prime}}=(TM_T/M)\,2\kB MT=2\kB M_TT^2$. For a single-component fluid this is Eq.~(2.19) of \cite{ChaudhriPRE2014}, $\noise\bQ=\sqrt{2\lambda\kB T^2}\bm{\mathcal W}^E$.

\paragraph{The noise conserves energy pathwise.} With \eqref{eq:noise_placement}, the stochastic contribution to $\partial_te_{\rm tot}$ is
\[
\bv\cdot\divv\noise\btau+\noise\btau:\nabla\bv-\divv\noise\bJ^q+\frac{\mu_E}\rho\divv\noise\bJ+\Big[\mu_E\Big(-\frac{\divv\noise\bJ}\rho\Big)+\divv\Big(\pd\eN{\nabla c}\Big(-\frac{\divv\noise\bJ}\rho\Big)\Big)\Big]
=\divv\Big[\noise\btau\cdot\bv-\noise\bJ^q-\frac{\divv\noise\bJ}\rho\pd\eN{\nabla c}\Big].
\]
This is a pure divergence. It is the continuum form of $\delta E/\delta\bx\cdot\Bop=0$.

\paragraph{Thermal drift.} As in the isothermal case, $\kB\partial_{\bx}\cdot\Mop$ is omitted in the Landau--Lifshitz form. Here it is non-zero even for constant transport coefficients, because $\Mop$ depends on $T$ (through $\bm T^\eta$ and $L$) and on $\mu_E$. It does not affect linearized fluctuations.

\subsection{Assembling the non-isothermal LLCHNS equations}\label{sec:noniso_final}
As in \cref{sec:iso_final} we now add the three parts of the stochastic GENERIC equation \eqref{eq:sgeneric}:
\[
\dd\bx=\underbrace{\Lop\cdot\fd E\bx\,\dd t}_{\eqref{eq:rev_noniso}}+\underbrace{\Mop\cdot\fd S\bx\,\dd t}_{\eqref{eq:irr_noniso}}+\underbrace{\Phiop^*\,\dd\bW}_{\eqref{eq:noise_placement}} ,
\]
again with the thermal drift omitted. For the conserved densities $\rho$, $\bg$ and $\rho c$ the result is immediate. For the energy we do not use the equation for the local internal energy $\varepsilon$ directly. Instead we combine it with the kinetic energy and with the change of the gradient energy $\eN$ caused by the motion and by diffusion. This gives a conservation law for the total energy $\rho E=e_{\rm tot}$, \eqref{eq:E_local_full}, with the interstitial fluxes in the energy flux. The resulting system, written in the conservative form of \cite{ChaudhriPRE2014}, is the following.

\begin{resultbox}[Non-isothermal, fully compressible fluctuating CHNS (LLCHNS) equations]\label{res:noniso_final}
Conservative form, in the notation and sign conventions of \cite{ChaudhriPRE2014}, with $\bQ\equiv-\bJq$ and $\rho E\equiv e_{\rm tot}=\tfrac12\rho v^2+\varepsilon+\eN$:
\begin{equation}
\frac{\partial\bm U}{\partial t}=-\divv\bm F_H-\divv\bm F_K-\divv\bm F_D-\divv\bm F_S,\qquad
\bm U=\begin{pmatrix}\rho\\ \rho\bv\\ \rho c\\ \rho E\end{pmatrix},\label{eq:U_noniso}
\end{equation}
\begin{equation}
\begin{gathered}
\bm F_H=\begin{pmatrix}\rho\bv\\ \rho\bv\bv^T+p\Id\\ \rho c\bv\\ (\rho E+p)\bv\end{pmatrix},\qquad
\bm F_K=-\begin{pmatrix}0\\ \bPi^K\\ 0\\ \bPi^K\!\cdot\bv-\rho(\divv\bv)\,\partial\eN/\partial\nabla\rho\end{pmatrix},\\[6pt]
\bm F_D=-\begin{pmatrix}0\\ \btau\\ -\bJ\\ \bQ+\btau\cdot\bv-\rho^{-1}(\divv\bJ)\,\partial\eN/\partial\nabla c\end{pmatrix},\\[6pt]
\bm F_S=-\begin{pmatrix}0\\ \noise\btau\\ -\noise\bJ\\ \noise\bQ+\noise\btau\cdot\bv-\rho^{-1}(\divv\noise\bJ)\,\partial\eN/\partial\nabla c\end{pmatrix}.
\end{gathered}\label{eq:fluxes_noniso_final}
\end{equation}
Written out:
\begin{align}
&\partial_t\rho+\divv(\rho\bv)=0,\label{eq:ni_mass}\\
&\partial_t(\rho\bv)+\divv(\rho\bv\bv^T)+\nabla p=\divv\big(\bPi^K+\btau+\noise\btau\big),\label{eq:ni_mom}\\
&\partial_t(\rho c)+\divv(\rho c\bv)=-\divv\big(\bJ+\noise\bJ\big),\label{eq:ni_c}\\
&\partial_t(\rho E)+\divv\big[(\rho E+p)\bv\big]\nonumber\\
&\qquad=\divv\Big[\bPi^K\!\cdot\bv-\rho(\divv\bv)\pd\eN{\nabla\rho}+(\btau+\noise\btau)\cdot\bv+\bQ+\noise\bQ-\frac{\divv(\bJ+\noise\bJ)}{\rho}\pd\eN{\nabla c}\Big].\label{eq:ni_E}
\end{align}
Closure:
\begin{itemize}[leftmargin=*,nosep]
\item $\varepsilon=\rho E-\tfrac12\rho v^2-\eN$; $T$ and $p$ from the local equation of state $s(\rho,c,\varepsilon)$ via \eqref{eq:s_derivs} and \eqref{eq:gibbs_rho_c}.
\item $\bPi^K=-\bPi^{\eN}+T\bPi^{\sN}$ from \eqref{eq:Pi_total}--\eqref{eq:PiK_noniso_CH}.
\item $\mut=\partial_cf+\rho^{-1}(\mu_E-T\mu_S)$.
\item $\btau=\eta[\nabla\bv+\nabla\bv^T]+(\zeta-\tfrac23\eta)(\divv\bv)\Id$.
\item $\bJ=-M\nabla\mut-M_T\nabla T$.
\item $\bQ=-\bJq=\hat\lambda\nabla T+TM_T\nabla\mut-\mut\bJ$.
\item Noises as in \eqref{eq:noise_noniso}--\eqref{eq:noise_construction_noniso}, with $\noise\bQ=-(\noise\bJ^{q\prime}+\mut\noise\bJ)$.
\end{itemize}
For the CH-only case, set $\partial\eN/\partial\nabla\rho=0$ and $\partial\eN/\partial\nabla c=\kappa_E\nabla c$. Then
\[
\mu_E-T\mu_S=\Big[\tfrac12\partial_c\kappa_E|\nabla c|^2-\divv(\kappa_E\nabla c)\Big]-T\Big[\tfrac12\partial_c\kappa_S|\nabla c|^2-\divv(\kappa_S\nabla c)\Big].
\]
The local $T$ multiplies the entropic bracket from outside; it is not differentiated. For constant coefficients this is $-\kappa\nabla^2c$ with $\kappa=\kappa_E-T\kappa_S$.
\end{resultbox}

\begin{notebox}[On the energy equation (2.12) of \cite{ChaudhriPRE2014}]
In \cite{ChaudhriPRE2014} the conserved energy variable is $\rho E=\tfrac12\rho v^2+u$, without the gradient energy, and the flux contains $\bPi\cdot\bv$ with the Korteweg tensor included. From \eqref{eq:E_local_rev} (with $\eN=\frac12\kappa|\nabla\rho|^2$), the exact equation for $\tfrac12\rho v^2+u$ implied by conservation of the total energy (2.2) of that paper is
\[
\partial_t(\tfrac12\rho v^2+u)+\divv\big[(\tfrac12\rho v^2+u)\bv+\bPi\cdot\bv\big]=-\bPi^K:\nabla\bv+(\text{irreversible terms}).
\]
Equivalently, it is \eqref{eq:ni_E} for $\rho E+\tfrac12\kappa|\nabla\rho|^2$ with the interstitial flux $\rho\kappa(\divv\bv)\nabla\rho$. Omitting $-\bPi^K:\nabla\bv$, as Eq.~(2.12) of \cite{ChaudhriPRE2014} does, conserves $\int(\tfrac12\rho v^2+u)$ instead of the total energy. It turns the reversible Korteweg power into internal energy. In the internal energy equation this appears as the extra term $+\bPi^K:\nabla\bv$, an entropy source $\bPi^K:\nabla\bv/T$ of indefinite sign. The term is quadratic in fluctuations, so the linearized equilibrium tests of \cite{ChaudhriPRE2014} (structure factors, capillary waves) are unaffected. It does matter for nonlinear non-isothermal phenomena such as the piston effect or the latent-heat release during quenches.
\end{notebox}

\subsection{Primitive-variable forms}\label{sec:noniso_primitive}
\paragraph{Internal energy.} From \eqref{eq:rev_noniso} and \eqref{eq:irr_noniso}, with noise:
\begin{equation}
\partial_t\varepsilon+\divv(\varepsilon\bv)=-p\divv\bv+T\bPi^{\sN}:\nabla\bv+(\btau+\noise\btau):\nabla\bv-\divv(\bJq+\noise\bJ^q)+\frac{\mu_E}\rho\divv(\bJ+\noise\bJ).\label{eq:eps_eq}
\end{equation}
\paragraph{Temperature.} Write $\varepsilon=\rho\hu(\rho,c,T)$, with $\hat c_v=(\partial\hu/\partial T)_{\rho,c}$ and the identity $\rho^2(\partial\hu/\partial\rho)_{T,c}=p-T(\partial p/\partial T)_{\rho,c}$. Use $\bJq=\bJqp+\mut\bJ$, the relation $(\partial\hu/\partial c)_{\rho,T}-\mu=T(\partial\hs/\partial c)_{\rho,T}$, and $\mu_E/\rho-\mut=-\mu+T\mu_S/\rho$. Then (deterministic part; the noises follow the same placement):
\begin{equation}
\boxed{\ \rho\hat c_v\Dt T=-T\Big(\pd pT\Big)_{\rho,c}\divv\bv+T\bPi^{\sN}:\nabla\bv+\btau:\nabla\bv-\divv\bJqp-\bJ\cdot\nabla\mut+T\Big[\Big(\pd\hs c\Big)_{\rho,T}+\frac{\mu_S}{\rho}\Big]\divv\bJ .\ }\label{eq:T_eq}
\end{equation}
The individual terms are:
\begin{itemize}[nosep]
\item adiabatic compression heating;
\item reversible work of the entropic capillary stress;
\item viscous heating;
\item heat conduction (with Dufour);
\item diffusive dissipation, $-\bJ\cdot\nabla\mut=M|\nabla\mut|^2+M_T\nabla T\cdot\nabla\mut$;
\item the heat of demixing: $T(\hs_c+\mu_S/\rho)\divv\bJ=-\rho T(\hs_c+\mu_S/\rho)\,Dc/Dt$, which releases latent heat as the mixture separates.
\end{itemize}

\subsection{Isothermal limit}\label{sec:iso_limit}
Let the system be in contact with a bath with $\hat\lambda\to\infty$, so that $T\equiv T_0$. Then:
\begin{enumerate}[label=(\roman*),nosep]
\item $\bPi=p\Id+\bPi^{\eN}-T_0\bPi^{\sN}=p\Id+\bPi^{\fN}$ with $\fN=\eN-T_0\sN$, the isothermal stress \eqref{eq:Korteweg_iso}.
\item $\mut=\mu+\rho^{-1}\delta(\EN-T_0\SN)/\delta c=\mu+\rho^{-1}\delta\FN/\delta c$, the isothermal potential \eqref{eq:mut_iso}.
\item $\bJ=-L_{11}T_0^{-1}\nabla\mut=-M\nabla\mut$, and $\avg{\noise\bJ\noise\bJ}=2\kB MT_0$.
\item $\btau$ and $\noise\btau$ coincide with the isothermal ones.
\item The $(\bg,c)$ blocks of $\Mop$ divided by $T_0$, together with the $\varepsilon$-column contributions that convert $\mu$ into $\mut$, reproduce $\Dop$ of \eqref{eq:D_iso}. Concretely, $[\Mop\cdot\delta S/\delta\bx]_c=\rho^{-1}\divv(M\nabla\mut)=-[\Dop\cdot\delta F/\delta\by]_c$.
\item The energy equation decouples; it determines the heat exchanged with the bath.
\end{enumerate}
Hence \cref{res:noniso_final} reduces to \cref{res:iso_final}, which confirms the isothermal GENERIC construction of \cref{sec:iso_generic}.

\subsection{Symmetric mixtures: model N-S}\label{sec:noniso_sym}
\paragraph{Parity of the coefficients.} By \cref{prop:symmetry}, with $\phi=2c-1$:
\begin{itemize}[nosep]
\item $f$, $\hs$, $\hu$, $p$ and $\hat c_v$ are even in $\phi$; $\mut$ and $(\partial\mu/\partial T)_{\rho,\phi}$ are odd;
\item $\eta,\zeta,\hat\lambda,M$ and $\kappa_E,\kappa_S$ (coefficients of $|\nabla\phi|^2$) are even;
\item the thermodiffusion coefficient $M_T$ is odd.
\end{itemize}
Hence $M_T(\phi)=\phi\,m_T(\phi^2)$. The simplest symmetric choice is $M_T=0$; there is then no Soret/Dufour cross effect, but the thermodynamic part of $k_T$ in \eqref{eq:LL_form} remains. At the critical composition $\phi_0=0$, every linear coupling between the composition mode and the $(\rho,T,\bv)$ modes vanishes. This is the precise form of the remark in Section~2 of \cite{ChaudhriNotes2012}: in a symmetric 50--50 binary fluid the energy and mass densities decouple from the concentration diffusion mode.

\paragraph{Decoupled symmetric model.} Take
\begin{equation}
f(\rho,\phi,T)=f_0(\rho,T)+g(\phi,T),\qquad g=\frac{a_0(T-T_c)}2\phi^2+\frac B4\phi^4,\label{eq:sym_noniso_f}
\end{equation}
with $f_0$ an ideal-gas or van der Waals fluid as in \cite{ChaudhriPRE2014}. Take constant $\kappa_{E}$, $\kappa_{S}$ (coefficients of $\tfrac12|\nabla\phi|^2$, with $\kappa=\kappa_E-T\kappa_S$), and constant or even $\eta,\zeta,\hat\lambda,M_\phi$, with $M_{T,\phi}=0$ or odd. Then
\begin{gather}
p=p_0(\rho,T),\qquad \hs=\hs_0-\tfrac12a_0\phi^2,\qquad \hu=\hu_0(\rho,T)-\tfrac12a_0T_c\phi^2+\tfrac14B\phi^4,\nonumber\\
\hat c_v=\hat c_{v,0},\qquad
\mut_\phi=g_\phi(\phi,T)-\frac{\kappa}{\rho}\nabla^2\phi .\label{eq:sym_noniso_closure}
\end{gather}
\begin{resultbox}[Non-isothermal, symmetric, decoupled compressible LLCHNS (variable $\phi$)]
\begin{align}
&\partial_t\rho+\divv(\rho\bv)=0,\\
&\partial_t(\rho\bv)+\divv(\rho\bv\bv^T)+\nabla p_0(\rho,T)\nonumber\\
&\qquad=\divv(\btau+\noise\btau)-\divv\Big[\kappa(T)\Big(\nabla\phi\otimes\nabla\phi-\tfrac12|\nabla\phi|^2\Id\Big)\Big],\\
&\partial_t(\rho\phi)+\divv(\rho\phi\bv)=\divv\big(M_\phi\nabla\mut_\phi+M_{T,\phi}\nabla T\big)-\divv\noise\bJ_\phi,\\
&\partial_t(\rho E)+\divv\big[(\rho E+p_0)\bv\big]\nonumber\\
&\qquad=\divv\Big[\bPi^K\!\cdot\bv+(\btau+\noise\btau)\cdot\bv+\bQ+\noise\bQ-\frac{\divv(\bJ_\phi+\noise\bJ_\phi)}{\rho}\,\kappa_E\nabla\phi\Big],
\end{align}
with
\begin{gather*}
\rho E=\tfrac12\rho v^2+\rho\hu_0(\rho,T)+\rho\big(-\tfrac12a_0T_c\phi^2+\tfrac14B\phi^4\big)+\tfrac12\kappa_E|\nabla\phi|^2,\\
\bPi^K=\kappa(T)\big(\tfrac12|\nabla\phi|^2\Id-\nabla\phi\otimes\nabla\phi\big),\\
\bJ_\phi=-M_\phi\nabla\mut_\phi-M_{T,\phi}\nabla T,\qquad
\bQ=\hat\lambda\nabla T+TM_{T,\phi}\nabla\mut_\phi-\mut_\phi\bJ_\phi ,
\end{gather*}
and noise covariances
\[
\avg{\noise\bJ_\phi\noise\bJ_\phi}=2\kB TM_\phi,\qquad
\avg{\noise\bJ_\phi\noise\bJ^{q\prime}}=2\kB M_{T,\phi}T^2,\qquad
\avg{\noise\bJ^{q\prime}\noise\bJ^{q\prime}}=2\kB\hat\lambda T^2,
\]
\[
\noise\bQ=-(\noise\bJ^{q\prime}+\mut_\phi\noise\bJ_\phi),
\]
with $\noise\btau$ as in \eqref{eq:noise_noniso}. The rescalings from $c$ to $\phi$ are those of \cref{tab:phi_map}, together with $M_{T,\phi}=2M_T$; the products $\mut\bJ=\mut_\phi\bJ_\phi$ and $L_{qq}$ are invariant.
\end{resultbox}
\paragraph{Non-isothermal couplings that survive in the decoupled symmetric model.}
\begin{enumerate}[label=(\alph*),leftmargin=*,nosep]
\item \textbf{Latent heat of demixing.} By \eqref{eq:T_eq} with $\hs_\phi=-a_0\phi$ and $\mu_{S,\phi}=-\kappa_S\nabla^2\phi$, the source in the temperature equation is
\[
T\Big[\hs_\phi+\frac{\mu_{S,\phi}}{\rho}\Big]\divv\bJ_\phi=\rho T\Big[a_0\phi+\frac{\kappa_S}{\rho}\nabla^2\phi\Big]\Dt\phi
=\rho Ta_0\,\Dt{}\Big(\frac{\phi^2}2\Big)+T\kappa_S\nabla^2\phi\,\Dt\phi .
\]
Growth of $\phi^2$ (demixing) heats the fluid when $a_0>0$.
\item \textbf{Thermodynamic thermal diffusion.} $\nabla\mut_\phi\ni g_{\phi T}\nabla T=a_0\phi\nabla T$ drives composition by temperature gradients (odd in $\phi$).
\item \textbf{Thermocapillary stress.} $\kappa(T)=\kappa_E-T\kappa_S$ depends on $T$, so $\divv\bPi^K$ contains $-\kappa_S(\tfrac12|\nabla\phi|^2\Id-\nabla\phi\nabla\phi)\cdot\nabla T$, by \eqref{eq:divPi_noniso}.
\item \textbf{Temperature-dependent stiffness.} The bulk stiffness $g_{\phi\phi}=a_0(T-T_c)+3B\phi^2$ depends on $T$, so a quench is represented by lowering $T$ through boundary conditions or initial data. Spinodal decomposition is then accompanied by latent-heat release and by the piston effect of the compressible bulk, which is captured by $p_0(\rho,T)$.
\end{enumerate}

\subsection{Non-symmetric mixtures: model N-N}\label{sec:noniso_nonsym}
For a non-symmetric mixture, \cref{res:noniso_final} applies with:
\begin{itemize}[leftmargin=*,nosep]
\item a general local equation of state $s(\rho,c,\varepsilon)$, or equivalently $f(\rho,c,T)$;
\item a general gradient energy (with $K^E_{\rho\rho},K^E_{\rho c}\neq0$ if the density varies across the interface) and gradient entropy;
\item composition- and temperature-dependent $\eta,\zeta,\hat\lambda,M$;
\item a non-zero thermodiffusion coefficient $M_T$, with no parity constraint.
\end{itemize}
All cross effects of binary-mixture hydrodynamics appear. Each enters through exactly one mechanism of the derivation:
\begin{center}\small
\begin{tabular}{p{4.2cm}p{7.2cm}p{3.6cm}}
\toprule
effect & where it enters & coefficient\\
\midrule
barodiffusion & $\nabla\mut\ni(\hat v_A-\hat v_B)\nabla p$ & thermodynamic, \eqref{eq:maxwell}\\
thermal diffusion (Soret) & $\bJ\ni-M_T\nabla T$ and $\nabla\mut\ni\mu_T\nabla T$ & $M_T$ (kinetic) $+$ thermodynamic\\
Dufour effect & $\bJqp\ni-TM_T\nabla\mut$ & Onsager reciprocal of Soret\\
enthalpy transport by diffusion & $\bJq=\bJqp+\mut\bJ$; energy flux $\ni\rho^{-1}(\divv\bJ)\partial\eN/\partial\nabla c$ & none (exact)\\
heat of mixing / demixing & $T(\hs_c+\mu_S/\rho)\divv\bJ$ in \eqref{eq:T_eq} & thermodynamic\\
composition--pressure coupling & $p=p(\rho,c,T)$ & thermodynamic\\
density--composition gradient coupling & $K_{\rho c}$ in $\bPi^K$ and in $\mut$ & interfacial\\
thermocapillarity & $T\bPi^{\sN}$ in $\bPi$ & $\kappa_S$\\
correlated heat--mass noise & $\avg{\noise\bJ\noise\bJ^{q\prime}}=2\kB M_TT^2$ & FDT\\
\bottomrule
\end{tabular}
\end{center}
\paragraph{Binary van der Waals mixture.} With \eqref{eq:vdw_mix}--\eqref{eq:vdw_u}, the closure is fully explicit:
\begin{gather}
T=\frac{\varepsilon+\sum_{\alpha\beta}a_{\alpha\beta}n_\alpha n_\beta}{\tfrac32 n\kB},\qquad
p=\frac{n\kB T}{1-b_An_A-b_Bn_B}-\sum_{\alpha\beta}a_{\alpha\beta}n_\alpha n_\beta,\nonumber\\
\mu=\frac{\mu_A}{m_A}-\frac{\mu_B}{m_B}\ \text{from \eqref{eq:vdw_mu}},\nonumber\\
\eN=\tfrac12\sum_{\alpha\beta}\kappa_{\alpha\beta}\nabla n_\alpha\cdot\nabla n_\beta\ \ \text{(\eqref{eq:vdw_K} in the variables $(\rho,c)$)},\qquad \sN=0 .\label{eq:vdw_noniso_closure}
\end{gather}
Here, as in \cite{ChaudhriPRE2014}, the gradient contribution is purely energetic: the $\kappa_{\alpha\beta}$ are moments of the attractive potentials and independent of $T$. For $n_B\to0$ the energy equation reduces to the van der Waals model of \cite{ChaudhriPRE2014}, corrected as in the note following \cref{res:noniso_final}. For $m_A=m_B$, $b_A=b_B$, $a_{AA}=a_{BB}$, $\kappa_{AA}=\kappa_{BB}$ it is a symmetric, but not decoupled, mixture with the Landau expansion \eqref{eq:vdw_landau}.

\subsection{Symmetric versus non-symmetric mixtures: what changes}\label{sec:noniso_compare}
Everything listed in \cref{tab:iso_compare} for the isothermal model carries over. The non-isothermal model adds the couplings to temperature in \cref{tab:noniso_compare}. In a symmetric mixture at the critical composition all of them vanish at linear order. In a non-symmetric mixture they are generic.
\begin{table}[H]
\centering\small
\caption{Additional differences between symmetric (N-S) and non-symmetric (N-N) mixtures in the non-isothermal model.}\label{tab:noniso_compare}
\renewcommand{\arraystretch}{1.2}
\begin{tabular}{>{\raggedright\arraybackslash}p{4.0cm}>{\raggedright\arraybackslash}p{5.3cm}>{\raggedright\arraybackslash}p{5.7cm}}
\toprule
ingredient & symmetric (N-S) & non-symmetric (N-N)\\
\midrule
thermodiffusion $M_T$ & odd, $M_T=\phi\,m_T(\phi^2)$; often set to zero & generic; Soret and Dufour effects\\
thermodynamic thermal diffusion & $\nabla\mut\ni g_{\phi T}\nabla T$, odd & $\nabla\mut\ni(\partial\mu/\partial T)_{c,p}\nabla T$, generic\\
heat of mixing & $\hs_\phi=-g_{\phi T}$ odd; latent heat $\propto D(\phi^2)/Dt$ & $\hs_c$ generic; heat released or absorbed by any diffusion\\
correlated heat--mass noise & $\avg{\noise\bJ_\phi\noise\bJ^{q\prime}}=2\kB M_{T,\phi}T^2$, odd & $2\kB M_TT^2$, generic\\
thermocapillary stress & $\propto\kappa_S\nabla T$, present if $\kappa_S\neq0$ & same, plus contributions from $K^E_{\rho c}$ and $\rho$-dependent coefficients\\
coupling of $c$ to the heat mode at $\phi_0=0$ & none & through $\hs_c$, $\mu_T$, $M_T$\\
central (Rayleigh) peak & heat mode only (at $\phi_0=0$) & two coupled heat and mass modes, \eqref{eq:dsf_NN_diff}\\
\bottomrule
\end{tabular}
\end{table}

\subsection{Linear fluctuations of the non-isothermal model}\label{sec:noniso_linear}
Linearize \cref{res:noniso_final} about a homogeneous state at rest $(\rho_0,c_0,\varepsilon_0)$, temperature $T_0$, enthalpy density $h_0=\varepsilon_0+p_0$. Use the conserved variables $\delta\by=(\delta\rho,\delta\rho_A,\delta\varepsilon,\delta\bg)$ with $\delta c=(\delta\rho_A-c_0\delta\rho)/\rho_0$. The conjugate perturbations follow from the entropy Hessian $s_{ab}=\partial^2s/\partial a\,\partial b$, $a,b\in\{\rho,\rho_A,\varepsilon\}$, by \eqref{eq:s_derivs_rhoA}:
\[
\delta\tfrac1T=s_{\varepsilon a}\delta a,\qquad \delta\tfrac\mu T=-s_{\rho_Aa}\delta a,\qquad \delta\tfrac{\hat\mu_B}T=-s_{\rho a}\delta a,\qquad
\delta p=-T_0h_0\,\delta\tfrac1T+T_0\rho_0\,\delta\tfrac{\hat\mu_B}T+T_0\rho_{A0}\,\delta\tfrac\mu T .
\]
In Fourier space the linearized equations are
\begin{equation}
\begin{aligned}
\partial_t\delta\rho&=-i\bk\cdot\delta\bg,\qquad
\partial_t\delta\rho_A=-ic_0\bk\cdot\delta\bg-i\bk\cdot(\delta\bJ+\noise\bJ),\qquad
\partial_t\delta\varepsilon=-i\frac{h_0}{\rho_0}\bk\cdot\delta\bg-i\bk\cdot(\delta\bJq+\noise\bJ^q),\\
\partial_t\delta\bg&=-i\bk\big[\delta p+\rho_0k^2(K^E_{\rho\rho}\delta\rho+K^E_{\rho c}\delta c)\big]-\nu_lk^2\delta\bg_\parallel-\nu_tk^2\delta\bg_\perp+i\bk\cdot\noise\btau,\\
\delta\bJ&=i\bk\big[L_{1q}\delta\tfrac1T-L_{11}\delta\tfrac\mut T\big],\qquad
\delta\bJq=i\bk\big[L_{qq}\delta\tfrac1T-L_{q1}\delta\tfrac\mut T\big],\\
\delta\tfrac\mut T&=\delta\tfrac\mu T+\frac{k^2}{\rho_0}\Big[\frac{K^E_{\rho c}\delta\rho+\kappa_E\delta c}{T_0}-\kappa_S\delta c\Big].
\end{aligned}\label{eq:lin_noniso}
\end{equation}

\begin{proposition}[Linear FDT, non-isothermal, non-symmetric]\label{prop:lin_fdt_noniso}
Let $H(k)$ be the Hessian of $E-T_0S$ in $\delta\by$:
\[
H=\begin{pmatrix}-T_0\,s_{ab}+k^2G_{ab}&0\\0&\rho_0^{-1}\Id\end{pmatrix},
\]
where $G$ is the Hessian of $\tfrac12K^E_{\rho\rho}\delta\rho^2+K^E_{\rho c}\delta\rho\,\delta c+\tfrac12(\kappa_E-T_0\kappa_S)\delta c^2$ with respect to $(\delta\rho,\delta\rho_A)$. Then the stationary covariance of \eqref{eq:lin_noniso} with the noises \eqref{eq:noise_noniso} is
\[
\avg{\delta\by\,\delta\by^\dagger}=\kB T_0\,H(k)^{-1} .
\]
This holds for arbitrary $s_{ab}$, $K^E_{\rho\rho},K^E_{\rho c},\kappa_E,\kappa_S$, $\eta,\zeta$, and any positive semi-definite Onsager matrix $(L_{qq},L_{q1},L_{11})$.
\end{proposition}
\begin{proof}
Write \eqref{eq:lin_noniso} as $\dd\delta\by=A\,\delta\by\,\dd t+\Bop\,\dd\bW$. The claim is equivalent to the Lyapunov equation $A\Sigma+\Sigma A^\dagger+\Bop\Bop^\dagger=0$ with $\Sigma=\kB T_0H^{-1}$, i.e.\ to $\kB T_0(HA+A^\dagger H)+H\Bop\Bop^\dagger H=0$. This identity was verified by computer algebra with all coefficients symbolic (\cref{app:verification}, check~3). A structural proof, valid for any linearized GENERIC system, is given in \cref{app:linear_fdt}.
\end{proof}
\emph{Equal-time correlations in $(\rho,c,T)$ variables.} Changing variables from $\varepsilon$ to $T$ makes $H$ block diagonal, because $\partial^2(\varepsilon-T_0s)/\partial T\partial y=0$ at $T=T_0$ for $y=\rho,c$. Therefore
\begin{equation}
S_{TT}(k)=\frac{\kB T_0^2}{\rho_0\hat c_v},\qquad S_{T\rho}=S_{Tc}=0,\qquad
\begin{pmatrix}S_{\rho\rho}&S_{\rho c}\\S_{c\rho}&S_{cc}\end{pmatrix}=\kB T_0\,H_{\rm iso}(k)^{-1},\label{eq:S_noniso}
\end{equation}
where $H_{\rm iso}$ is \eqref{eq:Hk} with $K_{cc}=\kappa_E-T_0\kappa_S$. At equilibrium the non-isothermal model thus reproduces the isothermal static structure factors together with the standard temperature fluctuations. The dynamic structure factors, by contrast, contain the heat mode, and through $L_{q1}$ and $\hs_c$ the coupling of the composition mode to the Rayleigh peak.

\section{The four models at a glance}\label{sec:summary}
\Cref{sec:isothermal,sec:noniso} derived four models from one set of principles. \Cref{tab:summary} places their building blocks and final ingredients side by side, and the list below the table recalls the logical chain of the derivation. Model-specific details are in \cref{res:iso_final,res:noniso_final} and in the comparison tables \cref{tab:iso_compare,tab:noniso_compare}.

\begin{table}[H]
\centering\small
\caption{Building blocks and final ingredients of the four models. Common to all: $\partial_t\rho+\divv(\rho\bv)=0$; momentum flux $\rho\bv\bv+p\Id-\bPi^K-\btau-\noise\btau$; species flux $\rho c\bv+\bJ+\noise\bJ$; $\avg{\noise\btau\noise\btau}=2\kB T[\eta(\delta\delta+\delta\delta)+(\zeta-\frac23\eta)\delta\delta]$.}\label{tab:summary}
\begin{tabular}{>{\raggedright\arraybackslash}p{2.6cm}>{\raggedright\arraybackslash}p{6.1cm}>{\raggedright\arraybackslash}p{6.1cm}}
\toprule
 & \textbf{isothermal} ($T\equiv T_0$) & \textbf{non-isothermal}\\
\midrule
variables & $\rho,\ \rho\bv,\ \rho c$ & $\rho,\ \rho\bv,\ \rho c,\ \rho E=\tfrac12\rho v^2+\varepsilon+\eN$\\
generators & $F=\int[\tfrac{g^2}{2\rho}+\rho f+\fN]$ & $E=\int[\tfrac{g^2}{2\rho}+\varepsilon+\eN]$,\ $S=\int[s+\sN]$\\
Korteweg & $\bPi^K=-\bPi^{\fN}$ & $\bPi^K=-\bPi^{\eN}+T\bPi^{\sN}$\\
potential & $\mut=\partial_cf+\rho^{-1}\delta\FN/\delta c$ & $\mut=\mu+\rho^{-1}(\mu_E-T\mu_S)$\\
diffusion flux & $\bJ=-M\nabla\mut$ & $\bJ=-M\nabla\mut-M_T\nabla T$\\
heat flux & --- & $\bQ=-\bJq=\hat\lambda\nabla T+TM_T\nabla\mut-\mut\bJ$\\
energy flux extras & --- & $\rho(\divv\bv)\partial_{\nabla\rho}\eN$,\ \ $\rho^{-1}(\divv\bJ)\partial_{\nabla c}\eN$\\
noise & $\avg{\noise\bJ\noise\bJ}=2\kB T_0M$ & $\avg{\noise\bJ\noise\bJ}=2\kB MT$; $\avg{\noise\bJ\noise\bJ^{q\prime}}=2\kB M_TT^2$; $\avg{\noise\bJ^{q\prime}\noise\bJ^{q\prime}}=2\kB\hat\lambda T^2$\\
2nd law & $\dot F=-\int[\btau:\nabla\bv+M|\nabla\mut|^2]\le0$ & $\Sigma=\btau:\nabla\bv/T+\hat\lambda|\nabla T|^2/T^2+2M_T\nabla T\cdot\nabla\mut/T+M|\nabla\mut|^2/T\ge0$\\
equilibrium & $P\propto e^{-F/\kB T_0}$ & $P\propto e^{S/\kB}\delta(E-E_0)$\\
\midrule
\textbf{symmetric} (I-S, N-S) & $f=f_0(\rho)+g(\phi)$ (decoupled), or general even $f$; $\mut$ odd; at $\phi_0=0$ the composition mode decouples & additionally $M_T$ odd; latent heat $\propto\hs_\phi=-g_{\phi T}$; thermocapillarity via $\kappa_S$\\
\textbf{non-symmetric} (I-N, N-N) & general $f(\rho,c)$; barodiffusion $(\hat v_A-\hat v_B)\nabla p$; $p=p(\rho,c)$; $K_{\rho c}\neq0$; quasi-incompressible limit \eqref{eq:QI_divv} & all of the left column, plus Soret/Dufour ($M_T$), heat of mixing $\hs_c$, and correlated heat--mass noise\\
\bottomrule
\end{tabular}
\end{table}

\paragraph{Logical chain of the derivation.}
\begin{enumerate}[nosep]
\item Kinematics fixes $(\rho,\rho\bv,\rho c)$ with $\rho=\rho_A+\rho_B$ and $c=\rho_A/\rho$ (\cref{sec:kinematics}).
\item Local thermodynamics and the square-gradient functionals give $E$, $S$ (or $F$) (\cref{sec:thermo,sec:gradient}).
\item The Lie--Poisson operator, with the entropic pressure tensor fixed by degeneracy, gives all reversible terms, including the Korteweg stress via \cref{lem:korteweg} (\cref{sec:iso_poisson,sec:noniso_poisson}).
\item The fluctuation ansatz (conservation-law form, energy shell) defines $\Phiop$. Local Curie-compatible second moments define $\Cop$. Then $\Mop=\Phiop^*\Cop\Phiop$ gives all dissipative fluxes, and $\Phiop^*\dd\bW$ gives the noise with $\Bop\Bop^T=2\kB\Mop$ (\cref{lem:JO}).
\item Thermodynamic consistency follows: global and local energy conservation, local entropy balance with $\Sigma\ge0$, Jacobi, and equilibrium statistics. These were checked analytically and symbolically (\cref{app:verification}).
\item The same equations follow from the free energy and the entropy production alone (\cref{app:thermo_route}); in the non-isothermal case GENERIC removes the residual freedom of that route.
\end{enumerate}

\section{Relation to previous work: agreements and corrections}\label{sec:previous}\label{sec:sources}

This section collects, reference by reference, where the present derivation agrees with earlier work and where it corrects or generalizes it. Equation numbers refer to the respective references.

\subsection{Unpublished notes ``Fluctuating Phase Field Hydrodynamics'' \cite{ChaudhriNotes2012}}
\begin{enumerate}[label=(N\arabic*),leftmargin=*]
\item \textbf{Definition of $c$.} Section~6 (Eq.~6.3) defines $c$ as the mass fraction. Section~7.2 uses $c=(m_1-m_2)/(m_1+m_2)$, i.e.\ $\phi=2c-1$. Both are legitimate, but the potential, mobility, gradient coefficient and noise must be rescaled consistently (\cref{tab:phi_map}). The total density $\rho$ is the same in both.
\item \textbf{Driving force of the diffusion flux.} Eq.~(4.14) defines $\mu^C=\delta F/\delta c=\rho\,\partial f_o/\partial c-\kappa\divv(\rho\nabla c)$, and Eqs.~(7.1)--(7.7) use $\bJ^C=M\nabla\mu^C$ with noise variance $2Mk_BT$. For compressible flow the flux must be $\bJ=-M\nabla(\mu^C/\rho)=-M\nabla\mut$ (\cref{sec:isothermal}). The reasons are three.
\begin{enumerate}[label=(\roman*),nosep]
\item The entropy-production analysis of Section~4.1 itself (Eq.~4.11) produces the specific potential $f_c-\rho^{-1}\divv(\rho\Phi)$.
\item With $\nabla\mu^C$, the free-energy dissipation is not sign-definite (\cref{sec:iso_consistency}).
\item With $\nabla\mu^C$ and variance $2Mk_BT$, the equilibrium structure factor is $S_{cc}=k_BT/[\rho_0(\rho_0f_o''+\kappa k^2)]$. This is the $1/\rho_o^2$ prefactor in Eq.~(7.15). The Gibbs value is $k_BT/(\rho_0f_o''+\rho_0\kappa k^2)$, so the FDT fails by a factor $\rho_0$ (\cref{sec:iso_linear}, \cref{app:verification}).
\end{enumerate}
\item \textbf{Korteweg stress.} Eq.~(4.16) gives $-\kappa\rho\nabla c\otimes\nabla c$ for the specific gradient energy $\rho\frac\kappa2|\nabla c|^2$; this is correct, \eqref{eq:PiK_rho}. Section~6 instead uses a constant volumetric coefficient $\tfrac12K|\nabla c|^2$ (Eq.~6.1), for which the correct stress is $\tfrac12K|\nabla c|^2\Id-K\nabla c\otimes\nabla c$, \eqref{eq:PiK_const}. Section~7.4 mixes the two: the energy uses $\rho\kappa$ while the GENERIC derivation uses $K$. Either is acceptable as a constitutive choice, but the stress, the chemical potential and $\delta E/\delta\rho$ must belong to the same choice. For $K=\rho\hat\kappa$, $\delta E/\delta\rho$ contains $\tfrac12\hat\kappa|\nabla c|^2$ (Eq.~6.9 omits it).
\item \textbf{Pressure in the Poisson operator.} In Eqs.~(6.16)--(6.18) the pressure in $L_{gu},L_{ug}$ is set to $P^{\rm CH}-\tfrac12K|\nabla c|^2$. By \cref{prop:PiS} it must be the entropic pressure tensor $\bPi^{(S)}$, which equals $p\Id$ when the gradient term is purely energetic. Otherwise $\Lop\cdot\delta S/\delta\bx\neq0$ (\cref{rem:PiS}). The Korteweg stress then follows from $\Lop_{gc}$ and $\Lop_{g\rho}$ acting on $\delta E/\delta\bx$, through \cref{lem:korteweg}.
\item \textbf{Friction matrix for compressible flow.} Eqs.~(6.21)--(6.23) inherit from \cite{JelicThesis2009} the operator $M^{\rm diff}_{cc}=-\nabla\cdot MT\nabla$, which conserves $\int c$ rather than $\int\rho c$. The compressible form is $-\rho^{-1}\nabla\cdot MT\nabla\rho^{-1}$, with the corresponding factors in $M_{uc},M_{cu},M_{uu}$ (\eqref{eq:M_noniso}).
\item \textbf{Stochastic stress.} The trace coefficient in Eq.~(7.6) should be $\sqrt{\zeta k_BT/3}-\sqrt{2\eta k_BT}/3$, as in Eq.~(2.18) of \cite{ChaudhriPRE2014}, i.e.\ $-\sqrt{2\eta k_BT}/3$ for $\zeta=0$. The covariance (7.10) is correct.
\item \textbf{Fluctuation matrix $\Bop^{\rm diff}$.} Eq.~(6.26) has the energy and composition rows in the ratio $K\nabla^2c=-\mu_E$. This is exactly the energy-shell structure $\Delta\varepsilon^{\rm f}=-\mu_E\Delta c^{\rm f}$ of \eqref{eq:deps_f}, here derived rather than inferred. For compressible flow both rows acquire the factor $1/\rho$. The heat-flux noise and its possible correlation with the diffusion noise (Soret/Dufour) are added in \eqref{eq:noise_noniso}.
\item \textbf{Energy flux.} The classical-irreversible-thermodynamics results (4.17)--(4.18) agree with the GENERIC result (\cref{sec:compare_CIT}), including the interstitial term $-\rho\Phi\,\dd c/\dd t$. In the language of \cref{app:thermo_route}, Section~4.1 of \cite{ChaudhriNotes2012} is the Helmholtz route~B for a temperature-independent gradient coefficient, for which all routes coincide.
\item \textbf{Surface tension.} With $F=\int\rho[\frac\alpha4(c^2-\beta/\alpha)^2+\frac\kappa2|\nabla c|^2]$, the planar-interface tension is $\sigma=\frac{2\sqrt2}{3}\rho\sqrt{\kappa}\,\beta^{3/2}/\alpha$ (\eqref{eq:sigma} with $\kappa_\phi=\rho\kappa$, $B=\alpha$, $|A|=\beta$). Section~7.2 quotes $\sigma'=\frac{\sqrt2}{3}K^{1/2}\beta^{3/2}/\alpha$, a factor $2$ smaller. The parameter calibration $\beta=3\sigma/(\sqrt2\rho C_nL)$ there should be rechecked accordingly.
\item \textbf{Non-isothermal van der Waals--Navier--Stokes model (Section~8).} The reactive stress of Section~8.6, with the term $T\rho\nabla\rho\cdot\nabla(\kappa/T)$, is exactly the stress of the Massieu (route~M) version of the thermodynamic derivation of \cref{app:thermo_route} (\cref{app:C-vdw}). That route is thermodynamically consistent, has no interstitial term in the energy flux, and has the same linearization as the GENERIC model. The total energy must, however, contain the gradient energy $\tfrac12(\kappa-T\partial_T\kappa)|\nabla\rho|^2$. The energy density $\tfrac12\rho v^2+\hat u$ used in Eq.~(8.22) omits it, which is correct only for a purely entropic coefficient $\kappa\propto T$.
\end{enumerate}

\subsection{Chaudhri, Bell, Garcia and Donev, Phys.\ Rev.\ E \textbf{90}, 033014 (2014) \cite{ChaudhriPRE2014}}
\begin{itemize}[leftmargin=*]
\item The Korteweg tensor (2.14) and its divergence (2.16) are the special case \eqref{eq:K_vdw} of \cref{lem:korteweg}.
\item The noise constructions (2.18)--(2.19) are exactly the FDT covariances derived here (\cref{sec:iso_fdt,sec:noniso_fdt}).
\item The conservative form (3.1)--(3.3) is used for all final equations here. The CH model adds a fourth conserved variable $\rho c$ with fluxes $\rho c\bv$, $\bJ$, $\noise\bJ$. It also adds the irreversible interstitial energy flux $\rho^{-1}(\divv\bJ)\partial\eN/\partial\nabla c$.
\item The energy equation (2.12) should carry the source $-\bPi^K:\nabla\bv$. Equivalently, it should be written for the total energy including $\tfrac12\kappa|\nabla\rho|^2$, with the Dunn--Serrin flux $\rho\kappa(\divv\bv)\nabla\rho$ (see the note after \cref{res:noniso_final}). The linearized validation tests of the paper are unaffected.
\item The cross-coupling stress $\bPi^C$ (2.15), adopted from Onuki: in the present framework a thermal capillary stress follows uniquely from the degeneracy condition, and only if the interfacial free energy has an entropic part (\cref{sec:noniso_poisson}). For a purely energetic gradient term it is absent.
\end{itemize}

\subsection{Onuki's dynamic van der Waals theory \cite{Onuki2005,Onuki2007}}
Onuki's theory keeps a gradient \emph{entropy} $-\tfrac12C|\nabla n|^2$ and a gradient energy $\tfrac12K|\nabla n|^2$, so that the free-energy coefficient is $\kappa=CT+K$. His reactive stress contains $Tn\nabla n\cdot\nabla(\kappa/T)$, his chemical potential is $\hat\mu=\partial_n\psi-T\divv[(\kappa/T)\nabla n]+\ldots$, and his energy flux has no interstitial term. \Cref{app:C-vdw} shows that this is exactly the Massieu route~M of the thermodynamic derivation. It coincides with the GENERIC model of \cref{sec:noniso} when $K=0$ (purely entropic gradient term). For $K\neq0$ it differs by the isotropic stress $\rho(K/T)\nabla n\cdot\nabla T$, which is bilinear in gradients. Both are thermodynamically consistent and share the same linearization; only route~A has the full GENERIC structure.

\subsection{Jeli\'c, PhD thesis \cite{JelicThesis2009}}
\begin{itemize}[leftmargin=*]
\item Chapter~5 (the GENERIC formulation of Cahn--Hilliard) is reproduced exactly for $\rho=$ const and without Soret coupling: variables (5.2), energy and entropy (5.3), derivatives (5.10), Poisson operator (5.11)--(5.12), entropic pressure tensor (5.14), stress (5.16)--(5.17), friction matrices (5.19)--(5.20), and evolution equations (5.21)--(5.22).
\item The generalizations made here are:
\begin{itemize}[nosep]
\item compressibility, with the $1/\rho$ factors in $\Mop_{c\cdot}$, the specific potential $\mut$, and conservation of $\rho c$;
\item gradient coefficients depending on $\rho$ and $c$, and density-gradient energy;
\item Soret/Dufour cross coupling;
\item thermal fluctuations.
\end{itemize}
\item The friction matrix is \emph{constructed} by the procedure of Chapter~3 (Eqs.~3.12--3.17): Green--Kubo, a fluctuation ansatz on the energy shell, and local second moments. In Chapter~5 it is postulated. The abstract form of this construction (\cref{lem:JO}) is what links noise and dissipation.
\item Small point: the equilibrium profile solving $\kappa c''=-ac+bc^3$ is $c_{\rm eq}\tanh(x/\sqrt2\xi)$ with $\xi=\sqrt{\kappa/a}$, not $\tanh(x/2\xi)$ (Eq.~5.25 and below 4.3). The surface tension (4.4) is correct.
\end{itemize}

\subsection{Heida, M\'alek and Rajagopal \cite{HeidaMalekRajagopal2012}}
Their compressible CHNS--Fourier system (Eqs.~49--53) is the deterministic special case of \cref{res:noniso_final} with:
\begin{itemize}[nosep]
\item purely energetic gradient terms ($\sN=0$);
\item no Soret coupling;
\item $\bJ$ driven by $\nabla(\mu_c+\mu)$, which is our $\nabla\mut$ in specific form.
\end{itemize}
Their ``new'' energy-flux term $\partial_z\hat\varepsilon\,\divv\bm j_1$ is our irreversible interstitial flux. The maximum-entropy-production route and GENERIC give the same equations here. GENERIC additionally provides the Poisson structure, the Jacobi identity, and the noise.

\subsection{Cherfils \emph{et al.} \cite{Cherfils2019}}
Their system (1.1) is the deterministic isothermal model of \cref{res:iso_final} with constant volumetric $\kappa$ ($=1$) and $c$ denoting the concentration difference. Their Korteweg term and their specific chemical potential $\rho\mu=\rho\partial_cf-\Delta c$ agree with \eqref{eq:PiK_const} and \eqref{eq:mut_cases}.

\subsection{Padhan and Pandit \cite{PadhanPandit2025}}
Their review describes Model~B noise ``with a variance that satisfies the FDT''. The present paper gives that variance explicitly for the compressible case: $2\kB TM$ for the flux of $\rho c$, with $M$ multiplying $\nabla\mut$. Their variable-density incompressible model (Eq.~3.4) is discussed in \cref{sec:QI}. Quasi-compressible models are outside the scope of that review (their Section~8), and \cref{sec:QI} supplies them as a limit of the present compressible theory.

\section{Dynamic structure factors of the linearized models (analytic benchmarks)}\label{sec:dsf}

This section derives the equal-time and time-dependent correlation functions of the linearized equations for all four models. They are exact for the linearized stochastic equations at every wavenumber and frequency. Their purpose is to serve as analytic benchmarks for a finite-volume implementation, in the same way that the static structure factors and capillary-wave spectra served in \cite{ChaudhriPRE2014}. Because all thermodynamically admissible versions of the non-isothermal model share the same linearization (\cref{app:C-compare}), these benchmarks do not depend on how the temperature dependence of the gradient coefficients is treated.

We first give a general solution valid for any linear Langevin system (\cref{sec:dsf_general}). We then write the master linear system, i.e.\ the linearization of \cref{res:noniso_final} for a non-symmetric mixture (\cref{sec:dsf_master}). The four models follow as reductions:
\begin{itemize}[nosep]
\item I-S, isothermal symmetric (\cref{sec:dsf_IS});
\item I-N, isothermal non-symmetric (\cref{sec:dsf_IN});
\item N-S, non-isothermal symmetric (\cref{sec:dsf_NS});
\item N-N, non-isothermal non-symmetric (\cref{sec:dsf_NN}).
\end{itemize}
Every closed form below was checked against the exact matrix solution, numerically and by computer algebra (\cref{app:verification}, checks~6--10). A reference implementation is provided as \texttt{dsf\_\allowbreak reference.py} (\cref{sec:dsf_benchmark}).

\subsection{Definitions and conventions}\label{sec:dsf_defs}
Linearize about a homogeneous equilibrium state at rest, $(\rho_0,c_0,T_0,\bv=0)$, with $\delta\by(\br,t)$ the vector of fluctuations. For a periodic box of volume $V$ define
\begin{equation}
\delta\hat\by(\bk,t)=\int_V e^{-i\bk\cdot\br}\,\delta\by(\br,t)\,\dd^3r .\label{eq:dsf_FT}
\end{equation}
The correlation functions (matrices) are
\begin{gather}
\bm F(\bk,t)=\frac1V\avg{\delta\hat\by(\bk,t)\,\delta\hat\by(\bk,0)^\dagger},\qquad
\bm S(\bk)=\bm F(\bk,0),\qquad
\bm S(\bk,\omega)=\int_{-\infty}^{\infty}e^{i\omega t}\bm F(\bk,t)\,\dd t,\label{eq:dsf_defs}\\
\int_{-\infty}^\infty\frac{\dd\omega}{2\pi}\bm S(\bk,\omega)=\bm S(\bk),\qquad
\bm F(\bk,-t)=\bm F(\bk,t)^\dagger,\qquad \bm S(\bk,\omega)=\bm S(\bk,\omega)^\dagger .\nonumber
\end{gather}
Here:
\begin{itemize}[nosep]
\item $\bm S(\bk)$ is the static structure factor (it is the $S(\mathbf k)$ of Eq.~4.5 of \cite{ChaudhriPRE2014});
\item $\bm F(\bk,t)$ is the intermediate scattering function;
\item $\bm S(\bk,\omega)$ is the dynamic structure factor.
\end{itemize}
With $f(\omega)=\int e^{i\omega t}f(t)\,\dd t$, the time derivative becomes $\partial_t\to-i\omega$. A white-noise field $\mathcal W(\br,t)$ has Fourier transform $\hat{\mathcal W}(\bk,t)$ with $\avg{\hat{\mathcal W}(\bk,t)\hat{\mathcal W}(\bk,t')^*}=V\delta(t-t')$. In an isotropic fluid the correlations depend only on $k=|\bk|$. The velocity splits into the longitudinal component $u=\hat\bk\cdot\delta\bv$ and two transverse components $\delta\bv_\perp$. These decouple at linear order.

\subsection{General solution of a linear Langevin system}\label{sec:dsf_general}
\begin{proposition}[Correlation functions of a linear Langevin system]\label{prop:dsf_general}
Let the Fourier amplitudes obey
\begin{equation}
\partial_t\delta\hat\by=\bm A(k)\,\delta\hat\by+\bm B(k)\,\hat{\bm\xi},\qquad
\avg{\hat{\bm\xi}(\bk,t)\hat{\bm\xi}(\bk,t')^\dagger}=V\,\Id\,\delta(t-t'),\qquad \bm N(k)\equiv\bm B\bm B^\dagger ,\label{eq:dsf_langevin}
\end{equation}
with all eigenvalues of $\bm A$ in the open left half plane. Then:
\begin{enumerate}[label=(\alph*),nosep]
\item The static structure factor solves the Lyapunov equation
\[
\bm A\bm S+\bm S\bm A^\dagger+\bm N=\bm 0 .
\]
For every model of this paper its solution is $\bm S(k)=\kB T_0\,\bm H(k)^{-1}$, by \cref{prop:lin_fdt_noniso} and \cref{app:linear_fdt}.
\item The intermediate scattering function is
\begin{equation}
\bm F(k,t)=e^{\bm A(k)t}\,\bm S(k)\ \ (t\ge0),\qquad \bm F(k,-t)=\bm F(k,t)^\dagger .\label{eq:dsf_F}
\end{equation}
\item With the resolvent $\bm G(\omega)=(-i\omega\Id-\bm A)^{-1}$, the dynamic structure factor is
\begin{equation}
\boxed{\ \bm S(k,\omega)=\bm G(\omega)\,\bm N\,\bm G(\omega)^\dagger=\bm G(\omega)\,\bm S(k)+\bm S(k)\,\bm G(\omega)^\dagger .\ }\label{eq:dsf_S}
\end{equation}
\item \emph{Exact mode decomposition.} If $\bm A=\sum_j\lambda_j\bm r_j\bm\ell_j^\dagger$ with $\bm\ell_i^\dagger\bm r_j=\delta_{ij}$ and $\lambda_j=-\Gamma_j+i\omega_j$, define the mode weights $\bm W_j=\bm r_j\bm\ell_j^\dagger\bm S$, with $\sum_j\bm W_j=\bm S$. Then
\begin{equation}
F_{aa}(k,t)=\sum_j(\bm W_j)_{aa}\,e^{\lambda_jt},\qquad
S_{aa}(k,\omega)=\sum_j\frac{2\,\Gamma_j\,{\rm Re}(\bm W_j)_{aa}-2(\omega+\omega_j)\,{\rm Im}(\bm W_j)_{aa}}{\Gamma_j^2+(\omega+\omega_j)^2} .\label{eq:dsf_modes}
\end{equation}
Each mode contributes a Lorentzian of width $\Gamma_j$ centred at $-\omega_j$, plus an antisymmetric dispersive part. This is the exact, all-$k$ version of the Landau--Placzek decomposition.
\end{enumerate}
\end{proposition}
\begin{proof}
(a) and (b) are the standard results for the Ornstein--Uhlenbeck process: $\delta\hat\by(t)=e^{\bm At}\delta\hat\by(0)+\int_0^te^{\bm A(t-s)}\bm B\hat{\bm\xi}(s)\,\dd s$, and the noise is independent of the past.

(c) Fourier transforming \eqref{eq:dsf_langevin} gives $\delta\hat\by(\omega)=\bm G(\omega)\bm B\hat{\bm\xi}(\omega)$, hence $\bm S(k,\omega)=\bm G\bm N\bm G^\dagger$. Inserting $\bm N=-\bm A\bm S-\bm S\bm A^\dagger=(-i\omega-\bm A)\bm S+\bm S(i\omega-\bm A^\dagger)$, the $\pm i\omega$ terms cancel, and $\bm G\bm N\bm G^\dagger=\bm S\bm G^\dagger+\bm G\bm S$.

(d) The relation $(\bm G\bm S)_{aa}=\sum_j(\bm W_j)_{aa}/(\Gamma_j-i(\omega+\omega_j))$ holds, and the diagonal entry of $\bm G\bm S+\bm S\bm G^\dagger$ is $2\,{\rm Re}(\bm G\bm S)_{aa}$.
\end{proof}
The second form in \eqref{eq:dsf_S} is preferable numerically. It needs only the static $\bm S(k)$, which is known in closed form, and one matrix inversion per frequency.

\subsection{The master linear system (model N-N)}\label{sec:dsf_master}
\paragraph{Reference-state coefficients.} All of the following are evaluated at $(\rho_0,c_0,T_0)$. They are the second derivatives of the specific free energy $f(\rho,c,T)$:
\begin{equation}
\begin{gathered}
p_\rho\equiv\Big(\pd p\rho\Big)_{T,c}=2\rho_0f_\rho+\rho_0^2f_{\rho\rho},\quad
p_c\equiv\Big(\pd pc\Big)_{\rho,T}=\rho_0^2f_{\rho c},\quad
p_T\equiv\Big(\pd pT\Big)_{\rho,c}=\rho_0^2f_{\rho T},\\
\mu_\rho=f_{\rho c}=\frac{p_c}{\rho_0^2},\quad
\mu_c=f_{cc},\quad
\mu_T=f_{cT},\quad
\hs_c\equiv\Big(\pd\hs c\Big)_{\rho,T}=-f_{cT}=-\mu_T,\quad
\hat c_v=-T_0f_{TT}.
\end{gathered}\label{eq:dsf_thermo}
\end{equation}
The remaining reference-state data are:
\begin{itemize}[nosep]
\item the gradient coefficients $K_{\rho\rho}$, $K_{\rho c}$ (energetic) and $K_{cc}=\kappa_E-T_0\kappa_S$;
\item the transport coefficients $\eta,\zeta,\hat\lambda,M,M_T$;
\item the kinematic viscosities $\nu_l=(\tfrac43\eta+\zeta)/\rho_0$ and $\nu_t=\eta/\rho_0$.
\end{itemize}
The Hessian \eqref{eq:Hk} enters through its entries
\begin{equation}
h_{\rho\rho}(k)=\frac{p_\rho}{\rho_0}+K_{\rho\rho}k^2,\qquad
h_{\rho c}(k)=\frac{p_c}{\rho_0}+K_{\rho c}k^2,\qquad
h_{cc}(k)=\rho_0\mu_c+K_{cc}k^2 .\label{eq:dsf_h}
\end{equation}

\paragraph{Longitudinal equations.} Linearizing \eqref{eq:ni_mass}--\eqref{eq:ni_c} and the temperature equation \eqref{eq:T_eq}, with the noises placed as in \eqref{eq:noise_placement}, gives for $\delta\by=(\delta\rho,\delta c,\delta T,u)$
\begin{equation}
\begin{aligned}
\partial_t\delta\rho&=-i\rho_0k\,u,\\
\partial_t\delta c&=-\frac{k^2}{\rho_0}\big(M\,\delta\mut+M_T\,\delta T\big)-\frac{ik}{\rho_0}\hat J,\\
\rho_0\hat c_v\,\partial_t\delta T&=-iT_0p_Tk\,u-k^2\big(\hat\lambda\,\delta T+T_0M_T\,\delta\mut\big)+T_0\hs_c\,k^2\big(M\,\delta\mut+M_T\,\delta T\big)-ik\hat J^{q\prime}+iT_0\hs_c\,k\hat J,\\
\partial_tu&=-ik\Big[h_{\rho\rho}\delta\rho+h_{\rho c}\delta c+\frac{p_T}{\rho_0}\delta T\Big]-\nu_lk^2u+\frac{ik}{\rho_0}\hat\tau_\parallel,
\end{aligned}\label{eq:dsf_lin}
\end{equation}
with the linearized generalized potential
\begin{equation}
\delta\mut=\frac{h_{\rho c}\,\delta\rho+h_{cc}\,\delta c}{\rho_0}+\mu_T\,\delta T .\label{eq:dsf_dmut}
\end{equation}
In \eqref{eq:dsf_lin}--\eqref{eq:dsf_dmut}:
\begin{itemize}[nosep]
\item the pressure and the linear Korteweg term combine into $\rho_0(h_{\rho\rho}\delta\rho+h_{\rho c}\delta c)+p_T\delta T$;
\item all products of fluctuations (viscous heating, $\bPi^{\sN}\!:\!\nabla\bv$, $-\bJ\cdot\nabla\mut$, $\mu_E\divv\bJ$, \ldots) drop out;
\item the noise amplitudes $\hat\tau_\parallel=\hat\bk\cdot\hat{\noise\btau}\cdot\hat\bk$, $\hat J=\hat\bk\cdot\hat{\noise\bJ}$ and $\hat J^{q\prime}=\hat\bk\cdot\hat{\noise\bJ}^{q\prime}$ have the covariances \eqref{eq:noise_noniso}, i.e.
\begin{equation}
\bm C\equiv\frac{1}{V}\avg{\begin{pmatrix}\hat\tau_\parallel\\ \hat J\\ \hat J^{q\prime}\end{pmatrix}\begin{pmatrix}\hat\tau_\parallel\\ \hat J\\ \hat J^{q\prime}\end{pmatrix}^\dagger}
=2\kB\begin{pmatrix}T_0(\tfrac43\eta+\zeta)&0&0\\0&MT_0&M_TT_0^2\\0&M_TT_0^2&\hat\lambda T_0^2\end{pmatrix}\delta(t-t').\label{eq:dsf_C}
\end{equation}
\end{itemize}

\paragraph{Explicit matrices.} With $\bm N=\bm B\bm C\bm B^\dagger$ (the $\delta(t-t')$ removed) and the rates
\begin{gather}
a_M=\frac{Mk^2}{\rho_0^2},\qquad a_{cT}=\frac{(M_T-M\hs_c)k^2}{\rho_0},\qquad a_{Tc}=\frac{T_0(M\hs_c-M_T)k^2}{\rho_0^2\hat c_v},\nonumber\\
a_T=\frac{(\hat\lambda+T_0M\hs_c^2-2T_0M_T\hs_c)k^2}{\rho_0\hat c_v},\qquad b_T=\frac{T_0p_T}{\rho_0\hat c_v},\label{eq:dsf_rates}
\end{gather}
the generator is
\begin{equation}
\bm A(k)=\begin{pmatrix}
0&0&0&-i\rho_0k\\
-a_Mh_{\rho c}&-a_Mh_{cc}&-a_{cT}&0\\
a_{Tc}h_{\rho c}&a_{Tc}h_{cc}&-a_T&-ib_Tk\\
-ikh_{\rho\rho}&-ikh_{\rho c}&-ikp_T/\rho_0&-\nu_lk^2
\end{pmatrix},\label{eq:dsf_A}
\end{equation}
\begin{equation}
\bm B(k)=\begin{pmatrix}0&0&0\\0&-\dfrac{ik}{\rho_0}&0\\0&\dfrac{ikT_0\hs_c}{\rho_0\hat c_v}&-\dfrac{ik}{\rho_0\hat c_v}\\ \dfrac{ik}{\rho_0}&0&0\end{pmatrix},\qquad
\bm S(k)=\kB T_0\begin{pmatrix}\begin{pmatrix}h_{\rho\rho}&h_{\rho c}\\h_{\rho c}&h_{cc}\end{pmatrix}^{-1}&\bm0&\bm0\\ \bm0&\dfrac{T_0}{\rho_0\hat c_v}&0\\ \bm0&0&\dfrac1{\rho_0}\end{pmatrix}.\label{eq:dsf_B_S}
\end{equation}
\begin{itemize}[leftmargin=*,nosep]
\item That \eqref{eq:dsf_A} is the linearization of the final equations, and that $\bm A\bm S+\bm S\bm A^\dagger+\bm N=\bm 0$ holds identically for arbitrary $f$-derivatives, gradient coefficients and Onsager coefficients, was verified symbolically (\cref{app:verification}, checks~6 and~9).
\item With \eqref{eq:dsf_S}, these two displays determine every longitudinal correlation function of every model.
\end{itemize}

\paragraph{Transverse velocity.} Each transverse component obeys
\[
\partial_tv_\perp=-\nu_tk^2v_\perp+\frac{ik}{\rho_0}\hat\tau_\perp,\qquad \frac1V\avg{\hat\tau_\perp\hat\tau_\perp^*}=2\kB T_0\eta .
\]
Therefore
\begin{equation}
S_{\perp}(k)=\frac{\kB T_0}{\rho_0},\qquad F_\perp(k,t)=\frac{\kB T_0}{\rho_0}e^{-\nu_tk^2|t|},\qquad S_\perp(k,\omega)=\frac{2\kB T_0\,\eta\,k^2/\rho_0^2}{\omega^2+(\nu_tk^2)^2}\qquad\text{(all models)}.\label{eq:dsf_transverse}
\end{equation}

\subsection{Model I-S: isothermal, symmetric (decoupled)}\label{sec:dsf_IS}
Use the variable $\phi$ and the decoupled model \eqref{eq:sym_iso_mass}--\eqref{eq:sym_iso_phi}. A density-gradient energy $\tfrac12K_{\rho\rho}|\nabla\rho|^2$ may be added if desired; $K_{\rho\phi}=0$. Then $h_{\rho\phi}=0$, the $3\times3$ system splits, and every correlation function has a closed form, exact at all $k$.

\paragraph{Composition.} With $h_{\phi\phi}(k)=\rho_0g''(\phi_0)+\kappa_\phi k^2$, the composition is an Ornstein--Uhlenbeck process:
\begin{equation}
\boxed{
\begin{aligned}
S_{\phi\phi}(k)&=\frac{\kB T_0}{h_{\phi\phi}(k)},\qquad
F_{\phi\phi}(k,t)=S_{\phi\phi}(k)\,e^{-\Gamma_\phi(k)|t|},\\
S_{\phi\phi}(k,\omega)&=\frac{2\kB T_0M_\phi k^2/\rho_0^2}{\omega^2+\Gamma_\phi(k)^2},\qquad
\Gamma_\phi(k)=\frac{M_\phi k^2h_{\phi\phi}(k)}{\rho_0^2}.
\end{aligned}}\label{eq:dsf_IS_phi}
\end{equation}
These hold for a stable homogeneous state, $h_{\phi\phi}>0$. Inside the spinodal ($g''<0$), $\Gamma_\phi<0$ for $k<k_c$ and no stationary state exists. The linear theory then gives the Cahn--Cook growth law $\avg{|\delta\hat\phi_k(t)|^2}/V=S^{\rm init}e^{-2\Gamma_\phi t}+\kB T_0(1-e^{-2\Gamma_\phi t})/h_{\phi\phi}$ instead. This is an additional benchmark for early-time spinodal decomposition.

\paragraph{Density and longitudinal velocity: damped acoustic oscillator.} Let
\[
\omega_s^2(k)=\rho_0h_{\rho\rho}(k)k^2=k^2(c_T^2+\rho_0K_{\rho\rho}k^2),\qquad \Gamma_d(k)=\nu_lk^2,\qquad \mathcal D(k,\omega)=(\omega_s^2-\omega^2)^2+\Gamma_d^2\omega^2 .
\]
Then
\begin{equation}
\boxed{
\begin{aligned}
S_{\rho\rho}(k)&=\frac{\rho_0\kB T_0}{c_T^2+\rho_0K_{\rho\rho}k^2},\qquad
S_{\rho\rho}(k,\omega)=S_{\rho\rho}(k)\,\frac{2\Gamma_d\,\omega_s^2}{\mathcal D},\\
S_{uu}(k)&=\frac{\kB T_0}{\rho_0},\qquad
S_{uu}(k,\omega)=S_{uu}(k)\,\frac{2\Gamma_d\,\omega^2}{\mathcal D},\qquad
S_{\rho u}(k,\omega)=\frac{2\kB T_0\,k\,\Gamma_d\,\omega}{\mathcal D},\\
F_{\rho\rho}(k,t)&=S_{\rho\rho}(k)\,e^{-\Gamma_d|t|/2}\Big[\cos\Omega t+\frac{\Gamma_d}{2\Omega}\sin\Omega|t|\Big],\qquad
\Omega=\sqrt{\omega_s^2-\Gamma_d^2/4},
\end{aligned}}\label{eq:dsf_IS_rho}
\end{equation}
with $\cos\to\cosh$, $\sin\to\sinh$ and $\Omega\to\sqrt{\Gamma_d^2/4-\omega_s^2}$ in the overdamped case. Also $S_{\rho\phi}=S_{u\phi}=0$. For $K_{\rho\rho}\neq0$ the static part reproduces Eq.~(4.7) of \cite{ChaudhriPRE2014}. The spectrum is a Brillouin doublet at $\omega\simeq\pm c_Tk$ of half-width $\Gamma_d/2$, with \emph{no} central (Rayleigh) peak. The isothermal one-component fluid has none, and in the decoupled mixture composition fluctuations do not modulate the density.

\paragraph{Off-critical, non-decoupled symmetric mixtures.} If the symmetric mixture is not decoupled (e.g.\ the symmetric van der Waals mixture \eqref{eq:vdw_landau}) and $\phi_0\neq0$, then $p_\phi$, $\mu_\rho$ and $K_{\rho\phi}$ are non-zero but odd in $\phi_0$. The I-N results below apply, and every cross effect scales with $\phi_0$.

\subsection{Model I-N: isothermal, non-symmetric}\label{sec:dsf_IN}
Delete the $\delta T$ row and column of \eqref{eq:dsf_A}--\eqref{eq:dsf_B_S} and set $M_T=0$. In the variables $(\delta\rho,\delta c,u)$:
\begin{equation}
\bm A_{\rm I}=\begin{pmatrix}0&0&-i\rho_0k\\ -\dfrac{Mk^2h_{\rho c}}{\rho_0^2}&-\dfrac{Mk^2h_{cc}}{\rho_0^2}&0\\ -ikh_{\rho\rho}&-ikh_{\rho c}&-\nu_lk^2\end{pmatrix},\qquad
\bm N_{\rm I}=2\kB T_0k^2\begin{pmatrix}0&0&0\\0&M/\rho_0^2&0\\0&0&\rho_0\nu_l/\rho_0^2\end{pmatrix},\label{eq:dsf_A_IN}
\end{equation}
\[
\bm S_{\rm I}(k)=\kB T_0\begin{pmatrix}h_{\rho\rho}&h_{\rho c}&0\\h_{\rho c}&h_{cc}&0\\0&0&\rho_0\end{pmatrix}^{-1}.
\]
The exact correlation functions follow from \eqref{eq:dsf_F}--\eqref{eq:dsf_S}. The characteristic polynomial is
\[
\det(\lambda-\bm A_{\rm I})=\lambda^3+a_2\lambda^2+a_1\lambda+a_0,
\]
with coefficients
\[
a_2=\nu_lk^2+\frac{Mk^2h_{cc}}{\rho_0^2},\qquad
a_1=\rho_0k^2h_{\rho\rho}+\frac{\nu_lMk^4h_{cc}}{\rho_0^2},\qquad
a_0=\frac{Mk^4}{\rho_0}\det\begin{pmatrix}h_{\rho\rho}&h_{\rho c}\\h_{\rho c}&h_{cc}\end{pmatrix}.
\]
\paragraph{Hydrodynamic modes ($k\to0$).} There are three modes. Two are sound modes $\lambda=\pm ic_Tk-\Gamma_sk^2$ and one is a diffusive mode $\lambda=-Dk^2$:
\begin{equation}
c_T^2=\Big(\pd p\rho\Big)_{T,c},\qquad
\Gamma_s=\frac12\Big[\nu_l+\frac{M\,p_c^2}{\rho_0^3\,p_\rho}\Big],\qquad
D=\frac{M}{\rho_0}\Big(\pd\mu c\Big)_{T,p}.\label{eq:dsf_IN_modes}
\end{equation}
Sound propagates at \emph{frozen} composition. Barodiffusion adds the attenuation $Mp_c^2/(2\rho_0^3p_\rho)$, which vanishes for density-matched components.

\paragraph{Composition spectrum.} Eliminating the fast (acoustic) variables at fixed generalized pressure $\rho_0(h_{\rho\rho}\delta\rho+h_{\rho c}\delta c)=0$ gives
\begin{equation}
S_{cc}(k,\omega)\simeq\frac{2\kB T_0Mk^2/\rho_0^2}{\omega^2+\Gamma_c(k)^2},\qquad
\Gamma_c(k)=\frac{Mk^2}{\rho_0^2}\,\frac{h_{\rho\rho}h_{cc}-h_{\rho c}^2}{h_{\rho\rho}},\qquad
\Gamma_c\to Dk^2\ (k\to0).\label{eq:dsf_IN_cc}
\end{equation}
This is the I-S form \eqref{eq:dsf_IS_phi} with $h_{\phi\phi}$ replaced by the ``composition stiffness at constant pressure'' $\det h/h_{\rho\rho}$. It integrates to the exact $S_{cc}(k)=\kB T_0h_{\rho\rho}/\det h$. Corrections are $O(\Gamma_c/(c_Tk))$; numerically they are $10^{-5}$--$10^{-4}$ for the parameters of \cref{fig:dsf}.

\paragraph{Density spectrum: Landau--Placzek decomposition.} At small $k$,
\begin{equation}
\begin{aligned}
S_{\rho\rho}(k,\omega)\simeq{}&I_c\,\frac{2Dk^2}{\omega^2+(Dk^2)^2}+I_B\Big[\frac{\Gamma_sk^2}{(\omega-c_Tk)^2+(\Gamma_sk^2)^2}+\frac{\Gamma_sk^2}{(\omega+c_Tk)^2+(\Gamma_sk^2)^2}\Big]\\
&+(\text{antisymmetric }O(k)\text{ term}),
\end{aligned}\label{eq:dsf_IN_LP}
\end{equation}
with intensities (verified numerically, \cref{app:verification}, check~8)
\begin{equation}
I_B=\frac{\rho_0\kB T_0}{(\partial p/\partial\rho)_{T,c}},\qquad
I_c=\rho_0\kB T_0\Big[\frac1{(\partial p/\partial\rho)_{T,\mu}}-\frac1{(\partial p/\partial\rho)_{T,c}}\Big]=\Big(\pd\rho c\Big)_{T,p}^2S_{cc}(0).\label{eq:dsf_IN_I}
\end{equation}
In a non-symmetric isothermal mixture the density spectrum therefore acquires a \emph{central} peak. It is the composition fluctuation seen through the density contrast $(\partial\rho/\partial c)_{T,p}$, and is absent in the I-S model (\cref{fig:dsf}a). Its width is the mass diffusivity, so the density field alone measures $D$. The exact spectrum, including the antisymmetric term, is given by \eqref{eq:dsf_S} or \eqref{eq:dsf_modes}.

\subsection{Model N-S: non-isothermal, symmetric (decoupled)}\label{sec:dsf_NS}
\paragraph{Critical composition $\phi_0=0$.} Here $\hs_\phi=-g_{\phi T}=0$, $\mu_T=0$, $M_T=0$ (odd), $p_\phi=0$ and $K_{\rho\phi}=0$. The composition decouples completely. Its correlations are \eqref{eq:dsf_IS_phi}, with the same expressions at temperature $T_0$. The $(\delta\rho,\delta T,u)$ block is the linearized Navier--Stokes--Fourier system of a one-component fluid. Its exact correlations follow from the $3\times3$ sub-block of \eqref{eq:dsf_A}. Its static correlations are
\[
S_{\rho\rho}(k)=\frac{\rho_0\kB T_0}{c_T^2+\rho_0K_{\rho\rho}k^2},\qquad S_{TT}=\frac{\kB T_0^2}{\rho_0\hat c_v},\qquad S_{\rho T}=0 .
\]
The modes are
\begin{equation}
c_s^2=\gamma\,c_T^2,\qquad \gamma=\frac{\hat c_p}{\hat c_v},\qquad \hat c_p=\hat c_v+\frac{T_0p_T^2}{\rho_0^2p_\rho},\qquad
D_T=\frac{\hat\lambda}{\rho_0\hat c_p},\qquad \Gamma=\frac12\big[\nu_l+(\gamma-1)D_T\big].\label{eq:dsf_NS_modes}
\end{equation}
These give the Rayleigh--Brillouin spectrum in Landau--Placzek form \cite{HansenMcDonald,BoonYip}:
\begin{equation}
\boxed{
\begin{aligned}
\frac{S_{\rho\rho}(k,\omega)}{S_{\rho\rho}(k)}&\simeq\frac{\gamma-1}{\gamma}\frac{2D_Tk^2}{\omega^2+(D_Tk^2)^2}
+\frac1\gamma\Big[\frac{\Gamma k^2}{(\omega+c_sk)^2+(\Gamma k^2)^2}+\frac{\Gamma k^2}{(\omega-c_sk)^2+(\Gamma k^2)^2}\Big]\\
&\quad+\frac{\Gamma+(\gamma-1)D_T}{\gamma}\,\frac{k}{c_s}\Big[\frac{\omega+c_sk}{(\omega+c_sk)^2+(\Gamma k^2)^2}-\frac{\omega-c_sk}{(\omega-c_sk)^2+(\Gamma k^2)^2}\Big],\\
\frac{F_{\rho\rho}(k,t)}{S_{\rho\rho}(k)}&\simeq\frac{\gamma-1}{\gamma}e^{-D_Tk^2|t|}+\frac1\gamma e^{-\Gamma k^2|t|}\Big[\cos(c_skt)+\frac{\Gamma+(\gamma-1)D_T}{c_s}\,k\,\sin(c_sk|t|)\Big].
\end{aligned}}\label{eq:dsf_RB}
\end{equation}
At $k=0.02$ (in the units of \cref{fig:dsf}) the relative deviation from the exact result is $10^{-7}$--$10^{-3}$, largest at the Brillouin peak (\cref{app:verification}, check~8). For a benchmark at finite $k$, use the exact \eqref{eq:dsf_S}. The Landau--Placzek ratio, i.e.\ central intensity over total Brillouin intensity, is $\gamma-1$. It is a direct check of the non-isothermal thermodynamics, i.e.\ of $p_T$ and $\hat c_v$, in the code.

\paragraph{Off-critical $\phi_0\neq0$.} Now $\hs_\phi=-a_0\phi_0$ and $\mu_T=a_0\phi_0$ for the Landau model, so composition couples to temperature through the heat of demixing and thermodynamic thermal diffusion. It still does not couple to density ($p_\phi=0$). The central peak then consists of two coupled diffusive modes. Their rates are given by \eqref{eq:dsf_NN_diff} with $p_c=0$.

\subsection{Model N-N: non-isothermal, non-symmetric}\label{sec:dsf_NN}
The full system \eqref{eq:dsf_A}--\eqref{eq:dsf_B_S} has four longitudinal modes. As $k\to0$ (agreement with the exact eigenvalues to better than $10^{-6}$ at $k=10^{-3}$, \cref{app:verification}, check~8):
\begin{itemize}[leftmargin=*]
\item \textbf{Sound}, $\lambda=\pm ic_sk-\Gamma_sk^2$, at fixed entropy and composition:
\begin{gather}
c_s^2=\Big(\pd p\rho\Big)_{s,c}=p_\rho+\frac{T_0p_T^2}{\rho_0^2\hat c_v},\label{eq:dsf_NN_cs}\\
\Gamma_s=\frac12\Big[\nu_l+\frac{\rho_0}{c_s^2}\Big(\frac{\hat\lambda\,\theta^2}{T_0}+2M_T\,\theta\,m+M\,m^2\Big)\Big],\nonumber\\
\theta=\Big(\pd T\rho\Big)_{s,c}=\frac{T_0p_T}{\rho_0^2\hat c_v},\qquad m=\Big(\pd\mu\rho\Big)_{s,c}=\mu_\rho+\mu_T\theta .\label{eq:dsf_NN_Gs}
\end{gather}
The attenuation is half the rate of entropy production \eqref{eq:Sigma} of the sound wave per unit acoustic energy. It contains viscous, thermal, cross (Soret) and barodiffusive dissipation. It reduces to \eqref{eq:dsf_IN_modes} when $p_T=0$, and to $\Gamma$ of \eqref{eq:dsf_NS_modes} when $M=M_T=0$.
\item \textbf{Two diffusive modes}, $\lambda=-d_{1,2}k^2$, of heat and mass at constant pressure. The rates $d_{1,2}$ are the eigenvalues of $\bm{\mathcal M}^{-1}\bm{\mathcal K}$, obtained from $\rho_0T_0\partial_t\delta\hs=-\divv\bJqp$ and $\rho_0\partial_t\delta c=-\divv\bJ$ at $\delta p=0$:
\begin{equation}
\bm{\mathcal M}=\begin{pmatrix}\rho_0&0\\ \rho_0T_0\hs_{c,p}&\rho_0\hat c_p\end{pmatrix},\qquad
\bm{\mathcal K}=\begin{pmatrix}M\mu_{c,p}&M_T-M\hs_{c,p}\\ T_0M_T\mu_{c,p}&\hat\lambda-T_0M_T\hs_{c,p}\end{pmatrix},\label{eq:dsf_NN_diff}
\end{equation}
with
\[
\mu_{c,p}=\Big(\pd\mu c\Big)_{T,p}=\mu_c-\frac{p_c^2}{\rho_0^2p_\rho},\qquad
\hs_{c,p}=\Big(\pd\hs c\Big)_{T,p}=\hs_c+\frac{p_Tp_c}{\rho_0^2p_\rho}=-\Big(\pd\mu T\Big)_{c,p},\qquad
\hat c_p=\hat c_v+\frac{T_0p_T^2}{\rho_0^2p_\rho}.
\]
For $M_T=0$ and $\hs_{c,p}=0$ the matrix is triangular, with $d_1=D$ and $d_2=D_T$. In general the two Rayleigh components are mixtures of heat and mass diffusion, the classical result for binary liquids \cite{MountainDeutch1969,deZarateSengers2006}.
\item \textbf{Intensities in} $S_{\rho\rho}$: the Brillouin doublet carries $I_B=\rho_0\kB T_0/c_s^2$; the two central modes together carry $I_c=\rho_0\kB T_0\big[1/(\partial p/\partial\rho)_{T,\mu}-1/c_s^2\big]$.
\item \textbf{Static correlations}: \eqref{eq:dsf_B_S}, i.e.\ \eqref{eq:S_noniso}.
\end{itemize}

\subsection{Correlations of the conserved variables}\label{sec:dsf_conserved}
A conservative code advances $\bm U=(\rho,\rho c,\rho E,\rho\bv)$. At linear order $\delta\bm U=\bm P\,\delta\by$, with
\begin{equation}
\bm P=\begin{pmatrix}1&0&0&0\\ c_0&\rho_0&0&0\\ \dfrac{h_0-T_0p_T}{\rho_0}&\rho_0(\mu_0+T_0\hs_c)&\rho_0\hat c_v&0\\ 0&0&0&\rho_0\end{pmatrix},\qquad
\bm S_U(k,\omega)=\bm P\,\bm S(k,\omega)\,\bm P^\dagger ,\label{eq:dsf_P}
\end{equation}
and the same for $\bm S_U(k)$ and $\bm F_U(k,t)$. Here $h_0=\varepsilon_0+p_0$ and $\mu_0$ is the reference exchange potential. The kinetic and gradient energies are quadratic and do not contribute at this order. For the isothermal models, delete the third row and column.

\subsection{Using the results as benchmarks for a finite-volume code}\label{sec:dsf_benchmark}
\paragraph{Discrete wavenumbers.} After spatial discretization, each continuum Fourier symbol in \eqref{eq:dsf_A}--\eqref{eq:dsf_B_S} is replaced by the symbol of the discrete operator actually used. There are three of them:
\begin{itemize}[leftmargin=*,nosep]
\item $k\to k_h(k)$ in the reversible terms: continuity, pressure gradient and compression work. For a face-based second-order scheme, $k_h=\frac2{\Delta x}\sin\frac{k\Delta x}2$. For the fourth-order face interpolation of \cite{ChaudhriPRE2014},
\[
k_h=\frac2{\Delta x}\sin\frac{k\Delta x}2\Big[\frac76\cos\frac{k\Delta x}2-\frac16\cos\frac{3k\Delta x}2\Big].
\]
\item $k^2\to\tilde k^2=\sum_\alpha\frac4{\Delta x_\alpha^2}\sin^2\frac{k_\alpha\Delta x_\alpha}2$ in all diffusive terms and in the gradient energies inside $h_{ab}$. This is Eq.~(4.8) of \cite{ChaudhriPRE2014}.
\item $k\to k_n(k)$ for the divergence of the stochastic fluxes. For face-centred noise, $k_n^2=\tilde k^2$.
\end{itemize}
If $k_h$ is real and the same symbol is used in the divergence (continuity) and in the gradient (momentum), and if $k_n^2=\tilde k^2$, then the discrete generator keeps the structure $(\bm L-\bm D)\bm H$ of \cref{app:linear_fdt}. This is the discrete FDT. The spatially discrete static structure factor is then \emph{exactly} $\kB T_0\tilde{\bm H}^{-1}$, and all formulas of this section hold with the symbols substituted. If a scheme violates one of these conditions, the stationary covariance is still the solution of $\bm A\bm S+\bm S\bm A^\dagger+\bm N=0$ with the discrete $\bm A,\bm N$, which the reference implementation computes. The deviation from $\kB T_0\tilde{\bm H}^{-1}$ then measures the discrete FDT error.

\paragraph{Time discretization.} The formulas are for continuous time. Temporal errors scale with $|\lambda_j(k)|\Delta t$. Following \cite{ChaudhriPRE2014,Donev2010}, use $\Delta t$ at about $10\%$ of the stability limit. The fastest rates at the grid scale are the sound frequency $c_sk_{\max}$ and the composition rate $\Gamma_c(k_{\max})\propto Mk_{\max}^4$. The latter is the stiff Cahn--Hilliard term and usually limits $\Delta t$.

\paragraph{Estimators.} From cell data $\delta\by_j$:
\begin{itemize}[nosep]
\item form $\delta\hat\by(\bk,t)=\Delta V\sum_je^{-i\bk\cdot\br_j}\delta\by_j$ with an FFT;
\item estimate $\bm S(\bk)$ by $V^{-1}\overline{\delta\hat\by\,\delta\hat\by^\dagger}$;
\item estimate $\bm F(\bk,t)$ by the time-lag average $V^{-1}\overline{\delta\hat\by(t_0+t)\,\delta\hat\by(t_0)^\dagger}$.
\end{itemize}
Average over $t_0$ and over the equivalent wavevectors of a shell. The intermediate scattering function is the recommended dynamic observable, because it avoids windowing and aliasing artefacts of a direct periodogram estimate of $\bm S(k,\omega)$. If $\bm S(k,\omega)$ is wanted, Fourier transform the estimated $\bm F$ with a window, or use Welch averaging. The relative statistical error of an $S(\bk)$ estimate from $N_{\rm ind}$ independent samples is about $1/\sqrt{N_{\rm ind}}$. Samples are independent over times of order $1/\min_j\Gamma_j(k)$, which for small $k$ is the slow diffusive time $1/(Dk^2)$.

\paragraph{Suggested test sequence.} Each test isolates one part of the implementation:
\begin{center}\small
\begin{tabular}{p{1.3cm}p{4.7cm}p{3.5cm}p{4.9cm}}
\toprule
model & observable & reference & tests\\
\midrule
all & $S_\perp(k)$, $F_\perp(k,t)$ & \eqref{eq:dsf_transverse} & viscous stress and its noise\\
I-S & $S_{\phi\phi}(k)$, $F_{\phi\phi}(k,t)$ & \eqref{eq:dsf_IS_phi} & CH flux, specific $\mut$, noise $2\kB TM$\\
I-S & spinodal growth of $\avg{|\delta\hat\phi_k|^2}$ & below \eqref{eq:dsf_IS_phi} & Cahn--Cook linear instability\\
I-S & $S_{\rho\rho}$, $S_{uu}$, $S_{\rho u}$, $F_{\rho\rho}$ & \eqref{eq:dsf_IS_rho} & acoustics, isothermal EOS\\
I-N & central peak in $S_{\rho\rho}$, $S_{\rho c}(k)$ & \eqref{eq:dsf_IN_I}, \eqref{eq:S0_nonsym} & barodiffusion, $p(\rho,c)$, $K_{\rho c}$\\
N-S & Rayleigh--Brillouin $S_{\rho\rho}$, $S_{TT}(k)$ & \eqref{eq:dsf_RB} & energy equation, $p_T$, $\hat c_v$, $\hat\lambda$\\
N-N & full $\bm F(k,t)$ incl.\ cross terms & \eqref{eq:dsf_F}, \eqref{eq:dsf_A} & Soret/Dufour, latent heat, interstitial fluxes\\
\bottomrule
\end{tabular}
\end{center}
The last test is the most stringent. The signs and magnitudes of the cross-correlations $F_{cT}(k,t)$ and $F_{\rho c}(k,t)$ at $t>0$ are fixed by $\hs_c$, $M_T$ and $p_c$ and vanish at $t=0$ (for $F_{cT}$). Sign errors in the coupling terms therefore show up even when every static structure factor is correct.

\begin{figure}[H]
\centering
\includegraphics[width=\textwidth]{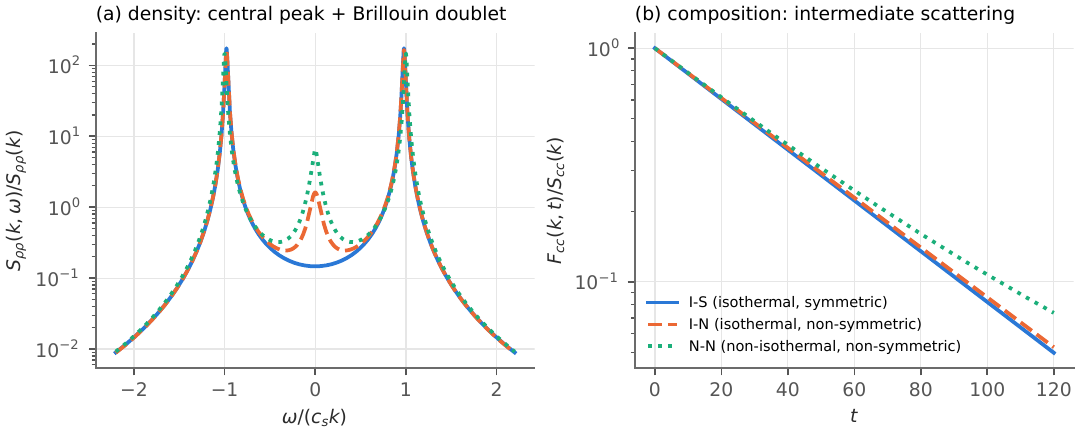}
\caption{Example spectra from the exact formulas \eqref{eq:dsf_F}--\eqref{eq:dsf_S}, at $k=0.25$, in units with $\rho_0=T_0=\kB=1$. Parameters: $f_\rho=1$, $f_{\rho\rho}=0.5$, $f_{\rho c}=0.3$ (0 for I-S), $f_{cc}=2$, $f_{\rho T}=0.4$, $f_{cT}=0.2$, $f_{TT}=-1.5$; $K_{\rho c}=0.005$, $K_{cc}=0.02$; $\eta=0.1$, $\zeta=0.05$, $M=0.2$, $\hat\lambda=0.3$, $M_T=0.05$ (N-N only).
(a) Density spectrum normalized by $S_{\rho\rho}(k)$. All three models show the Brillouin doublet at $\omega\simeq\pm c_sk$. The isothermal non-symmetric model adds the barodiffusive central peak \eqref{eq:dsf_IN_I}. The non-isothermal model adds heat-mode and mass-mode central components.
(b) Normalized composition intermediate scattering function: a single exponential for I-S, very nearly one for I-N, and visibly a sum of two exponentials (heat and mass modes, \eqref{eq:dsf_NN_diff}) for N-N.}\label{fig:dsf}
\end{figure}

\paragraph{Reference implementation.} The module \texttt{dsf\_\allowbreak reference.py}, distributed with the paper as an ancillary file, implements the material of this section, for the continuum and for any discrete symbols:
\begin{itemize}[nosep]
\item $\bm A(k)$, $\bm N(k)$, $\bm H(k)$ of \eqref{eq:dsf_A}--\eqref{eq:dsf_B_S} (or their isothermal reductions);
\item $\bm S(k)$ from the Lyapunov equation;
\item $\bm S(k,\omega)$ from \eqref{eq:dsf_S} and $\bm F(k,t)$ from \eqref{eq:dsf_F};
\item the transverse correlations;
\item the transformation \eqref{eq:dsf_P};
\item the hydrodynamic-mode formulas.
\end{itemize}
\Cref{fig:dsf} was produced with it.

\section{Conclusions and outlook}\label{sec:conclusions}
We have derived, from first principles, a thermodynamically consistent and fully compressible fluctuating hydrodynamic model of a Cahn--Hilliard binary fluid mixture. It comes in four versions: isothermal and non-isothermal, each for symmetric and non-symmetric mixtures. The main conclusions are the following.

\begin{enumerate}[leftmargin=*]
\item \textbf{The variables are not a matter of taste.} The momentum equation must use the total mass density $\rho=\rho_A+\rho_B$ and the barycentric velocity. The Cahn--Hilliard variable is the mass fraction $c=\rho_A/\rho$, and the conserved quantity is the partial density $\rho c$. The diffusion flux is driven by the \emph{specific} exchange potential $\mut=\rho^{-1}\delta F/\delta c$. Using the volumetric derivative $\delta F/\delta c$ instead, a common choice inherited from incompressible models, destroys the sign of the dissipation and violates the fluctuation--dissipation theorem in a compressible fluid.
\item \textbf{One identity gives all capillary stresses.} The generalized Korteweg identity (\cref{lem:korteweg}) produces the reversible stress for any isotropic square-gradient energy in density and composition. It is applied to the free energy in the isothermal model and separately to the gradient energy and the gradient entropy in the non-isothermal model.
\item \textbf{Noise and dissipation come from the same structure.} In the Jeli\'c--\"Ottinger construction the deterministic fluxes and the stochastic fluxes are generated by the same operator and the same correlation matrix. The FDT therefore holds by construction, and every noise amplitude ($2\kB TM$, $2\kB T^2M_T$, $2\kB T^2\hat\lambda$, and the viscous ones) is read off from a transport coefficient. Correlated heat and mass noises accompany the Soret and Dufour effects.
\item \textbf{The non-isothermal model has genuinely new structure.} The pressure tensor of the Poisson operator is fixed by the degeneracy condition. It produces a thermal capillary stress whenever the interfacial free energy has an entropic part. The total energy, including the gradient energy, is conserved locally with reversible and irreversible interstitial-working fluxes. The isothermal limit reproduces the isothermal model exactly.
\item \textbf{Symmetric and non-symmetric mixtures differ only in their constitutive input.} Barodiffusion, composition--pressure coupling, density--composition gradient coupling, heat of mixing and thermodiffusion are all contained in the general equations. In a symmetric mixture they are odd in the order parameter and vanish at the critical composition.
\item \textbf{Analytic benchmarks.} The static, intermediate and dynamic structure factors of the linearized equations are available in closed form for all four models, together with the hydrodynamic modes and Landau--Placzek intensities. A reference implementation accompanies the paper.
\item \textbf{Pure thermodynamics agrees, up to a gauge.} The free energy and the entropy production alone reproduce all the isothermal models uniquely. In the non-isothermal case they leave a gauge freedom of terms bilinear in $\nabla T$ and the density or composition gradients, which GENERIC removes. All admissible versions share the same linearization (\cref{app:thermo_route}).
\end{enumerate}

\paragraph{Outlook.} The equations are written in the conservative flux form of \cite{ChaudhriPRE2014}, which fits finite-volume, method-of-lines discretizations with staggered stochastic fluxes \cite{Donev2010}. The natural next step is such an implementation, tested first against the structure factors of \cref{sec:dsf} and then against nonlinear benchmarks. Several physical questions can then be addressed with the same model:
\begin{itemize}[nosep]
\item the role of fluctuations in the nucleation and coarsening of mixtures with density contrast;
\item latent-heat release and the piston effect during deep quenches;
\item thermocapillary migration of droplets;
\item the crossover between liquid--liquid and liquid--vapour phase separation in the binary van der Waals mixture.
\end{itemize}
Extensions to multicomponent mixtures, to chemically reacting mixtures and to the low-Mach and quasi-incompressible limits derived in \cref{app:C-gibbs} follow the same construction.

\section*{Acknowledgements}\label{sec:acknowledgements}
\addcontentsline{toc}{section}{Acknowledgements}
Assistance on some details in the paper were provided by Claude Opus 5.5. Opus used the author's unpublished notes on using GENERIC to derive a thermodynamically consistent framework for a fully compressible fluctuating hydrodynamics model for symmetric mixtures and extended it to non-symmetric mixtures including deriving the noise terms that go with it. The calculation of structure factors was extended to these models and Opus provided the python scripts to test them. The correctness of the models and the equations lie solely with the author.

\appendix
\section{Proofs and verification}\label{app:proofs}

\subsection{Jacobi identity of the non-isothermal Poisson operator}\label{app:jacobi}
Introduce new variables $\bx'=(\rho,\bg,\sigma,c)$ with $\sigma=s(\rho,c,\varepsilon)+\sN(\rho,c,\nabla c)$. The Jacobi identity is invariant under (invertible) changes of variables. It therefore suffices to show that the transformed operator $\Lop'=\mathcal J\Lop\mathcal J^*$, with $\mathcal J=\partial\bx'/\partial\bx$, is the canonical Lie--Poisson operator in which $\sigma$ is an advected density:
\[
(\Lop'\cdot\bm b')_{\bg}=-\rho\nabla b'_\rho-\sigma\nabla b'_\sigma+b'_c\nabla c+(\text{the }\bg\bg\text{ part}),\qquad (\Lop'\cdot\bm b')_\sigma=-\divv(\sigma\bm b'_g).
\]
The $\rho$, $\bg$ and $c$ rows of $\Lop$ act only on the $\bg$-component; the $\varepsilon$-row also acts only on $\bm a_g$; and $\Lop_{\rho\rho}=\Lop_{\rho\varepsilon}=\Lop_{\varepsilon\varepsilon}=\dots=0$. Hence every entry of the $\sigma$-row of $\Lop'$ vanishes except $\Lop'_{\sigma\bg}$. It equals the reversible change of $\sigma$ under a velocity field $\bm b_g$, which is $-\divv(\sigma\bm b_g)$ by the computation leading to \eqref{eq:S_local_rev} (and \cref{app:verification}, item~4).

For the $\bg$-row, the pull-back of $\bm b'$ is
\[
b_\rho=b'_\rho+\big(\partial_\rho s+\partial_\rho\sN\big)b'_\sigma,\qquad
b_\varepsilon=b'_\sigma/T,\qquad
b_c=b'_c+\big(\partial_cs+\partial_c\sN\big)b'_\sigma-\divv\big(\partial_{\nabla c}\sN\,b'_\sigma\big).
\]
Insert this into the $\bg$-row of \eqref{eq:L_noniso_action} and collect terms.
\begin{itemize}[leftmargin=*,nosep]
\item Terms with $b'_\sigma$ undifferentiated: their coefficient is exactly the left side of the degeneracy condition, which vanishes by \cref{prop:PiS}.
\item Terms with $\nabla b'_\sigma$: using $\bPi^{(S)}/T=(p/T)\Id-[\rho\partial_\rho\sN-\sN]\Id-\nabla c\otimes\partial_{\nabla c}\sN$, they are
\[
\nabla b'_\sigma\Big[-\rho\partial_\rho s-\rho\partial_\rho\sN-\frac\varepsilon T-\frac pT+\rho\partial_\rho\sN-\sN\Big]
+\Big[\nabla c\,(\partial_{\nabla c}\sN\cdot\nabla b'_\sigma)-\nabla c\,(\partial_{\nabla c}\sN\cdot\nabla b'_\sigma)\Big]=-(s+\sN)\nabla b'_\sigma ,
\]
since $-\rho\partial_\rho s-\varepsilon/T-p/T=(\rho\mub-\varepsilon-p)/T=-s$.
\end{itemize}
Hence $(\Lop'\cdot\bm b')_{\bg}=-\rho\nabla b'_\rho-\sigma\nabla b'_\sigma+b'_c\nabla c+(\bg\bg\text{ part})$. This is the canonical Lie--Poisson operator for $(\rho,\bg,\sigma,c)$ \cite{MorrisonGreene1980,HolmMarsdenRatiuWeinstein1985,Ottinger2005}, which satisfies Jacobi. \hfill$\square$

\emph{Remark.} If $\sN$ depended on $\nabla\rho$, the pull-back $b_\rho$ would contain $-\divv(\partial_{\nabla\rho}\sN\,b'_\sigma)$. The $\bg$-row would then acquire second derivatives of $b'_\sigma$ that cannot cancel, and the transformed operator would no longer be canonical. This is the reason for the restriction in \cref{sec:gradient}. (In the classical thermodynamic route of \cref{app:C-routeA}, which does not require the Jacobi identity, a $\nabla\rho$-dependence of $\sN$ is admissible.)

\subsection{Linear FDT for any linearized GENERIC system}\label{app:linear_fdt}
Let $\bx_0$ be an equilibrium state at rest. Define the availability $\mathcal A=E-T_0S-\sum_I\lambda_II$. It is stationary at $\bx_0$, $\mathcal A_{\bx}(\bx_0)=0$, for suitable multipliers $\lambda_I$ of the conserved quantities $I$ (masses), where $T_0$ is the equilibrium temperature. Using $\Lop\cdot S_{\bx}=0$, $\Mop\cdot E_{\bx}=0$, $\Lop\cdot I_{\bx}=\Mop\cdot I_{\bx}=0$:
\[
\Lop\cdot E_{\bx}+\Mop\cdot S_{\bx}=\Big(\Lop-\frac{\Mop}{T_0}\Big)\cdot\mathcal A_{\bx} .
\]
Linearizing about $\bx_0$, all terms involving derivatives of $\Lop,\Mop$ multiply $\mathcal A_{\bx}(\bx_0)=0$. Therefore
\[
\dd\,\delta\bx=A\,\delta\bx\,\dd t+\Bop_0\dd\bW,\qquad A=\Big(\Lop_0-\frac{\Mop_0}{T_0}\Big)H,\qquad H=\mathcal A_{\bx\bx}(\bx_0),\qquad \Bop_0\Bop_0^\dagger=2\kB\Mop_0 .
\]
With $\Sigma=\kB T_0H^{-1}$:
\[
A\Sigma+\Sigma A^\dagger=\kB T_0\Big[\Lop_0+\Lop_0^\dagger-\frac2{T_0}\Mop_0\Big]=-2\kB\Mop_0=-\Bop_0\Bop_0^\dagger .
\]
This holds because $\Lop_0^\dagger=-\Lop_0$ and $\Mop_0$ and $H$ are Hermitian. Hence $\Sigma=\kB T_0H^{-1}$ is the stationary covariance, which is the Einstein formula. If the conserved densities are among the variables, the $I$ are linear and do not contribute to $H$. In the isothermal case replace $E-T_0S$ by $F$ and $\Mop/T_0$ by $\Dop$. The It\^o drift does not contribute at this order. \hfill$\square$

\subsection{Symbolic verification}\label{app:verification}
All identities used in this paper were checked with SymPy. The scripts are distributed as ancillary files with the paper. Unless stated otherwise:
\begin{itemize}[nosep]
\item all fields are arbitrary functions of $(x,y)$ in two dimensions;
\item all coefficient functions ($K^E_{\rho\rho}$, $K^E_{\rho c}$, $\kappa_E$, $\kappa_S$, $K_{ab}$, $\psi$, \dots) are arbitrary functions of their arguments;
\item the local entropy $s(\rho,c,\varepsilon)$ and the specific free energy $f(\rho,c,T)$ are arbitrary;
\item every residual simplifies to zero, except where a failure is the point of the check.
\end{itemize}
The checks are numbered as the scripts.
\begin{enumerate}[leftmargin=*,label=\textbf{Check \arabic*.},ref=\arabic*]
\item \texttt{check1\_\allowbreak korteweg\_\allowbreak energy.py}: reversible structure of the GENERIC model.
\begin{itemize}[nosep]
\item The Korteweg identity \eqref{eq:korteweg_identity} for $w=\eN$, and the symmetry of $\bPi^w$.
\item The degeneracy $\Lop\cdot\delta S/\delta\bx=0$ with $\bPi^{(S)}=p\Id-T\bPi^{\sN}$, using $T=1/s_\varepsilon$, $\mub=-Ts_\rho$, $p=Ts-\varepsilon+\rho\mub$.
\item Reversible energy conservation \eqref{eq:E_local_rev}, including the interstitial flux $\rho(\divv\bv)\partial\eN/\partial\nabla\rho$, for arbitrary fields $p$ and $T$.
\item Reversible entropy conservation \eqref{eq:S_local_rev} for $\sigma=s+\sN$.
\end{itemize}
\item \texttt{check2\_\allowbreak irreversible.py}: irreversible energy conservation \eqref{eq:E_local_full} with arbitrary $\btau$, $\bJq$ and $\bJ$, and the entropy balance \eqref{eq:entropy_flux}--\eqref{eq:Sigma} in the form $\Sigma=\btau:\nabla\bv/T+\bJq\cdot\nabla(1/T)-\bJ\cdot\nabla(\mut/T)$.
\item \texttt{check3\_\allowbreak linear\_\allowbreak fdt\_\allowbreak noniso.py}: linear FDT, $\kB T_0(HA+A^\dagger H)+H\Bop\Bop^\dagger H=0$, for the non-isothermal model \eqref{eq:lin_noniso}. The entropy Hessian $s_{ab}$, the enthalpy $h_0$, $K^E_{\rho\rho}$, $K^E_{\rho c}$, $\kappa_E$, $\kappa_S$, $\eta$, $\zeta$ and the Onsager coefficients $(L_{qq},L_{q1},L_{11})$, including Soret and Dufour, are all symbolic.
\item \texttt{check4\_\allowbreak linear\_\allowbreak fdt\_\allowbreak iso.py}: the same for the isothermal non-symmetric model \eqref{eq:lin_nonsym_iso} with arbitrary $\psi$-Hessian and $K_{ab}$. Also: with $\bJ=-M\nabla(\rho\mut)$ and noise $2\kB T_0M$, the volumetric-potential variant leaves residuals proportional to $(1-\rho_0)$, i.e.\ it violates the FDT.
\item \texttt{check5\_\allowbreak thermo.py}: thermodynamics and noise construction.
\begin{itemize}[nosep]
\item The binary van der Waals mixture: the pressure \eqref{eq:vdw_p}, the chemical potentials \eqref{eq:vdw_mu}, the gradient coefficients \eqref{eq:vdw_K}, the ideal-mixing derivative \eqref{eq:ideal_dmudc} and the Landau expansion \eqref{eq:vdw_landau}.
\item The noise tensor \eqref{eq:noise_construction_iso}: $4ab+6b^2-2\kB T(\zeta-\frac23\eta)=0$. For the coefficient of Eq.~(7.6) of \cite{ChaudhriNotes2012} with $\zeta=0$ the residual is $\frac83\eta\kB T(2+\sqrt3)\neq0$.
\end{itemize}
\item \texttt{check6\_\allowbreak primitive\_\allowbreak fdt.py}: the primitive-variable linearization \eqref{eq:dsf_lin} in $(\delta\rho,\delta c,\delta T,u)$, with $f$-derivatives, gradient and Onsager coefficients all symbolic, satisfies $\kB T_0(\bm H\bm A+\bm A^\dagger\bm H)+\bm H\bm N\bm H=0$ with $\bm S=\kB T_0\bm H^{-1}$ of \eqref{eq:dsf_B_S}.
\item \texttt{check7\_\allowbreak modes.py}: symbolic perturbation of the characteristic polynomial gives $c_T^2$, $c_s^2$, $\Gamma_s$ and $D$ of \eqref{eq:dsf_IN_modes} and \eqref{eq:dsf_NN_cs}, and the diffusive polynomial of model N-N.
\item \texttt{check8\_\allowbreak dsf\_\allowbreak numeric.py}: numerical checks of
\begin{itemize}[nosep]
\item the frequency sum rule $\int S\,\dd\omega/2\pi=\bm S(k)$ (error $10^{-14}$);
\item the eigenvalues against \eqref{eq:dsf_IN_modes}, \eqref{eq:dsf_NN_cs}, \eqref{eq:dsf_NN_Gs} and \eqref{eq:dsf_NN_diff};
\item the Landau--Placzek intensities \eqref{eq:dsf_IN_I} and those of model N-N;
\item the Rayleigh--Brillouin formula \eqref{eq:dsf_RB} against the exact spectrum.
\end{itemize}
\item \texttt{check9\_explicit\_A.py}: the matrix \eqref{eq:dsf_A} coincides with the symbolic linearization, and the characteristic-polynomial coefficients of model I-N are as stated.
\item \texttt{check10\_\allowbreak closed\_\allowbreak forms.py}: the closed forms \eqref{eq:dsf_IS_phi} and \eqref{eq:dsf_IS_rho} (including $S_{\rho u}$ and $F_{\rho\rho}$), the identity $I_c=(\partial\rho/\partial c)^2_{T,p}S_{cc}(0)$ and \eqref{eq:dsf_IN_cc}, against the reference implementation. Also checked are the equality of the two forms of \eqref{eq:dsf_S} and the discrete FDT for the face-based symbols and those of \cite{ChaudhriPRE2014}.
\item \texttt{check11\_\allowbreak helmholtz\_\allowbreak route.py}: thermodynamic route (\cref{app:thermo_route}), with $K=K^E-TK^S$ for both $\nabla\rho$ and $\nabla c$.
\begin{itemize}[nosep]
\item The route-B entropy balance \eqref{eq:C-Sigma_B} and the local Gibbs relation \eqref{eq:C-gibbs}.
\item The differences \eqref{eq:C-BA} between routes B and A.
\item The route-A entropy balance \eqref{eq:C-fluxes_A}--\eqref{eq:C-Sigma_A}, with $\sN$ depending on $\nabla\rho$.
\end{itemize}
\item \texttt{check12\_\allowbreak gibbs\_\allowbreak route.py}: the Maxwell relations and the expressions for $\hat c_p-\hat c_v$ and $\cs^2$ in \eqref{eq:C-dictionary}, and the pressure and temperature equations \eqref{eq:C-p_eq}--\eqref{eq:C-T_eq}.
\item \texttt{check13\_\allowbreak massieu\_\allowbreak route.py}: the relations \eqref{eq:C-MA}, and the route-M entropy balance \eqref{eq:C-fluxes_M} with no interstitial energy flux.
\item \texttt{check14\_\allowbreak vdw\_\allowbreak routes.py}: for one component with $\kappa(\rho,T)$:
\begin{itemize}[nosep]
\item route~M reproduces the reactive stress of Section~8.6 of \cite{ChaudhriNotes2012}, Onuki's stress and Onuki's $\hat\mu$;
\item route~B gives $\Lambda_{\rm B}=\partial_T\kappa$, and route~A has no $\nabla T$ term;
\item the differences are $\rho K^S\nabla\rho\cdot\nabla T$ and $\rho(K^E/T)\nabla\rho\cdot\nabla T$.
\end{itemize}
\item \texttt{check15\_helmholtz\_general\_T.py}: the route-B entropy balance and the Gibbs relation for gradient coefficients $K_{ab}(\rho,c,T)$ with \emph{arbitrary} temperature dependence.
\item \texttt{check16\_isothermal\_and\_QI.py}: the isothermal free-energy imbalance \eqref{eq:C-F_imbalance} for a general $\Psi(\rho,c,\nabla\rho,\nabla c)$, and the quasi-incompressible dissipation \eqref{eq:C-QI} with $\mut_{\rm QI}=\rho^{-1}\delta F/\delta c+p\,\dvh$.
\end{enumerate}
In addition, \texttt{dsf\_\allowbreak reference.py} implements the structure factors of \cref{sec:dsf}, and \texttt{make\_\allowbreak dsf\_\allowbreak figure.py} produces \cref{fig:dsf}. The file \texttt{verification\_\allowbreak output.txt} records the output of all checks.

\section{Remarks on discretization}\label{app:numerics}
These remarks follow the finite-volume framework of \cite{ChaudhriPRE2014}.
The conservative form \eqref{eq:U_noniso}--\eqref{eq:fluxes_noniso_final} is designed to fit the finite-volume, method-of-lines framework of \cite{ChaudhriPRE2014,Donev2010}.

\begin{enumerate}[leftmargin=*]
\item \textbf{Noise amplitudes.} A white-noise field is replaced by independent standard normal variables on the grid locations of the corresponding flux, multiplied by $\mathfrak S=1/\sqrt{\Delta V\Delta t}$. Thus
\begin{itemize}[nosep]
\item $\noise J_{i+1/2}=\sqrt{2\kB(MT)_{i+1/2}}\,\mathfrak S\,\mathcal W^{(c,x)}_{i+1/2}$;
\item $\noise J^{q\prime}_{i+1/2}=(TM_T/M)_{i+1/2}\noise J_{i+1/2}+\sqrt{2\kB(\lambda T^2)_{i+1/2}}\,\mathfrak S\,\mathcal W^{(E,x)}_{i+1/2}$;
\item $\noise\btau$ as in Section~3.1 of \cite{ChaudhriPRE2014}.
\end{itemize}
\item \textbf{Discrete FDT.} The discrete Lemma~\ref{lem:JO} holds if three conditions are met:
\begin{itemize}[nosep]
\item the discrete divergence applied to stochastic fluxes is the negative adjoint of the discrete gradient used in the deterministic fluxes ($\mathbf D=-\mathbf G^{*}$);
\item the face-averaged coefficients ($MT$, $\lambda T^2$, $\ldots$) are the same in both;
\item the composition equation is discretized for the conserved $\rho c$, with $\mut$ evaluated at cell centres and $\bJ=-M\mathbf G\mut$ on faces.
\end{itemize}
The factor $1/\rho$ of the continuum theory is then automatic.
\item \textbf{Capillary terms.} Two discretizations of the Korteweg stress are available.
\begin{itemize}[nosep]
\item The conservative tensor form $\divv\bPi^K$ conserves momentum exactly. This is the form used for $\kappa|\nabla\rho|^2$ in \cite{ChaudhriPRE2014}, Eq.~(3.8).
\item The potential form \eqref{eq:potential_form}, $-\rho\nabla\mut_\rho+\rho\mut\nabla c$, preserves the discrete free-energy identity $\int\bv\cdot(\ldots)=-\int\mut_\rho\,\partial_t\rho-\ldots$ and suppresses parasitic currents at equilibrium interfaces.
\end{itemize}
Either is consistent in the continuum limit. A discretization that preserves the discrete version of \eqref{eq:iso_force_identity} has both properties.
\item \textbf{Interstitial energy flux.} The term $\rho^{-1}(\divv\bJ)\,\kappa_E\nabla c$ in the energy flux contains the divergence of a noise field when $\bJ\to\bJ+\noise\bJ$. It is best handled by not discretizing \eqref{eq:ni_E} directly. Instead discretize \eqref{eq:eps_eq} for $\varepsilon$ with the term $(\mu_E/\rho)\divv(\bJ+\noise\bJ)$, and reconstruct $\rho E$. Both forms conserve the total energy in the continuum. Only the latter avoids differentiating white noise twice.
\item \textbf{Grid resolution and filtering.} As discussed in \cite{ChaudhriPRE2014}, near the critical point the noise amplitude per cell grows as $\Delta V^{-1/2}$. The coarse-graining length implicit in $F$ (the cell size in the Cahn--Hilliard sense) should not be smaller than the interface width. Filtering the stochastic fluxes \cite{Donev2014} separates the discretization length from the physical coarse-graining length.
\end{enumerate}

\section{The thermodynamic route: all models from the free energy and the entropy production}\label{app:thermo_route}
\subsection{Purpose and summary}\label{app:C-intro}
In the main text the models were derived with GENERIC. A large part of the diffuse-interface literature uses a more elementary route instead, that of classical irreversible thermodynamics (CIT) \cite{deGrootMazur}. Section~4.1 of \cite{ChaudhriNotes2012} is an example. The route has five steps:
\begin{enumerate}[nosep]
\item take a free energy, here $\Psi(\rho,c,T,\nabla\rho,\nabla c)$;
\item write the local Gibbs relation $\dd\varepsilon=\dd\Psi+T\dd s+s\dd T$;
\item insert the balance laws with unknown stress, heat and entropy fluxes;
\item rearrange the result into ``divergence $+$ entropy production'';
\item choose the unknowns so that the production is non-negative for every process.
\end{enumerate}
Section~8 of \cite{ChaudhriNotes2012} then writes down a fluctuating van der Waals--Navier--Stokes (vdW-NS) model with a Korteweg stress containing $T\rho\nabla\rho\cdot\nabla(\kappa/T)$.

This appendix carries the CIT procedure through for the \emph{most general} free energy of this paper:
\begin{itemize}[nosep]
\item compressible;
\item binary, with $c$ the mass fraction and $\rho$ the total density;
\item gradients of both $\rho$ and $c$;
\item temperature-dependent gradient coefficients;
\item Soret/Dufour cross-effects.
\end{itemize}
We ask which of the models of \cref{sec:isothermal,sec:noniso} it reproduces, and what role the Helmholtz and Gibbs free energies play.

\begin{resultbox}[Summary of the thermodynamic route]
\begin{enumerate}[leftmargin=*,nosep]
\item \textbf{It works.} The procedure fixes the reversible stress, the diffusion potential, the interstitial fluxes, the entropy flux and the form of all dissipative fluxes. It also fixes the noise through the Landau--Lifshitz prescription. It needs only:
\begin{itemize}[nosep]
\item a free energy $\Psi(\rho,c,T,\nabla\rho,\nabla c)$;
\item the balance laws;
\item the second law;
\item Curie's principle.
\end{itemize}
\item \textbf{Isothermal.} The Helmholtz route is complete and unique, and coincides with the GENERIC model of \cref{app:C-isothermal} (\cref{app:C-isothermal}).
\item \textbf{Non-isothermal.} The procedure has a gauge-like freedom. Terms of the form $(\text{gradient})\times\nabla T$ can be moved between the reversible stress or chemical potential and the energy or entropy fluxes without changing the sign of the entropy production (\cref{app:C-family}). Three natural choices are: hold $T$ fixed (route~B, Helmholtz), hold $1/T$ fixed (route~M, Massieu/Planck), or hold $\varepsilon$ fixed (route~A, entropy representation). Route~A is the GENERIC model of \cref{sec:noniso}.
\item \textbf{Linear level.} All routes have the same linearization. The dynamic structure factors of \cref{sec:dsf} are therefore route-independent and remain valid analytic benchmarks.
\item \textbf{Gibbs.} The Gibbs energy gives the $(p,T,c)$ form and the low-Mach and quasi-incompressible limits (\cref{app:C-gibbs}). It does not by itself give the capillary terms.
\end{enumerate}
\end{resultbox}

\paragraph{Notation and conventions.} These follow \cref{sec:organization}. We recall those used most here.
\begin{itemize}[leftmargin=*,nosep]
\item $\rho=\rho_A+\rho_B$ is the total mass density and $\bv$ the barycentric velocity.
\item $c=\rho_A/\rho$ is the mass fraction. $\bJ$ is the diffusion flux of species $A$ relative to $\bv$.
\item $D/Dt=\partial_t+\bv\cdot\nabla$, $\theta\equiv\divv\bv$ and $\cdt\equiv Dc/Dt$.
\item $\bD=(\nabla\bv)^{\sym}$, and $\mathring\bD$ is its traceless part.
\item A colon denotes full contraction, $\bm A:\bm B=A_{ij}B_{ij}$, and $(\nabla\bv)_{ij}=\partial_jv_i$.
\item Volumetric densities carry no hat. Specific (per-mass) quantities carry a hat or are written in lower case as in $f=\psi/\rho$.
\item The sign convention of Chaudhri \emph{et al.}\ \cite{ChaudhriPRE2014} is $\bPi=p\Id-\bPi^K$: $\bPi$ is the reversible (pressure-like) momentum flux and $\bPi^K$ the capillary stress.
\end{itemize}

\subsection{Balance laws with unknown fluxes}\label{app:C-balance}
The only kinematic input is the conservation of the two masses, momentum and energy:
\begin{align}
&\partial_t\rho+\divv(\rho\bv)=0, \label{eq:C-mass}\\
&\partial_t(\rho c)+\divv(\rho c\bv+\bJ)=0\quad\Longleftrightarrow\quad \rho\cdt=-\divv\bJ, \label{eq:C-species}\\
&\partial_t(\rho\bv)+\divv(\rho\bv\bv)=\divv\bm\sigma, \label{eq:C-momentum}\\
&\partial_t\mathcal E+\divv\big(\mathcal E\bv-\bm\sigma\cdot\bv+\bm q\big)=0,\qquad \mathcal E=\tfrac12\rho v^2+\etot, \label{eq:C-energy}\\
&\partial_t\stot+\divv\big(\stot\bv+\bJ_s\big)=\Sigma\ \ge0 . \label{eq:C-entropy}
\end{align}
The Cauchy stress $\bm\sigma$ (symmetric, by angular momentum), the non-convective energy flux $\bm q$, the entropy flux $\bJ_s$ and the entropy production $\Sigma$ are unknown. The diffusion flux $\bJ$ is an unknown constitutive flux. The energy density $\etot$ and the entropy density $\stot$ are fixed by the free energy (\cref{app:C-thermo}). The kinetic-energy theorem, $\partial_t(\tfrac12\rho v^2)+\divv(\tfrac12\rho v^2\bv)=\bv\cdot\divv\bm\sigma$, turns \eqref{eq:C-energy} into the internal-energy balance
\begin{equation}
\partial_t\etot+\divv(\etot\bv)=\bm\sigma:\nabla\bv-\divv\bm q .\label{eq:C-internal}
\end{equation}
The stochastic fluxes are added in \cref{app:C-noise}. Until then everything is deterministic.

\subsection{Local thermodynamics of a non-uniform mixture}\label{app:C-thermo}
\subsubsection{The Helmholtz free energy}
The constitutive input is a Helmholtz free-energy density (per volume)
\begin{equation}
\Psi(\rho,c,T,\nabla\rho,\nabla c)=\psi(\rho,c,T)+\psi_\nabla,\qquad \psi=\rho f(\rho,c,T),\label{eq:C-Psi}
\end{equation}
with a square-gradient part
\begin{equation}
\psi_\nabla=\tfrac12K_{\rho\rho}|\nabla\rho|^2+K_{\rho c}\nabla\rho\cdot\nabla c+\tfrac12K_{cc}|\nabla c|^2,\qquad K_{ab}=K_{ab}(\rho,c,T).\label{eq:C-psiN}
\end{equation}
Thermodynamics of the inhomogeneous state is \emph{defined} by $\Psi$:
\begin{equation}
\stot\equiv-\pd\Psi T,\qquad \etot\equiv\Psi+T\stot .\label{eq:C-s_e}
\end{equation}
\begin{itemize}[leftmargin=*,nosep]
\item \textbf{Bulk quantities.}
\[
p=\rho\psi_\rho-\psi=\rho^2f_\rho,\qquad \mu=\psi_c/\rho=f_c,\qquad \hg\equiv\psi_\rho=f+p/\rho,\qquad s=-\psi_T=\rho\hs .
\]
Here $\mu$ is the specific exchange chemical potential $\mu_A-\mu_B$ and $\hg$ the specific Gibbs energy. Gibbs--Duhem reads $\dd p=\rho\,\dd\hg-\rho\mu\,\dd c+s\,\dd T$.
\item \textbf{Energy--entropy split.} When the gradient coefficients are affine in $T$,
\begin{equation}
K_{ab}=K^E_{ab}-TK^S_{ab},\qquad \psi_\nabla=\eN-T\sN,\quad \eN=\tfrac12K^E_{ab}\nabla a\cdot\nabla b,\quad \sN=\tfrac12K^S_{ab}\nabla a\cdot\nabla b ,\label{eq:C-split}
\end{equation}
($a,b\in\{\rho,c\}$, summed). Then $\stot=s+\sN$ and $\etot=\varepsilon+\eN$, with $\varepsilon=\psi+Ts$. Gradients lower the entropy, so $K^S\le0$ typically. This is the split of \cref{sec:gradient} ($\kappa_E=K^E_{cc}$, $\kappa_S=K^S_{cc}$), but here $\sN$ may also depend on $\nabla\rho$. Onuki's vdW theory \cite{Onuki2007} has $K^E_{\rho\rho}=K$ and $K^S_{\rho\rho}=-C$.
\item \textbf{Two common choices.} The model of \cite{ChaudhriPRE2014} and of Espa\~nol--Thieulot \cite{EspanolThieulot2003} has $K^S=0$ (purely energetic gradients). The choice $K^E=0$ is purely entropic, $\kappa\propto T$.
\end{itemize}
Route~B below needs no split and holds for arbitrary $K_{ab}(\rho,c,T)$.

\subsubsection{Functional derivatives at fixed temperature field}
For a density $w(\rho,c,T,\nabla\rho,\nabla c)$ define
\begin{equation}
w_{\nabla\rho}\equiv\pd w{\nabla\rho},\qquad w_{\nabla c}\equiv\pd w{\nabla c},\qquad
\fd W\rho\Big|_{c,T}\equiv\pd w\rho-\divv w_{\nabla\rho},\qquad
\fd Wc\Big|_{\rho,T}\equiv\pd wc-\divv w_{\nabla c}. \label{eq:C-fd}
\end{equation}
The temperature is a given field $T(\br)$, so the divergences act on it as well: for example $\divv(K\nabla c)=K\nabla^2c+\nabla c\cdot(\partial_\rho K\nabla\rho+\partial_cK\nabla c+\partial_TK\nabla T)$. We abbreviate
\begin{equation}
\Psi_\rho\equiv\fd F\rho\Big|_{c,T},\qquad \Psi_c\equiv\fd Fc\Big|_{\rho,T},\qquad \mutB\equiv\frac{\Psi_c}{\rho},\qquad F[\rho,c;T]=\int\Psi\,\dd^3r .\label{eq:C-PsiR_PsiC}
\end{equation}
$\Psi_\rho$ is the local (generalized) specific Gibbs energy. $\mutB$ is the specific exchange potential of \cref{sec:isothermal}, $\rho^{-1}\delta F/\delta c|_\rho=\delta F/\delta\rho_A|_\rho$, now evaluated with $T$ inside the divergence.

\subsubsection{Two exact identities}
\begin{lemma}[Local Gibbs relation]\label{lem:C-gibbs}
For arbitrary time-dependent fields $\rho,c,T$:
\begin{equation}
\partial_t\etot=T\,\partial_t\stot+\Psi_\rho\,\partial_t\rho+\Psi_c\,\partial_tc+\divv\big(\Pgr\,\partial_t\rho+\Pgc\,\partial_tc\big).\label{eq:C-gibbs}
\end{equation}
The same holds with $\partial_t$ replaced by $D/Dt$, provided $\nabla(\partial_ta)$ is replaced by $D\nabla a/Dt=\nabla\dot a-(\nabla\bv)^T\cdot\nabla a$.
\end{lemma}
\begin{proof}
$\partial_t\etot=\partial_t\Psi+T\partial_t\stot+\stot\partial_tT$. The chain rule gives
\[
\partial_t\Psi=\psi_\rho^{\partial}\partial_t\rho+\psi_c^{\partial}\partial_tc-\stot\partial_tT+\Pgr\cdot\nabla\partial_t\rho+\Pgc\cdot\nabla\partial_tc,
\]
where $\Psi^\partial_a$ are partial derivatives. The $\partial_tT$ terms cancel. Integrating the last two terms by parts converts $\Psi^\partial_a$ into $\Psi_a$ of \eqref{eq:C-fd}. Check~11(b) and check~15 confirm this.
\end{proof}

\begin{lemma}[Korteweg identity at non-uniform temperature]\label{lem:C-korteweg}
Define the Helmholtz capillary tensor
\begin{equation}
\PiB\equiv\big[\rho\Psi_\rho-\Psi\big]\Id+\nabla\rho\otimes\Pgr+\nabla c\otimes\Pgc
= p\,\Id-\bPi^K_{\rm B}.\label{eq:C-PiB}
\end{equation}
Then
\begin{equation}
\divv\PiB=\rho\nabla\Psi_\rho-\Psi_c\nabla c+\stot\nabla T .\label{eq:C-korteweg}
\end{equation}
\end{lemma}
\begin{proof}
$\nabla\Psi=\Psi^\partial_\rho\nabla\rho+\Psi^\partial_c\nabla c-\stot\nabla T+(\nabla\nabla\rho)\cdot\Pgr+(\nabla\nabla c)\cdot\Pgc$. Also $\divv(\nabla a\otimes\Psi_{\nabla a})=(\nabla\nabla a)\cdot\Psi_{\nabla a}+\nabla a\,\divv\Psi_{\nabla a}$ and $\nabla(\rho\Psi_\rho)=\rho\nabla\Psi_\rho+\Psi_\rho\nabla\rho$. Adding, and using $\Psi_a=\Psi_a^\partial-\divv\Psi_{\nabla a}$, all second-derivative terms cancel and \eqref{eq:C-korteweg} remains.
\end{proof}
For uniform $T$ this is the Korteweg identity of \cref{lem:korteweg} with $w=\Psi$. The extra $\stot\nabla T$ is the thermal force on the mixture, the gradient analogue of $\nabla p=\rho\nabla\hg-\rho\mu\nabla c+s\nabla T$. Note that \eqref{eq:C-korteweg} immediately gives the ``Gibbs'' (potential) form of the reversible force:
\begin{equation}
-\divv\PiB=-\rho\nabla\Psi_\rho+\rho\mutB\nabla c-\stot\nabla T .\label{eq:C-potential_form}
\end{equation}

\subsection{Route B: Helmholtz free energy, temperature held fixed}\label{app:C-routeB}
This is the procedure of Section~4.1 of \cite{ChaudhriNotes2012}, carried out for the general $\Psi$.

\paragraph{Step 1: entropy rate from the Gibbs relation.} Use \cref{lem:C-gibbs} in material form, together with $D\rho/Dt=-\rho\theta$, $Dc/Dt=\cdt$ and $\Psi_{\nabla a}\cdot D\nabla a/Dt=\divv(\Psi_{\nabla a}\dot a)-\dot a\,\divv\Psi_{\nabla a}-(\nabla a\otimes\Psi_{\nabla a}):\nabla\bv$:
\[
T\frac{D\stot}{Dt}=\frac{D\etot}{Dt}+\rho\theta\,\Psi_\rho-\Psi_c\,\cdt+(\nabla\rho\otimes\Pgr+\nabla c\otimes\Pgc):\nabla\bv-\divv\big(-\rho\theta\Pgr+\cdt\,\Pgc\big).
\]
\paragraph{Step 2: insert the balance laws.} With \eqref{eq:C-internal} in material form, $D\etot/Dt=-\etot\theta+\bm\sigma:\nabla\bv-\divv\bm q$, and $\etot=\Psi+T\stot$:
\begin{equation}
T\big[\partial_t\stot+\divv(\stot\bv)\big]=(\bm\sigma+\PiB):\nabla\bv-\divv\hQ+\mutB\divv\bJ,\label{eq:C-TdS_B}
\end{equation}
where we have \emph{defined} the heat flux $\hQ$ by subtracting the interstitial working from $\bm q$:
\begin{equation}
\bm q=\hQ+\rho\theta\,\Pgr-\cdt\,\Pgc .\label{eq:C-q_B}
\end{equation}
We used $\rho\Psi_\rho\theta-\Psi\theta=(\rho\Psi_\rho-\Psi)\Id:\nabla\bv$ and $-\Psi_c\cdt=\mutB\divv\bJ$ (from \eqref{eq:C-species}).

\paragraph{Step 3: separate divergence and production.} Since $\mutB\divv\bJ=\divv(\mutB\bJ)-\bJ\cdot\nabla\mutB$ and $-\divv\hQ=-T\divv(\hQ/T)+T\hQ\cdot\nabla(1/T)$:
\begin{equation}
\boxed{\ \bJ_s=\frac{\hQ-\mutB\bJ}{T},\qquad
\Sigma=\frac{(\bm\sigma+\PiB):\nabla\bv}{T}+\hQ\cdot\nabla\frac1T-\bJ\cdot\nabla\frac{\mutB}{T}\ .}\label{eq:C-Sigma_B}
\end{equation}
This is checked symbolically in check~11(a) and, for arbitrary $K_{ab}(\rho,c,T)$, in check~15.

\paragraph{Step 4: reversible stress (Coleman--Noll argument).} The production must vanish for every rigid motion and for every process in which the dissipative mechanisms are switched off. Hence the part of $\bm\sigma$ that is not a dissipative response must cancel $\PiB$:
\begin{equation}
\bm\sigma=-\PiB+\btau ,\qquad \bm\sigma:\nabla\bv+\PiB:\nabla\bv=\btau:\nabla\bv .\label{eq:C-sigma}
\end{equation}
The reversible stress is the generalized Korteweg tensor of the \emph{Helmholtz} functional. Its isotropic part contains the thermodynamic pressure. By \eqref{eq:C-potential_form}, the reversible force is $-\rho\nabla\Psi_\rho+\rho\mutB\nabla c-\stot\nabla T$.

\paragraph{Step 5: flux--force pairs and Curie closure.} With the reduced heat flux $\hQp\equiv\hQ-\mutB\bJ$ the production is
\begin{equation}
\Sigma=\frac{\btau:\mathring\bD+\tfrac13\tr\btau\,\theta}{T}+\hQp\cdot\nabla\frac1T-\bJ\cdot\frac{\nabla\mutB}T\ \ge0 .\label{eq:C-Sigma_B2}
\end{equation}
The tensorial, scalar and vectorial forces do not couple in an isotropic fluid (Curie). The linear Onsager laws are
\begin{equation}
\btau=2\eta\mathring\bD+\zeta\theta\Id,\qquad
\bJ=-M\nabla\mutB-M_T\nabla T,\qquad
\hQp=-\hat\lambda\nabla T-TM_T\nabla\mutB,\label{eq:C-closure_B}
\end{equation}
with Onsager symmetry built in. They give
\[
T\Sigma=2\eta\mathring\bD:\mathring\bD+\zeta\theta^2+\frac{\hat\lambda}T|\nabla T|^2+2M_T\nabla T\cdot\nabla\mutB+M|\nabla\mutB|^2\ \ge0
\]
if and only if $\eta,\zeta,M\ge0$ and $\lambda\equiv\hat\lambda-TM_T^2/M\ge0$. These are the same coefficients and positivity conditions as in the GENERIC model, \eqref{eq:positivity}.

\begin{resultbox}[Route B: non-isothermal compressible CHNS from the Helmholtz free energy]
\begin{align}
&\partial_t\rho+\divv(\rho\bv)=0,\\
&\partial_t(\rho\bv)+\divv(\rho\bv\bv^T)+\nabla p=\divv\big(\bPi^K_{\rm B}+\btau\big),\\
&\partial_t(\rho c)+\divv(\rho c\bv)=-\divv\bJ,\qquad \bJ=-M\nabla\mutB-M_T\nabla T,\\
&\partial_t\mathcal E+\divv\big[(\mathcal E+p)\bv\big]=\divv\Big[\bPi^K_{\rm B}\cdot\bv+\btau\cdot\bv-\hQ-\rho\theta\,\Pgr-\frac{\divv\bJ}{\rho}\Pgc\Big],
\end{align}
with
\begin{itemize}[leftmargin=*,nosep]
\item $\mathcal E=\tfrac12\rho v^2+\Psi-T\partial_T\Psi$;
\item $\bPi^K_{\rm B}=p\Id-\PiB$ from \eqref{eq:C-PiB};
\item $\mutB=\rho^{-1}\delta F/\delta c|_{\rho,T}$;
\item $\hQ=\hQp+\mutB\bJ$ with $\hQp$ from \eqref{eq:C-closure_B}.
\end{itemize}
\end{resultbox}
\begin{itemize}[leftmargin=*,nosep]
\item The interstitial flux $\cdt\,\Pgc=-(\divv\bJ/\rho)\Pgc$ is the term $-\rho\Phi\,\dd c/\dd t$ of Eq.~(4.17) of \cite{ChaudhriNotes2012} with $\rho\Phi=\Pgc$.
\item The heat-flux law $\hQp=-\lambda\nabla T$ is Eq.~(4.17) there (with $M_T=0$).
\item The entropy flux $\bJ_s=\hQp/T$ is Eq.~(4.18) there, rewritten.
\end{itemize}
In other words, Section~4.1 of \cite{ChaudhriNotes2012} is route~B for $\Psi=\rho f_0(c)+\frac\kappa2|\nabla c|^2$ (they use $f$ independent of $T$). In that case $\sN=0$, and all routes coincide (\cref{app:C-compare}).

\subsection{Route A: entropy representation (the GENERIC model)}\label{app:C-routeA}
Now treat the gradient energy as energy and the gradient entropy as entropy. Assume the split \eqref{eq:C-split}. Write the Gibbs relation for the \emph{bulk} part, $T\dd s=\dd\varepsilon-\hg\,\dd\rho-\rho\mu\,\dd c$ with $\varepsilon=\etot-\eN$, and differentiate $\eN$ and $\sN$ separately with their own functional derivatives. Then $T$ multiplies $\delta\SN$ but never enters a divergence. Repeating Steps~1--3 gives (check~11(e)):
\begin{align}
&\bm q=\bJq+\rho\theta\,\pd\eN{\nabla\rho}-\cdt\,\pd\eN{\nabla c},\qquad
\bJ_s=\frac{\bJq-\mutA\bJ}{T}+\rho\theta\,\pd\sN{\nabla\rho}-\cdt\,\pd\sN{\nabla c},\label{eq:C-fluxes_A}\\
&\Sigma=\frac{(\bm\sigma+\PiA):\nabla\bv}{T}+\bJq\cdot\nabla\frac1T-\bJ\cdot\nabla\frac{\mutA}{T},\label{eq:C-Sigma_A}
\end{align}
with
\begin{gather}
\mutA=\mu+\frac1\rho\Big(\fd\EN c-T\fd\SN c\Big),\qquad
\PiA=p\Id+\bPi^{\eN}-T\,\bPi^{\sN},\label{eq:C-PiA}\\
\bPi^{w}=\Big[\rho\fd W\rho-w\Big]\Id+\nabla\rho\otimes w_{\nabla\rho}+\nabla c\otimes w_{\nabla c}.\nonumber
\end{gather}
Here the functional derivatives of $\EN$ and $\SN$ do \emph{not} see $T$. The closure is identical to \eqref{eq:C-sigma}--\eqref{eq:C-closure_B} with $(\hQ,\mutB,\PiB)\to(\bJq,\mutA,\PiA)$.

For $K^S_{\rho\rho}=K^S_{\rho c}=0$ these are exactly the equations of \cref{sec:noniso}:
\begin{itemize}[nosep]
\item \eqref{eq:PiS}--\eqref{eq:Pi_total}: $\bPi=p\Id+\bPi^{\eN}-T\bPi^{\sN}$;
\item \eqref{eq:forces_noniso}: $\mut$;
\item \eqref{eq:E_local_full}--\eqref{eq:Sigma}: the energy and entropy fluxes and the production.
\end{itemize}
These are the equations obtained in \cref{sec:noniso} from the GENERIC building blocks. At the CIT level a $\nabla\rho$-dependence of $\sN$ is allowed and check~11(e) includes it. In the GENERIC derivation it was excluded because the Poisson operator of \cref{sec:noniso_poisson} then violates the Jacobi identity (\cref{app:jacobi}).

\subsection{Route M: Massieu (Planck) potential, $1/T$ held fixed}\label{app:C-routeM}
The entropy-representation analogue of the Helmholtz function is the Massieu--Planck density $-\Psi/T$. It is natural in the variables $(\rho,c,\beta=1/T)$, with $\partial(\beta\Psi)/\partial\beta=\etot$. The local Gibbs relation reads $\partial_t\stot=\beta\partial_t\etot-\delta(\beta F)/\delta\rho\,\partial_t\rho-\dots$. It is \cref{lem:C-gibbs} with $\Psi_a$ replaced by $T\,\delta(\int\beta\Psi)/\delta a|_{\beta}$, and with the divergence term divided by $T$ \emph{inside} the divergence. Define
\begin{equation}
\mutM\equiv\frac T\rho\fd{}{c}\int\frac\Psi T\,\dd^3r\Big|_{\rho,1/T},\qquad
\PiM\equiv T\,\bPi^{\Psi/T}=\Big[\rho T\fd{}{\rho}\int\frac\Psi T\Big|_{c,1/T}-\Psi\Big]\Id+\nabla\rho\otimes\Pgr+\nabla c\otimes\Pgc .\label{eq:C-PiM}
\end{equation}
Because $T\divv(\bm X/T)=\divv\bm X-\bm X\cdot\nabla T/T$, these are related to the Helmholtz quantities by
\begin{equation}
\mutM=\mutB+\frac{\Pgc\cdot\nabla T}{\rho T},\qquad \PiM=\PiB+\rho\,\frac{\Pgr\cdot\nabla T}{T}\,\Id .\label{eq:C-M_vs_B}
\end{equation}
The same Steps~1--3, with the divergence now kept inside the factor $1/T$, give (check~13):
\begin{equation}
\boxed{\begin{gathered}\bm q=\hQ,\qquad \bJ_s=\frac{\hQ-\mutM\bJ}{T}-\frac{\rho\theta\,\Pgr-\cdt\,\Pgc}{T},\\
\Sigma=\frac{(\bm\sigma+\PiM):\nabla\bv}{T}+\hQ\cdot\nabla\frac1T-\bJ\cdot\nabla\frac{\mutM}T\end{gathered}}\label{eq:C-fluxes_M}
\end{equation}
The interstitial working has moved \emph{entirely} from the energy flux into the entropy flux. The closure is again \eqref{eq:C-sigma}--\eqref{eq:C-closure_B} with $(\mutB,\PiB)\to(\mutM,\PiM)$.

\subsection{Comparison of the routes and the source of non-uniqueness}\label{app:C-compare}
\begin{proposition}[Route differences]\label{prop:C-diff}
With the split \eqref{eq:C-split} (checks~11(c,d), 13(a,b)):
\begin{align}
\mutB-\mutA&=\frac{1}{\rho}\pd\sN{\nabla c}\cdot\nabla T, &
\PiB-\PiA&=\rho\Big(\pd\sN{\nabla\rho}\cdot\nabla T\Big)\Id, \label{eq:C-BA}\\
\mutM-\mutA&=\frac{1}{\rho T}\pd\eN{\nabla c}\cdot\nabla T, &
\PiM-\PiA&=\frac{\rho}{T}\Big(\pd\eN{\nabla\rho}\cdot\nabla T\Big)\Id . \label{eq:C-MA}
\end{align}
Consequently:
\begin{enumerate}[nosep]
\item $\text{A}=\text{B}$ if $\sN=0$, i.e.\ purely energetic gradient terms (temperature-independent $K_{ab}$). This is the case in \cite{ChaudhriPRE2014}, \cite{EspanolThieulot2003} and Section~4.1 of \cite{ChaudhriNotes2012}.
\item $\text{A}=\text{M}$ if $\eN=0$, i.e.\ purely entropic gradient terms ($K_{ab}\propto T$).
\item $\text{A}=\text{B}=\text{M}$ whenever $\nabla T=0$, in particular in all isothermal models.
\item All differences are bilinear in $\nabla T$ and $\nabla\rho$ or $\nabla c$. The same is true of the differences in the energy flux: $\theta\nabla\rho$ and $\cdt\nabla c$ in the interstitial terms. Hence \emph{the three routes have identical linearizations about any uniform equilibrium state.}
\end{enumerate}
\end{proposition}
\begin{proof}
For \eqref{eq:C-BA}, write $\Psi=\psi+\eN-T\sN$. Then $\divv\Psi_{\nabla c}=\divv\partial_{\nabla c}\eN-T\divv\partial_{\nabla c}\sN-\partial_{\nabla c}\sN\cdot\nabla T$. The first two terms are those of route~A. Since $\Psi_c$ contains $-\divv\Psi_{\nabla c}$, the last term contributes $+\partial_{\nabla c}\sN\cdot\nabla T=\rho(\mutB-\mutA)$. The $\rho$-derivative is treated the same way and enters $\PiB$ through $\rho\Psi_\rho$. Then \eqref{eq:C-MA} follows from \eqref{eq:C-M_vs_B} and $\Psi_{\nabla a}=\partial_{\nabla a}\eN-T\partial_{\nabla a}\sN$.
\end{proof}

\subsubsection{Why pure CIT cannot decide}\label{app:C-family}
Let $\bm\chi_\rho$, $\bm\chi_c$ be any vector fields built from the gradients. Shift the reversible stress and the exchange potential by
\[
\delta\bPi=\rho(\bm\chi_\rho\cdot\nabla T)\Id,\qquad \delta\mut=\frac{\bm\chi_c\cdot\nabla T}{\rho},
\]
and set
\[
\bm b=\rho\theta\bm\chi_\rho-\cdt\,\bm\chi_c .
\]
Shift the energy flux by $-T\bm b$ and the entropy flux by $-\bm b$. Then $\bm b\cdot\nabla T$ exactly compensates the extra power $-\delta\bPi:\nabla\bv$ and the extra diffusive work $\cdt\,\bm\chi_c\cdot\nabla T$. As a result the production $\Sigma$ keeps the \emph{same} bilinear form in the shifted forces. This is the algebra behind \eqref{eq:C-fluxes_M}: route~M is route~B shifted by $\bm\chi_a=\Psi_{\nabla a}/T$.

Every member of this family satisfies all requirements of CIT:
\begin{itemize}[nosep]
\item mass, momentum and energy are conserved;
\item there is a local entropy balance with $\Sigma\ge0$;
\item Onsager symmetry and Curie's principle hold.
\end{itemize}
The freedom is the classical non-uniqueness of the entropy flux and of the ``heat flux'' in non-local thermodynamics. Neither is fixed by the balance laws, a point made by M\"uller \cite{Muller1967}, Dunn--Serrin \cite{DunnSerrin1985} and Heida--M\'alek--Rajagopal \cite{HeidaMalekRajagopal2012}.

\begin{notebox}[What selects route A]
GENERIC adds two structural requirements that CIT does not have.
\begin{itemize}[leftmargin=*,nosep]
\item The reversible dynamics must be generated by the energy through a Poisson operator satisfying the Jacobi identity. It must also conserve the entropy functional $S=\int(s+\sN)$ \emph{as a functional of the state}, through degeneracy, independently of how heat flows.
\item The dissipative part must be generated by the entropy.
\end{itemize}
Routes B and M put temperature-gradient forces into the reversible stress that cannot come from an energy-generated Hamiltonian flow. Route~A is selected. This is why the GENERIC derivation of \cref{sec:noniso} produced exactly route~A. In practice:
\begin{itemize}[leftmargin=*,nosep]
\item For isothermal simulations, and for any model with temperature-independent gradient coefficients, the question does not arise.
\item For non-isothermal models with $\kappa=\kappa(T)$, route~A is recommended. The others are admissible alternatives that differ at nonlinear order in $\nabla T$.
\end{itemize}
\end{notebox}

\subsection{Isothermal models: the free-energy imbalance}\label{app:C-isothermal}
At uniform, constant $T=T_0$ the second law reduces to a free-energy imbalance. Because $\nabla T=0$, all routes coincide, so the result is unique. Combine \cref{lem:C-gibbs} (with $\partial_tT=0$, so $\partial_t\Psi=\partial_t\etot-T_0\partial_t\stot$), the kinetic-energy theorem, \eqref{eq:C-mass}, \eqref{eq:C-species} and \cref{lem:C-korteweg}:
\begin{equation}
\boxed{\ \partial_t\big(\tfrac12\rho v^2+\Psi\big)+\divv\Big[\big(\tfrac12\rho v^2+\Psi\big)\bv-\bm\sigma\cdot\bv+\rho\theta\,\Pgr-\cdt\,\Pgc+\mut\bJ\Big]=-(\bm\sigma+\bPi):\nabla\bv+\bJ\cdot\nabla\mut\ }\label{eq:C-F_imbalance}
\end{equation}
with $\bPi=\PiB$ and $\mut=\rho^{-1}\delta F/\delta c|_\rho$ at $T_0$. The second law in the isothermal setting is $\mathcal D\equiv T_0\Sigma=-\text{RHS}\ge0$. As in Step~4, this gives $\bm\sigma=-\bPi+\btau$, and then
\begin{equation}
\mathcal D=\btau:\nabla\bv-\bJ\cdot\nabla\mut\ \ge0\quad\Longrightarrow\quad \btau=2\eta\mathring\bD+\zeta\theta\Id,\qquad \bJ=-M\nabla\mut .\label{eq:C-D_iso}
\end{equation}
Check~16(a) verifies \eqref{eq:C-F_imbalance} for general $\Psi(\rho,c,\nabla\rho,\nabla c)$.

\begin{resultbox}[Isothermal compressible LLCHNS from the Helmholtz free energy (identical to the GENERIC result)]
\begin{align*}
&\partial_t\rho+\divv(\rho\bv)=0,\\
&\partial_t(\rho\bv)+\divv(\rho\bv\bv^T)+\nabla p=\divv\big(\bPi^K+\btau+\noise\btau\big),\qquad \bPi^K=p\Id-\bPi,\\
&\partial_t(\rho c)+\divv(\rho c\bv)=\divv\big(M\nabla\mut\big)-\divv\noise\bJ,\qquad \mut=\frac1\rho\fd Fc\Big|_\rho,\\
&\avg{\noise\bJ(\br,t)\noise\bJ(\br',t')}=2\kB T_0M\,\Id\,\delta(\br-\br')\delta(t-t'),\\
&\avg{\noise\tau_{ij}\noise\tau_{kl}}=2\kB T_0\big[\eta(\delta_{ik}\delta_{jl}+\delta_{il}\delta_{jk})+(\zeta-\tfrac23\eta)\delta_{ij}\delta_{kl}\big]\delta(\br-\br')\delta(t-t') .
\end{align*}
This is \cref{res:iso_final}, i.e.\ the conservative form of \cite{ChaudhriPRE2014}, Eq.~(3.1), with the corrections discussed in \cref{sec:previous}.
\end{resultbox}

\begin{notebox}[What the free-energy imbalance says about $c$ versus $\rho$]
In \eqref{eq:C-F_imbalance} the composition enters through $\Psi_c\,\cdt=\rho\mut\,\cdt=-\mut\divv\bJ$. This term is the only place where the composition dynamics meets the free energy. The force conjugate to $\bJ$ is therefore $-\nabla\mut$, the gradient of the \emph{specific} potential $\rho^{-1}\delta F/\delta c=\delta F/\delta\rho_A|_\rho$.

The volumetric $\mu^C=\delta F/\delta c=\rho\mut$ used in Eq.~(4.14) of \cite{ChaudhriNotes2012} would give $\mathcal D\ni M\nabla(\rho\mut)\cdot\nabla\mut$, which is not sign-definite when $\rho$ varies. This is the CIT counterpart of the failure of the FDT found in \cref{sec:iso_linear} and in check~4 of \cref{app:verification}. The thermodynamic route thus settles the ``$c$ versus $\rho$'' question independently of GENERIC:
\begin{itemize}[nosep]
\item $\rho$ is the total density;
\item $c$ is the mass fraction;
\item $\rho c$ is the conserved partial density;
\item the driving potential is specific.
\end{itemize}
\end{notebox}

\paragraph{Symmetric and non-symmetric mixtures.} The derivation used no property of $\Psi$ other than its existence. The four models of the main text follow by specializing $\Psi$ (see \cref{tab:C-models}).
\begin{itemize}[leftmargin=*,nosep]
\item \textbf{Symmetric} (models I-S and N-S). Use $\phi=2c-1$ and $\psi$, $K_{ab}$ invariant under $\phi\to-\phi$. In the decoupled case, $\psi=\rho f_0(\rho,T)+\rho g(\phi,T)$ and $\psi_\nabla=\frac12\kappa_\phi|\nabla\phi|^2$.
\item \textbf{Non-symmetric} (models I-N and N-N). Use a general $\psi(\rho,c,T)$, for example the binary van der Waals mixture \eqref{eq:vdw_mix}, together with all three $K_{ab}$.
\end{itemize}
For $\psi_\nabla=\frac12\kappa|\nabla c|^2$ the capillary stress is $\bPi^K=\frac12(\kappa-\rho\partial_\rho\kappa)|\nabla c|^2\Id-\kappa\nabla c\otimes\nabla c$. For $\psi_\nabla=\frac12\kappa|\nabla\rho|^2$ it is $\bPi^K=(\kappa\rho\nabla^2\rho+\frac12\kappa|\nabla\rho|^2)\Id-\kappa\nabla\rho\otimes\nabla\rho$, which is Eq.~(2.14) of \cite{ChaudhriPRE2014}.

\subsection{The Gibbs free energy}\label{app:C-gibbs}
\subsubsection{Gibbs as a change of variables for the bulk}
The specific Gibbs energy $\hg(p,T,c)=f+p/\rho$ is the Legendre transform of $f(\rho,c,T)$ in $1/\rho$:
\begin{equation}
\dd\hg=-\hs\,\dd T+\frac{\dd p}{\rho}+\mu\,\dd c .\label{eq:C-dg}
\end{equation}
It is a pointwise transform. It exists wherever $p_\rho\equiv(\partial p/\partial\rho)_{T,c}>0$, so it is not single-valued inside the vdW spinodal. The gradient part cannot be Legendre-transformed pointwise, because $\psi_\nabla$ depends on $\nabla\rho$, not on $\rho$. Writing $\nabla\rho=\rho(\beta_T\nabla p-\alpha_p\nabla T+\gamma_c\nabla c)$ is legitimate, but it produces a ``free energy'' depending on $\nabla p$ and $\nabla T$, which is not a local-equilibrium state function. The practical rule is therefore:
\begin{notebox}[Rule]
Compute $\bPi^K$, $\mut$ and the interstitial fluxes from the Helmholtz functional (\cref{app:C-routeB,app:C-routeA,app:C-isothermal}). Then express the \emph{bulk} thermodynamics in Gibbs variables. The resulting equations are identical; only the primitive variables change.
\end{notebox}
In one sense the Gibbs energy is already present in the Helmholtz route. $\Psi_\rho=\delta F/\delta\rho|_c$ is the generalized specific Gibbs energy, and \eqref{eq:C-potential_form} is the Gibbs--Duhem form of the route-B momentum equation, $\rho\,D\bv/Dt=-\rho\nabla\Psi_\rho+\rho\mut\nabla c-\stot\nabla T+\divv\btau$.

\subsubsection{Dictionary of Gibbs coefficients}
From $f(\rho,c,T)$, with subscripts denoting partial derivatives at fixed remaining $(\rho,c,T)$:
\begin{equation}
\begin{gathered}
\alpha_p=-\frac1\rho\Big(\pd\rho T\Big)_{p,c}=\frac{p_T}{\rho p_\rho},\qquad
\beta_T=\frac1\rho\Big(\pd\rho p\Big)_{T,c}=\frac1{\rho p_\rho},\qquad
\gamma_c=\frac1\rho\Big(\pd\rho c\Big)_{T,p}=-\frac{p_c}{\rho p_\rho},\\
\hat c_v=T\Big(\pd\hs T\Big)_{\rho,c},\qquad \hat c_p=\hat c_v+\frac{Tp_T^2}{\rho^2p_\rho},\qquad
\cs^2=\Big(\pd p\rho\Big)_{\hs,c}=p_\rho+\frac{Tp_T^2}{\rho^2\hat c_v},\\
\hs_{c,p}\equiv\Big(\pd\hs c\Big)_{T,p}=-\Big(\pd\mu T\Big)_{c,p},\qquad
\dvh\equiv\Big(\pd{(1/\rho)}c\Big)_{T,p}=\Big(\pd\mu p\Big)_{T,c}=-\frac{\gamma_c}\rho,\qquad
\mu_{c,p}\equiv\Big(\pd\mu c\Big)_{T,p}.
\end{gathered}\label{eq:C-dictionary}
\end{equation}
The two Maxwell relations and the expressions for $\hat c_p-\hat c_v$ and $\cs^2$ are verified in check~12. The partial specific enthalpy difference is $\hat h_c\equiv(\partial\hat h/\partial c)_{T,p}=\mu+T\hs_{c,p}$.

\subsubsection{Forces in Gibbs variables: barodiffusion and thermodiffusion}
Split $\mut=\mu(p,T,c)+\mut_\nabla$, where $\mut_\nabla$ is the gradient part. By \eqref{eq:C-dg},
\begin{equation}
\nabla\mut=\mu_{c,p}\nabla c+\dvh\,\nabla p-\hs_{c,p}\nabla T+\nabla\mut_\nabla .\label{eq:C-grad_mu_gibbs}
\end{equation}
The diffusion flux of \eqref{eq:C-closure_B} becomes the Landau--Lifshitz form with barodiffusion:
\begin{gather}
\bJ=-\rho D\Big[\nabla c+\frac{k_T}{T}\nabla T+\frac{k_p}{p}\nabla p\Big]-M\nabla\mut_\nabla,\nonumber\\
\rho D=M\mu_{c,p},\quad k_T=\frac{T(M_T/M-\hs_{c,p})}{\mu_{c,p}},\quad k_p=\frac{p\,\dvh}{\mu_{c,p}} .\label{eq:C-J_gibbs}
\end{gather}
This agrees with \eqref{eq:LL_form}. The barodiffusion ratio $k_p$ vanishes for a symmetric mixture at $\phi=0$ and for equal partial specific volumes.

\subsubsection{Pressure--temperature (primitive) form}
Take $\sN=0$, so that all routes coincide. Let $\mathcal Q\equiv\btau:\nabla\bv-\divv\hQp-\bJ\cdot\nabla\mut$ be the heating rate per unit volume. The entropy balance becomes $\rho T\,D\hs/Dt=\mathcal Q$, and $\rho\,Dc/Dt=-\divv\bJ$. Eliminating $\rho$ and $\hs$ in favour of $p$ and $T$ gives (check~12):
\begin{align}
\frac{Dp}{Dt}&=-\rho\cs^2\theta+\cs^2\Big[\frac{\alpha_p}{\hat c_p}\big(\mathcal Q+T\hs_{c,p}\divv\bJ\big)+\gamma_c\divv\bJ\Big],\label{eq:C-p_eq}\\
\rho\hat c_p\frac{DT}{Dt}&=\mathcal Q+T\alpha_p\frac{Dp}{Dt}+T\hs_{c,p}\divv\bJ ,\label{eq:C-T_eq}\\
\rho\frac{Dc}{Dt}&=-\divv\bJ,\qquad \rho\frac{D\bv}{Dt}=-\nabla p+\divv\big(\bPi^K+\btau\big).\label{eq:C-cv_eq}
\end{align}
When $\mut_\nabla=0$, \eqref{eq:C-T_eq} is the textbook mixture temperature equation \cite{deGrootMazur}:
\[
\rho\hat c_pDT/Dt=-\divv\bJq+\hat h_c\divv\bJ+T\alpha_pDp/Dt+\btau:\nabla\bv .
\]
With gradient entropy ($\sN\neq0$, route~A), add $T\bPi^{\sN}:\nabla\bv$ and $T(\mu_S/\rho)\divv\bJ$ to $\mathcal Q$, as in \eqref{eq:T_eq}.

\subsubsection{Low-Mach limit}\label{app:C-lowmach}
For flows slow compared with sound, split $p=P_0(t)+\pi$ with $\pi=O(\mathrm{Ma}^2)$. Evaluate all thermodynamics at $(P_0,T,c)$ and retain $\nabla\pi$ only in the momentum equation. Since $\nabla P_0=0$, \eqref{eq:C-p_eq} turns into a constraint on the velocity divergence:
\begin{equation}
\divv\bv=\frac1\rho\Big[\frac{\alpha_p}{\hat c_p}\big(\mathcal Q+T\hs_{c,p}\divv\bJ\big)+\gamma_c\divv\bJ\Big]-\frac{1}{\rho\cs^2}\frac{\dd P_0}{\dd t},\qquad
\rho\frac{D\bv}{Dt}=-\nabla\pi+\divv\big(\bPi^K+\btau\big).\label{eq:C-lowmach}
\end{equation}
In the isothermal case with $\dd P_0/\dd t=0$ this is $\divv\bv=\gamma_c\divv\bJ/\rho=-\dvh\,\divv\bJ$. With the ideal-volume mixing rule $1/\rho=c/\bar\rho_A+(1-c)/\bar\rho_B$ we have $\dvh=\bar\rho_A^{-1}-\bar\rho_B^{-1}$, which is the constraint of Donev \emph{et al.}\ \cite{Donev2014}. The Gibbs representation is the natural one here: the thermodynamic pressure is a parameter, not a field.

The low-Mach reduction applies to binary-mixture (CH) phase separation at fixed total pressure. It does not apply to liquid--vapour (vdW) phase separation, where $p_\rho<0$ inside the interface and the Gibbs transform is multivalued.

\subsubsection{Quasi-incompressible (Lowengrub--Truskinovsky) limit}\label{app:C-QI}
If both components are incompressible, $\rho=\rho(c)$ exactly, and the pressure becomes a Lagrange multiplier for $\divv\bv=-\dvh(c)\divv\bJ$. Take $\Psi=\rho(c)\hat f(c,T_0)+\psi_\nabla(c,\nabla c)$ and $\bm\sigma=-p\Id-\bPi^c+\btau$ with $\bPi^c=\nabla c\otimes\Pgc-\Psi\Id$. The free-energy imbalance (check~16(b)) gives
\begin{equation}
\mathcal D=\btau:\nabla\bv-\bJ\cdot\nabla\mut_{\rm QI},\qquad \mut_{\rm QI}=\frac1\rho\fd Fc+p\,\dvh .\label{eq:C-QI}
\end{equation}
The pressure enters the chemical potential exactly as the Gibbs energy $\hg=\hat f+p/\rho(c)$ predicts: $\mu=(\partial\hg/\partial c)_{T,p}=\hat f_c+p\,\dvh$. This is the quasi-incompressible model of \cite{LowengrubTruskinovsky1998}. In this limit the Gibbs energy \emph{is} the complete thermodynamic potential, because the gradient energy depends only on $\nabla c$.

\subsection{Fluctuations: noise from the entropy production}\label{app:C-noise}
\subsubsection{Landau--Lifshitz prescription}
In every route the entropy production has the bilinear form $\Sigma=\sum_\alpha\bm X_\alpha\cdot\bm F_\alpha$, with linear laws $\bm X_\alpha=\sum_\beta L_{\alpha\beta}\bm F_\beta$. The Landau--Lifshitz--Fox--Uhlenbeck prescription \cite{LandauLifshitzFM,deZarateSengers2006} adds a random flux $\noise{\bm X}_\alpha$ to each dissipative flux, with
\begin{equation}
\avg{\noise X_{\alpha}(\br,t)\,\noise X_{\beta}(\br',t')}=2\kB L_{\alpha\beta}\,\delta(\br-\br')\delta(t-t') .\label{eq:C-LL}
\end{equation}
For $(\btau,\bJ,\hQp)$ with forces $(\nabla\bv/T,-\nabla\mut/T,\nabla(1/T))$, the Onsager matrix of \eqref{eq:C-closure_B} gives
\begin{equation}
\begin{aligned}
\avg{\noise\tau_{ij}\noise\tau_{kl}}&=2\kB T\big[\eta(\delta_{ik}\delta_{jl}+\delta_{il}\delta_{jk})+(\zeta-\tfrac23\eta)\delta_{ij}\delta_{kl}\big]\delta\delta,\\
\avg{\noise J_i\noise J_j}&=2\kB TM\,\delta_{ij}\delta\delta,\qquad
\avg{\noise J_i\,\noise Q'_j}=2\kB T^2M_T\,\delta_{ij}\delta\delta,\qquad
\avg{\noise Q'_i\,\noise Q'_j}=2\kB T^2\hat\lambda\,\delta_{ij}\delta\delta .
\end{aligned}\label{eq:C-noise_cov}
\end{equation}
These are exactly the covariances \eqref{eq:noise_noniso}, obtained in \cref{sec:noniso_fdt} from the Jeli\'c--\"Ottinger construction. The noises enter wherever the deterministic fluxes enter.
\begin{itemize}[leftmargin=*,nosep]
\item $\noise\btau$ enters the momentum and energy equations.
\item $\noise\bJ$ enters the species equation, and also the energy flux through the interstitial term $\cdt\Pgc$ (routes A, B), since $\cdt$ contains $-\divv\noise\bJ/\rho$.
\item $\noise{\hQ}=\noise{\hQ}'+\mut\noise\bJ$ enters the energy flux.
\end{itemize}
Total energy is then conserved pathwise.

\subsubsection{Why this is the correct noise}
\begin{itemize}[leftmargin=*]
\item \textbf{Linear FDT (all routes).} Linearize about a uniform equilibrium $\bx_0$. By \cref{prop:C-diff}(4) all routes give the GENERIC linearization
\[
\dd\delta\bx=(\Lop_0-\Mop_0/T_0)H\delta\bx\,\dd t+\Bop_0\dd\bW,\qquad
H=\partial^2_{\bx}(E-T_0S-\textstyle\sum\lambda_IN_I),\qquad \Bop_0\Bop_0^\dagger=2\kB\Mop_0,
\]
with $\Mop_0=\Phiop^*L\Phiop$ built from \eqref{eq:C-LL}. Its stationary covariance is the Einstein result $\kB T_0H^{-1}$ (\cref{app:linear_fdt}). Equivalently, the static structure factors are those of the Gibbs--Boltzmann distribution $\propto\exp(-F[\delta\rho,\delta c]/\kB T_0)$ in the isothermal case. The dynamic structure factors, Landau--Placzek ratios and Rayleigh--Brillouin spectra of \cref{sec:dsf} therefore hold unchanged for every route. The symbolic check~6 of \cref{app:verification} (the primitive-variable linear FDT) covers all routes at once.
\item \textbf{Nonlinear level.} For the stationary measure $\exp(S/\kB)$ of the \emph{nonlinear} stochastic equations one needs, in addition:
\begin{itemize}[nosep]
\item a reversible drift that conserves $S$ and is phase-space-volume preserving (degeneracy and Jacobi);
\item the It\^o drift $\kB\partial_{\bx}\cdot\Mop$.
\end{itemize}
Route~A has this structure (GENERIC). For routes B and M it is not guaranteed by construction. This is a further, statistical-mechanical, argument for route~A in non-isothermal fluctuating simulations.
\end{itemize}

\subsection{All the models from the single thermodynamic idea}\label{app:C-models}
\subsubsection{Recipe}
\begin{enumerate}[leftmargin=*,nosep]
\item Choose $\Psi(\rho,c,T,\nabla\rho,\nabla c)$, i.e.\ $f(\rho,c,T)$ and $K_{ab}(\rho,c,T)$.
\item In the isothermal case, compute $\bPi^K=p\Id-\bPi$ and $\mut=\rho^{-1}\delta F/\delta c$ from \eqref{eq:C-PiB}. Otherwise choose the route; route~A is recommended. It is irrelevant if $K_{ab}$ is $T$-independent.
\item Close with Onsager coefficients $\eta,\zeta,M,M_T,\hat\lambda$ satisfying the positivity conditions.
\item Add noise with covariance $2\kB\times$ Onsager \eqref{eq:C-noise_cov}.
\item Optionally, change to Gibbs variables for the pressure--temperature form, the low-Mach form or the quasi-\allowbreak incompressible form.
\end{enumerate}

\begin{table}[H]\centering\small
\caption{Models obtained from the recipe. ``$=$GENERIC'' means identical to the corresponding GENERIC model of the main text.}\label{tab:C-models}
\renewcommand{\arraystretch}{1.25}
\begin{tabular}{@{}>{\raggedright\arraybackslash}p{3.2cm}>{\raggedright\arraybackslash}p{5.4cm}>{\raggedright\arraybackslash}p{6.6cm}@{}}
\toprule
Model & Free energy input & Outcome of the thermodynamic route\\
\midrule
I-S (isothermal, symmetric) & $\psi=\rho f_0(\rho)+\rho g(\phi)$, $g$ even; $\tfrac12\kappa_\phi|\nabla\phi|^2$ & Unique; $=$GENERIC; $\mut_\phi=g'-\rho^{-1}\divv(\kappa_\phi\nabla\phi)$\\
I-N (isothermal, non-symmetric) & general $\psi(\rho,c)$, $K_{\rho\rho},K_{\rho c},K_{cc}$ & Unique; $=$GENERIC\\
N-S, N-N with $T$-independent $K$ & $\psi(\rho,c,T)$, $K^S=0$ & Unique (A$=$B$=$M); $=$GENERIC\\
N-S, N-N with $K(T)$ & $K=K^E-TK^S$ & Route A $=$GENERIC; B, M differ at $O(\nabla T\,\nabla\rho,\nabla T\,\nabla c)$\\
vdW-NS, isothermal & $\rho f(\rho)+\tfrac12\kappa|\nabla\rho|^2$ & Unique; Korteweg stress Eq.~(2.14) of \cite{ChaudhriPRE2014}\\
vdW-NS, non-isothermal & $\rho f(\rho,T)+\tfrac12\kappa(\rho,T)|\nabla\rho|^2$ & \cite{ChaudhriNotes2012} Sec.~8.6 / Onuki $=$ route M; see \cref{app:C-vdw}\\
Low-Mach & Gibbs $\hg(P_0,T,c)$ & \eqref{eq:C-lowmach}; $\divv\bv=-\dvh\divv\bJ$ (isothermal)\\
Quasi-incompressible & $\rho=\rho(c)$, $\hg=\hat f+p/\rho(c)$ & \eqref{eq:C-QI}; Lowengrub--Truskinovsky\\
\bottomrule
\end{tabular}
\end{table}

\subsubsection{The fluctuating van der Waals--Navier--Stokes model}\label{app:C-vdw}
For a one-component fluid, $\Psi=\rho f(\rho,T)+\frac12\kappa(\rho,T)|\nabla\rho|^2$ with $\kappa=K^E-TK^S$. Check~14 gives the three reversible stresses in the sign convention $\bPi=p\Id-\bPi^K$:
\begin{equation}
\bPi^K_{X}=\Big[\kappa\rho\nabla^2\rho+\tfrac12\big(\kappa+\rho\partial_\rho\kappa\big)|\nabla\rho|^2+\rho\,\Lambda_X\,\nabla\rho\cdot\nabla T\Big]\Id-\kappa\nabla\rho\otimes\nabla\rho,\label{eq:C-vdw_PiK}
\end{equation}
\[
\Lambda_{\rm A}=0,\qquad \Lambda_{\rm B}=\partial_T\kappa=-K^S,\qquad \Lambda_{\rm M}=T\partial_T(\kappa/T)=-K^E/T .
\]
\begin{itemize}[leftmargin=*,nosep]
\item \textbf{Route M is the form of \cite{ChaudhriNotes2012}, Section~8.6, and of Onuki.} With $\Lambda_{\rm M}$, \eqref{eq:C-vdw_PiK} is exactly
\[
\Pi_{\rm notes}=\big[\kappa\rho\nabla^2\rho+\tfrac12(\kappa+\kappa'\rho)|\nabla\rho|^2+T\rho\nabla\rho\cdot\nabla(\kappa/T)\big]\Id-\kappa\nabla\rho\otimes\nabla\rho
\]
of Section~8.6 of \cite{ChaudhriNotes2012}. Here $\nabla(\kappa/T)$ means $\partial_T(\kappa/T)\nabla T$; the $\kappa'$-term is written separately. It is also the stress of Onuki's dynamic vdW theory \cite{Onuki2005,Onuki2007} (Eq.~12 of \cite{Onuki2005}) with $\kappa=M_{\rm Onuki}=CT+K$. The chemical potential is $\mutM=\partial_\rho\psi+\tfrac12\partial_\rho\kappa|\nabla\rho|^2-T\divv[(\kappa/T)\nabla\rho]$, which is Onuki's $\hat\mu$ (Eq.~6 there). Consistently with \eqref{eq:C-fluxes_M}, the energy flux contains \emph{no} interstitial term, again as in \cite{Onuki2007} and Eq.~(8.22) of the notes.
\item \textbf{Required energy density.} Thermodynamic consistency requires the total energy density
\[
\rho E=\tfrac12\rho v^2+\rho u(\rho,T)+\tfrac12K^E|\nabla\rho|^2,\qquad K^E=\kappa-T\partial_T\kappa .
\]
The energy density $\rho E=\frac12\rho v^2+\hat u$ of \cite{ChaudhriNotes2012}, Eq.~(8.22), omits the gradient energy. That is correct only for a purely entropic coefficient, $\kappa=CT$, for which routes A and M coincide and $\nabla(\kappa/T)=0$. For a constant $\kappa$ (purely energetic) the energy must contain $\frac12\kappa|\nabla\rho|^2$. Then routes A and B coincide and give
\begin{align*}
&\partial_t(\rho\bv)+\divv(\rho\bv\bv^T+p\Id)=\divv\big(\bPi^K+\btau+\noise\btau\big),\qquad \bPi^K=\big(\kappa\rho\nabla^2\rho+\tfrac12\kappa|\nabla\rho|^2\big)\Id-\kappa\nabla\rho\otimes\nabla\rho,\\
&\partial_t(\rho E)+\divv\big[(\rho E+p)\bv\big]=\divv\big[\bPi^K\cdot\bv-\rho\kappa\theta\nabla\rho+(\btau+\noise\btau)\cdot\bv+\lambda\nabla T+\noise\bQ\big],
\end{align*}
with $\avg{\noise\bQ\noise\bQ}=2\kB\lambda T^2\delta\delta$. Route~M for the same constant $\kappa$ has the extra isotropic stress $-\rho(\kappa/T)\nabla\rho\cdot\nabla T$ in $\bPi^K$ and drops the term $-\rho\kappa\theta\nabla\rho$. Both are consistent, and they are equal at linear order.
\item \textbf{Isothermal.} All routes give the Korteweg stress of \cite{ChaudhriPRE2014}, Eq.~(2.14).
\end{itemize}

\subsection{Conclusions of the thermodynamic route}\label{app:C-conclusions}
\begin{enumerate}[leftmargin=*]
\item \textbf{Yes, for every model.} The Helmholtz free energy, the balance laws, the second law and Curie's principle produce all the isothermal and non-isothermal, symmetric and non-symmetric, CH and vdW models of the main text. The Landau--Lifshitz rule (noise $=2\kB\times$ Onsager) supplies the fluctuations. The derivation is short and needs no Poisson brackets.
\item \textbf{The isothermal theory is unique} and identical to GENERIC. It also settles the question of $c$ versus $\rho$ independently: the diffusion potential is $\rho^{-1}\delta F/\delta c$.
\item \textbf{The non-isothermal theory is unique up to a gauge.} The gauge is a family of $\nabla T$-dependent capillary terms that can be traded against energy or entropy fluxes. Three natural members exist: Helmholtz (B), Massieu (M, $=$ Section~8.6 of \cite{ChaudhriNotes2012} and Onuki) and entropy representation (A, $=$ GENERIC). They coincide unless the gradient coefficients depend on temperature and $\nabla T\neq0$, and they always share the same linearization. GENERIC (Jacobi, degeneracy, nonlinear FDT) selects route~A.
\item \textbf{The Gibbs free energy} is the right potential for the pressure-based form, for the low-Mach model and for the quasi-incompressible model. It must be combined with capillary terms computed from the Helmholtz functional.
\item \textbf{For the finite-volume benchmark} nothing changes. The linearized equations, the static covariance $\kB T_0H^{-1}$ and all dynamic structure factors of \cref{sec:dsf} are route-independent. Any of the routes can be discretized and tested against them. The route-dependent terms are nonlinear and can be tested separately, for example by the energy and entropy budgets of a planar interface in a weak imposed temperature gradient.
\end{enumerate}

\clearpage
\phantomsection

\end{document}